\documentclass[a4paper]{elsarticle}
\usepackage{array}
\usepackage{bm}
\usepackage{color}
\usepackage{graphicx}
\usepackage{amsmath,amssymb,amstext}
\numberwithin{equation}{section}   
\usepackage{multirow}
\usepackage{rotating}
\usepackage{float}
\usepackage{mathptmx}
\usepackage[numbers]{natbib}
\usepackage{soul,xcolor} % Strike out the text \st{text}
\setul{-.6ex}{.3ex} 
\setstcolor{red}
\def\bra#1{\bigl\langle{ #1} \bigr|}
\def\ket#1{\bigl|{ #1} \bigr\rangle}

\def\ovlp#1#2{\bigl\langle{ #1}\big|{#2} \bigr\rangle}

\def\rvec {{\bf r}}
\def\pvec {{\bf p}}

\def\hvec {{\bf h}}
\def\qvec {{\bf q}}

\def\kvec {{\bf k}}
\def\lvec {{\bf l}}
\def\mvec {{\bf m}}
\def\nvec {{\bf n}}
\def\NO {{\left\langle \hat N \right\rangle_0}}

\def\creat#1{a_{#1}^{\dagger}}                    % Creation
\def\annil#1{a_{#1}^{\phantom{\dagger}}}           % Annihilation
\def\he#1{$^{#1}$He}
\def\EF{e_{\rm F}}
\def\mcf{m c_{\rm F}^2}
\def\KF{k_{\rm F}}
\def\SF{S_{\rm F}}
\def\a0{a_0}
\def\I{{\rm i}}

\def\etal{{\em et al.\/}\ }
\def\ie{{\em i.e.\/}\ }
\def\cf{{\em cf.\/}\ }

\DeclareMathOperator{\sgn}{sgn}

\usepackage{tikz} %TikZ can produce portable graphics in both PDF and PostScript formats using either plain (pdf)TEX, (pdf)Latex or ConTEXt.

\usetikzlibrary{positioning}
\usetikzlibrary{decorations.text}
\usetikzlibrary{decorations.pathmorphing}
\usetikzlibrary{decorations.markings}
\usetikzlibrary{shapes.misc}
\tikzset{->-/.style={decoration={
  markings,
  mark=at position .5 with {\arrow[scale=1.2,black]{>}}},postaction={decorate}}}
\tikzset{-->--/.style={decoration={
  markings,
  mark=at position .4 with {\arrow[scale=1.4,black]{>}}},postaction={decorate}}}

\tikzset{cross/.style={cross out, draw=black, fill=none, minimum size=2*(#1-\pgflinewidth), inner sep=0pt, outer sep=0pt}, cross/.default={2pt}}

\begin{document}
\begin{frontmatter}
  
\title{Variational and Parquet-diagram theory for strongly correlated
  normal and superfluid systems}

\author{H.-H. Fan$^{\dagger}$ and E.~Krotscheck$^{\dagger\ddagger}$}
%\email[]{eckhardk@buffalo.edu}
%\homepage[]{Your web page}
%\thanks{}

\address{$\dagger$Department of Physics, University at Buffalo, SUNY
Buffalo NY 14260}
\address{$^\ddagger$Institut f\"ur Theoretische Physik, Johannes
Kepler Universit\"at, A 4040 Linz, Austria}

\begin{abstract}

We develop the variational and correlated basis
functions/parquet-diagram theory of strongly interacting normal and
superfluid systems. The first part of this contribution is devoted to
highlight the connections between the Euler equations for the
Jastrow-Feenberg wave function on the one hand side, and the ring,
ladder, and self-energy diagrams of parquet-diagram theory on the
other side. We will show that these subsets of Feynman diagrams are
contained, in a local approximation, in the variational wave function.

In the second part of this work, we derive the fully optimized
Fermi-Hypernetted Chain (FHNC-EL) equations for a superfluid system.
Close examination of the procedure reveals that the na\"ive
application of these equations exhibits spurious unphysical properties
for even an infinitesimal superfluid gap. We will conclude that it is
essential to go {\em beyond\/} the usual Jastrow-Feenberg
approximation and to include the exact particle-hole propagator to
guarantee a physically meaningful theory and the correct stability
range.

We will then implement this method and apply it to neutron matter and
low density Fermi liquids interacting via the Lennard-Jones model
interaction and the P\"oschl-Teller interaction. While the
quantitative changes in the magnitude of the superfluid gap are
relatively small, we see a significant difference between
applications for neutron matter and the Lennard-Jones and P\"oschl-Teller
systems. Despite the fact that the gap in neutron matter can be as
large as half the Fermi energy, the corrections to the gap
are relatively small. In the Lennard-Jones and P\"oschl-Teller models,
the most visible consequence of the self-consistent calculation
is the change in stability range of the system.

\end{abstract}

\ead{eckhardk@buffalo.edu}

\end{frontmatter}
\newpage
\tableofcontents
\newpage

\section{Introduction}
\label{sec:intro}

The repertoire of methods for the quantitative microscopic description
of {\em normal\/} quantum many-body systems has condensed, over the
past few decades, to a relatively small number of techniques. These
can be roughly classified on the one hand side as various shades of
large-scale numerical simulation methods and, on the other hand,
diagrammatic approaches summing, in some approximation, the parquet
class of Feynman diagrams. Numerical simulations are capable of high
precision but are computationally demanding and limited to relatively
simple Hamiltonians and mostly ground state properties. They also need
the physical intuition of the user about the possible state of the
system. Semi-analytic, diagrammatic approaches lead to a better
understanding of the underlying physical mechanisms, but have limited
accuracy due to some necessary approximations. These diagrammatic
approaches are versions of Green's function methods
\cite{FetterWalecka}, Coupled Cluster theory \cite{KLZ}, and
variational methods \cite{FeenbergBook}. The interconnections between
the various methods are well understood for Bose systems
\cite{parquet1,BishopValencia}; the level and the details of
implementation for different systems varies, however, vastly.

For normal systems, Jackson \etal make compelling arguments
\cite{parquet1} that the summation of the so-called parquet diagrams
is a {\em minimum requirement\/} for a microscopic treatment of the
many-body problem that that treats both the short-ranged structure and
the long-wavelengths excitations on equal footings. We will review
these arguments further below. There are presently basically two
theoretical approaches that have the diagrammatic completeness of the
parquet diagrams, these are the Jastrow-Feenberg variational method
\cite{FeenbergBook} and the local parquet-diagram summation of
Refs. \citenum{parquet1} and \citenum{parquet2}. These methods have
led, for boson systems, to exactly the same equations.

The situation is also intuitively clear for fermions, although
technically more complicated due to the multitude of exchange diagrams
generated by the antisymmetry of the fermion wave function,
Unfortunately the fermion version \cite{FermiParquet} has so far not
led to practical applications.

Among the many-body methods that sum, in some approximation, the
parquet class of diagrams, the Jastrow-Feenberg method has been
developed farthest.  Both, local parquet theory and the
Jastrow-Feenberg method are ``robust'' in the sense that exactly the
same equations can be used for very different interactions like
electrons, nucleons, and quantum fluids.  Coupled Cluster theory
\cite{KLZ} has also been very successful for electrons
\cite{BiL78,Bishop} and nuclear systems \cite{Day78,Day81,DayZab81}
but it requires different truncation schemes for these two classes of
many-body system. It lacks, therefore, the robustness of the
Jastrow-Feenberg method. It was also less successful in predicting the
ground state properties of the helium fluids.  For bosons, a version
of coupled cluster theory - the so-called ``super-SUB-2''
approximation has been developed \cite{BishopValencia} that is
equivalent to the local parquet or Jastrow-Feenberg theory.

All of the above statements refer to {\em normal\/} systems. However,
{\em pairing phenomena\/} are ubiquitous in the physics of many-body
systems. Sixty years ago, the proof of the Cooper theorem
\cite{BCS,BCS50book} provided the key to understanding the pairing
phenomenon in interacting many-fermion systems, triggering the
creation of the BCS theory of electronic superconductivity. It was
quickly understood \cite{BMP} that this work also has significant
implications for nuclear phenomena, basically explaining the energy
gap between the ground state and the first excited state for certain
classes of nuclei. Since that landmark development, the basic paradigm
of BCS theory has been extended to diverse fermionic systems with
considerable success. We refer to two recent excellent review articles
\cite{SchuckBCS2018,SedrakianClarkBCSReview} for a very complete
account of the present situation and a comprehensive survey of the
relevant literature.  The work to be presented in this paper is meant
to be complementary to these papers. We will, as far as
justifiable, avoid overlaps. Rather, we shall focus on the technical
aspects of many-body theory and spend very little space reviewing and
comparing specific calculations.

A theory for superfluid many-body systems of the same diagrammatic
completeness that was achieved for normal systems is presently
unavailable, although specific partial summations of the perturbation
series have been carried out
\cite{PhysRevA.74.042717,Steiner2009,TohyamaSchuck2015,Vitali2017}.  A
version of Coupled Cluster theory for BCS-type wave functions has also
been developed \cite{PhysRevB.30.2049}.

In systems where the pairing is due to the underlying many-body
Hamiltonian, one often relies on effective interaction approximations
which have the useful feature to permit the examination of mechanism
and dependencies, but come with all the uncertainties involved in
constructing effective interactions.

One of the intentions of our paper is therefore to develop a theory
superfluid systems that is diagrammatically, as far as justified by
the problems at hand, equivalent to the Jastrow-Feenberg or parquet
theory for normal systems.

Our paper is organized as follows: In the following section, we will
first give a pedagogical review of the motivations behind the
correlated wave function method and the parquet-diagram theory. We
will begin with the optimized hypernetted chain (HNC-EL)
method for bosons which has proven to be the preferred systematic
method of summing infinite classes of cluster diagrams.

As mentioned above, the boson version of the theory is identical to
the summation of local parquet diagrams or to a specific version of
coupled cluster theory. A priori, all two-particle vertices in a
Feynman-diagram based theory are functions of four energy/momentum
variables. Energy and momentum conservation and isotropy reduce that
to ten variables which is still too much for diagram summation
methods. {\em Local\/} parquet theory then introduces specific
approximations to make all vertices functions of the momentum transfer
only, and derives a procedure to replace the energy dependence by an
average energy.

The situation is much more complicated for fermions for two reasons:
One is that the antisymmetry requirement for the wave function leads
to a multitude of exchange diagrams, the other is that the Fermi sea
provides a natural frame of reference, whereas the Jastrow-Feenberg
theory makes the approximation that all correlations depend only on
the distance between particles. That requires futher approximations.

We will here examine what approximations must be made to go
from a specific set of Feynman diagrams to a corresponding set of
Jastrow-Feenberg diagrams.  It will turn out that exactly the same
procedure of defining an average energy that has been established for
Bose systems \cite{parquet1} can be carried over to Fermi systems. The
existence of a preferred frame of reference will require an additional
Fermi-sea averaging procedure to generate vertices that depend only on
the momentum transfer. We will see that {\em exactly the same\/}
procedures for defining an average energy and an average momentum
apply in the three channels, particle-hole, particle-particle, and
single-particle propagators.

Realizing the relevant correspondences we will be led to formulate a
hybrid theory that has the same diagrammatic content as the
variational theory but avoids some important approximations.

We then develop the generalization of the Jastrow-Feenberg method for
superfluids. We derive the diagrammatic expansions, carry out the FHNC
summations for a correlated superfluid state, and derive the Euler
equation for optimizing the correlations. By examining the Euler
equation, we will unveil a severe problem of the Jastrow-Feenberg wave
function: We will demonstrate that the Euler equation for the pair
distribution function displays spurious instabilities for
net-attractive interactions, characterized by an attractive Landau
parameter $F_0^s < 0$. We will demonstrate that these problems are
caused by the so-called ``collective'' or ``single-pole''
approximation \cite{Rip79,polish} for the particle-hole propagator,
which is implicit to the Jastrow-Feenberg wave function and will be
discussed at length in Section \ref{sec:parquet}. We are therefore
lead to conclude that {\em the Jastrow-Feenberg wave function for a
  superfluid system does not permit a sensible optimization of the
  pair correlations.}  Hence, one must go beyond the simple
Jastrow-Feenberg method and to implement the correlated-basis
functions (CBF) or parquet-diagram theory for superfluid states. We
will show that the instabilities are then removed.

The need for developing CBF/parquet-diagram methods for the superfluid
state has also a quantitative reason.  We have argued\, --\ and
demonstrated\ --\ many years ago that local correlations of the kind
(\ref{eq:Jastrow}) are inadequate to deal with pairing phenomena. The
qualitative explanation for that is quite simple: local correlation
functions treat all particles within the Fermi sea in the same
way. BCS-pairing occurs at the Fermi surface; correlations that are
independent of the location of the particle within the Fermi seas
should therefore be particularly poor to describe pairing
phenomena. {\em Quantitative\/} evidence for this was provided in our
neutron matter calculations of Refs.  \citenum{shores} and
\citenum{CCKS86}. This is another reason that one must go {\em
  beyond\/} the Jastrow-Feenberg theory to deal with pairing phenomena
reliably.

In Secs. \ref{ssec:NeutronMatter}, \ref{ssec:LJium}, and
\ref{ssec:ptpot}, we apply our theoretical methods to a few physically
interesting cases: model Fermi gases at low densities, and neutron
matter. With that, we follow up on previous works
\cite{HNCBCS,CBFPairing} which was partly motivated by the interest in
the BCS-BEC crossover in cold Fermi gases (see
Refs. \citenum{duinePhysRep04} and \citenum{chenPhysRep05} for review
articles) and superfluidity in neutron matter (see
Ref. \citenum{ECTSI} for a collection of review papers). We have
recently examined pairing phenomena in both model Fermi systems
\cite{cbcs} and neutron matter \cite{ectpaper}.  We extend these
calculations, which have assumed a small superfluid gap and treated
the BCS correlations perturbatively, within the much more advanced
theory to be developed in this paper.

In applications to neutron matter, we demonstrate that the inclusion
of the full superfluid propagators in both the density and the
spin-channel have a rather visible consequence for the superfluid gap.

The second case to be discussed are many-particle systems interacting
via a family of Lennard-Jones model interaction. The attractive
Lennard-Jones liquid has a more interesting phase diagram than neutron
matter since it can have two spinodal points at which the speed of
sound vanishes. One spinodal point appears at negative pressure at
about 60 percent of the equilibrium density. This point can be reached
by gradually lowering the density; it is characterized by the fact
that the Fermi liquid Landau parameter $F_0^s\searrow -1$.  A second
spinodal point appears at very low density when the attractive
interaction begins to dominate over the Pauli pressure. The appearance
of this instability is obvious from the equation of state, it has
already been observed by Owen \cite{OwenVar}.

We have studied BCS pairing for the Lennard-Jones interactions
extensively in Ref. \citenum{cbcs} in an approximation that assumed
that the BCS correlations are weak. Going beyond this approximation,
we face the aforementioned spurious instabilities of locally
correlated wave functions which have a drastic effect in the
Lennard-Jones liquid: In the whole density regime where the weakly
coupled theory predicted a superfluid transition, FHNC-EL equations
for local correlations have no physically acceptable solutions. We
solve this problem by including the correct superfluid particle-hole
propagator which has a quite visible quantitative effect on both the
phase diagram and pairing properties.

Computations in the vicinity of the spinodal points become very
demanding since the correlations become very long-ranged. In
Ref. \citenum{cbcs}, we have observed that it is rather easy to come
close to the upper spinodal point. However, getting close to the lower
spinodal point turned out to be impossible since the solutions to the
FHNC-EL equations diverge already a distance from the limit
$F_0^s\searrow -1$.  This divergence was identified as a divergence of
the {\em in-medium scattering length\/}. We find exactly the same
property in our much more advanced calculations to be presented in
this paper. In fact, our inclusion of exchange diagrams, which
improves the predictions of the Fermi-liquid parameter $F_0^s$
significantly, hardly changes the location of the instability.

Computations for the purely attractive P\"oschl-Teller potential,
which has due to the absence of a repulsive hard core no stable
high-density phase and only the lower spinodal point confirm our
conclusions.

We conclude this paper with a brief summary of our results.

\section{Motivation: Variational and local parquet theory for bosons}

\subsection{Rationalization of parquet diagram summations}

To describe the physics in the interaction-dominated short-distance
region within diagrammatic perturbation theory, short-ranged
correlations must be dealt with properly. These are treated, in
perturbation theory, by summing the ladder diagrams which determine,
among others, the pair distribution function $g(r)$ at small
distances. The description of generally long-ranged effects, in
particular phonons or plasmons and the behavior of the static
structure function $S(q)$ at long wavelengths $q$, requires the
summation of chain diagrams. The simultaneously correct treatment of
both short- and long-ranged effects requires, therefore, the
self-consistent summation of ring- and ladder-diagrams which defines
the set of parquet diagrams \cite{parquet1,parquet2,parquet3}.

The resulting two-body vertices are still functions of three
independent four-mo\-men\-ta. To make execution of the theory
practical, approximations must be made. {\em Local parquet theory\/}
localizes the vertices by choosing an average energy such that the
contribution to the pair distribution of the full vertex is the same
as the that of the localized one. That way, the iterative procedure
is, in every step, connected to a physical observable. Moreover, we
shall see below that the pair distribution function can indeed be
considered the only necessary independent variable that determines the
properties of the system.

Carrying out this procedure for bosons, it turns out that one arrives
at a set of equations that had been known for many years
\cite{FeenbergBook,PPA1,Woo70,PPA2}, namely the HNC-EL equations
\cite{parquet2} (dubbed ``Paired-Phonon Analysis at that time) and the
HNC-EL energy functional \cite{parquet3}. In fact, the analogy goes
farther in the sense that the inclusion of the leading {\em
  non-parquet\/} diagrams \cite{TripletParquet} is the same as the
inclusion of optimized three-body correlations in the wave function
(\ref{eq:Jastrow}) \cite{bosegas}.

The situation is more complicated for fermions, mostly due to the
multitude of additional diagrams. Of course, since the correspondence
between parquet diagrams and the HNC-EL method is clear for bosons, a
similar correspondence should be expected for fermions.

In the nexy section we shall demonstrate the analogy between fermion
parquet theory and FHNC-EL for important classes of diagrams, namely
the ring- and ladder-diagrams. This is the essence of the simplest
version of FHNC-EL, referred to FHNC-EL//0 \cite{Johnreview}. More
complete implementations \cite{polish} also include RPA-exchange
diagrams, self-energy corrections, and mixture of all of these.  These
versions are referred to as FHNC//n where n is the level of
higher-order exchange diagrams retained. As we shall see, the
prescription of making all vertices energy-independent by choosing a
well-defined average energy can be carried over from the boson parquet
theory. The existence of a preferred reference frame, the Fermi sea,
still causes the simplest vertices depend on three momenta; turning
these into functions of momentum transfer will require the
introduction of an additional specific averaging procedure of
single-particle energies over the occupied states in the Fermi sea.

Both of these localizing procedures might seem {\em ad-hoc\/} from the
point of view of conventional perturbation theory, and without further
consideration other procedures might look equally well justified.
They are rationalized by the fact that these localization
prescriptions lead to the Jastrow-Feenberg wave function
(\ref{eq:Jastrow}). The optimization prescription (\ref{eq:optu2})
makes sure that one has, that way, constructed the best wave function
that can be represented in terms of local functions. In fact, a generalization
of the Hohenberg-Kohn theorem to two-body functions \cite{PairDFT,lowdens}
shows that the pair distribution  $g(r)$ is indeed quite generally
defined by a variational problem.

The Jastrow-Feenberg wave function is known to reproduce the
properties of both helium fluids with better than 90 percent accuracy
\cite{JordiQFSBook}; below about 25 percent of the helium saturation
density the accuracy of the FHNC//0 approximation is better than 1
percent, and Coulomb systems, both bosons \cite{bosegas} and fermions
\cite{annals} are generally reproduced at the percent level or better.

\subsection{Hypernetted Chain and Euler equations}
\label{ssec:bhnc}

We choose the Jastrow-Feenberg approach here to derive what we shall
refer to as ``generic'' many-body method because it requires
relatively little formal input.  It is suitable for a non-relativistic
many-body Hamiltonian
\begin{equation}
H = -\sum_{i}\frac{\hbar^2}{2m}\nabla_i^2 + \sum_{i<j}
v(i,j)\,.
\label{eq:Hamiltonian}
\end{equation}

The method starts with an {\em ansatz\/} for the $N$-body wave
function,
\begin{eqnarray}
\ket{\Psi_{\bf o}^{(N)}} &=& \frac{1}{\sqrt{I^{(N)}_{\bf o}}}
        F_N({\bf r}_1,\ldots,{\bf r}_N)\ket{{\bf o}}
        \label{eq:wavefunction},\\
F_N({\bf r}_1,\ldots,{\bf r}_N) &=& \exp\frac{1}{2}
\left[\sum_{i<j}  u_2({\bf r}_i,{\bf r}_j) + \cdot\cdot +
\sum_{i_1<\ldots<i_n}u_n({\bf r}_{i_1},.., {\bf r}_{i_n})
+ \cdot\cdot \right]\,,
\label{eq:Jastrow}
\end{eqnarray}
where ${I_{\bf o}} = \bra{{\bf o}} F_N^\dagger F_N \ket{{\bf o}}$ is
the normalization constant. Here $\ket{{\bf o}}$ is a model state,
which is normally a Slater determinant for normal Fermi systems, and
$F_N$ is an $N$-body correlation operator. The explicit reference to
the particle number $N$ will be necessary later when we generalize the
method to BCS states, we shall omit it from now on for the normal
system which has a fixed particle number.

When truncated at the two-body term $u_2(\rvec_i,\rvec_j)$,
Eq. (\ref{eq:Jastrow}) defines the standard Jastrow theory.
Historically \cite{Jastrow55} the theory was developed as a ``quick
and dirty'' way to deal with the strong, short-ranged forces prevalent
in nuclei. The task of the function $f(r) =
\exp\left(\frac{1}{2}u_2(r)\right)$ is to bend the wave function to
zero inside the regime of the repulsive hard core. The energy
expectation value
\begin{equation}
  H_{{\bf o}} = \bra{\Psi_{\bf o}} H \ket{\Psi_{\bf o}}\label{eq:Hoo}
\end{equation}
and other physically interesting quantities like the pair density
\begin{equation}
  \rho_2(\rvec,\rvec') =
      \bra{\Psi_0}\sum_{i\ne j}\delta(\rvec_i-\rvec)\delta(\rvec_j-\rvec')
        \ket{\Psi_0}\,,
      \label{eq:rho2}
\end{equation}
the pair distribution function
\begin{equation}
  g(\rvec,\rvec') = \frac{\rho_2(\rvec,\rvec')}{\rho^2}\,,
  \label{eq:gofr}
  \end{equation}
and the static structure function
\begin{equation}
	S(k) = 1 + \rho\int d^3r
		e^{\I{\bf k}\cdot{\bf r}}\left[g(r)-1\right],
\label{eq:sofk}
\end{equation}
are then calculated by cluster expansion and resummation techniques.
We will deal in this paper exclusively with translationally invariant
and isotropic systems, hence $g(\rvec,\rvec') = g(|\rvec-\rvec'|)$.

\begin{figure}[H]
      \centerline{
        \includegraphics[width=0.65\columnwidth,angle=-90]{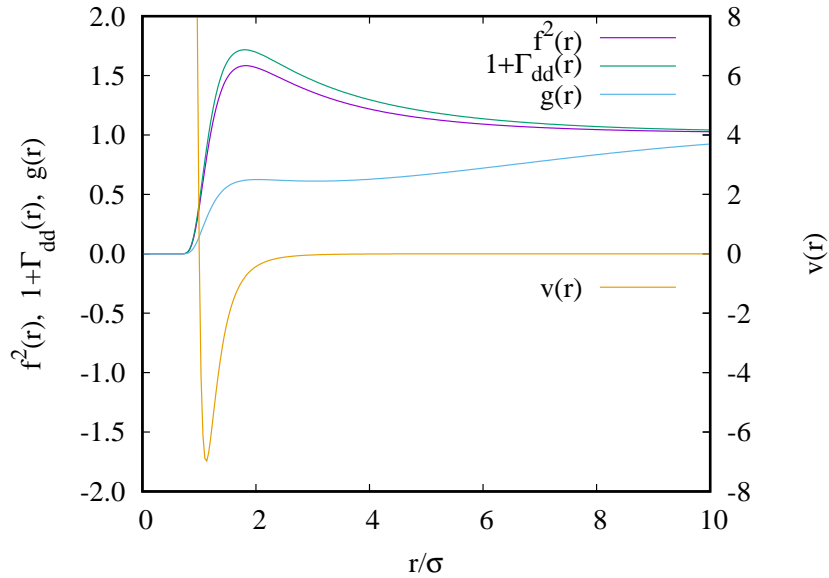}
      }
      \caption{(color online) The figure shows a sample bare
        interaction with a strong repulsive core (ocre line), an
        optimized Jastrow correlation function (purple line), and the
        corresponding pair distribution function $g(r)$ (blue
        line). The specific case is for the Lennard-Jones potential
        discussed below with strength $V_0 = 7.0$ and a Fermi wave
        number $\KF\sigma = 0.3$ where $\sigma$ is the radius of the
        repulsice core. The figure also shows the dynamic correlation
        function $\Gamma_{\!\rm dd}(r)$ which will be introduced in
        Section \ref{ssec:FHNC-EL} and the bare potential.
        \label{fig:v_and_g_and_u}
}
  \end{figure}

A typical situation is seen in Fig. \ref{fig:v_and_g_and_u} where we
examine the example of an interaction with a strong repulsive core and
an attractive well. We show correlation and distribution functions for
that interaction. The specific example is for one of the Lennard-Jones
models to be discussed in Section \ref{ssec:LJium}. At higher
densities, the pair distribution function begins to develop
oscillations typical for strong short-ranged order. The correlation
function $f^2(r)$ and the ``dressed'' correlation function
$1+\Gamma_{\!\rm dd}(r)$ to be introduced in Section
\ref{ssec:FHNC-EL} describe the dynamic correlations induced by the
interaction whereas the pair distribution function $g(r)$ is
determined by both dynamic and statistical correlations.
    
The evaluation of physical quantities for the wave function
(\ref{eq:Jastrow}) requires approximations.  It was quickly realized
\cite{FeenbergBook} that the hypernetted chain summation (HNC) and
truncation scheme has the advantage over other integral equation
techniques like Born-Green-Yvon \cite{BGY46} or Percus
Yevick\cite{PeY58} that it facilitates the unconstrained functional
optimization of the pair correlations by solving the Euler equation
\begin{equation}
  \frac{\delta H_{\bf o}}{\delta u_2}(\rvec_1,\rvec_2)
  = 0\,.
  \label{eq:optu2}
\end{equation}

To set the scene for the further discussions, and to demonstrate both
the simplicity and the power of the method, as well as its physical
content, let us first discuss the simpler case of a Bose liquid. In
that case, the model state is $\ket{{\bf o}} = 1$. The HNC scheme for
bosons is known from the theory of imperfect gases where the Jastrow
correlation function $u_2(r)$ is replaced by $-\beta v(r)$, $\beta$
being the inverse temperature \cite{Morita58,MOH60,LGB}. The equations
are
\begin{eqnarray}
  g(r) &=& \exp\left(u_2(r) + N(r) + E(r)\right)\nonumber\\
  X(r) &\equiv& g(r) - 1 - N(r)\nonumber\\
  \tilde N(k) &=& \frac{\tilde X(k)}{1-\tilde X(k)}\,.
  \label{eq:BoseHNC}
\end{eqnarray}
Above, $X(r)$ and $N(r)$ are the sums of ``non-nodal'' and ``nodal''
diagrams, and $E(r)$ is the sum of ``elementary'' diagrams which can
be expressed in terms of the pair distribution function $g(r)$. We
define, as usual in this field, the dimensionless Fourier transform by
including a density factor $\rho$:
\begin{equation}
  \tilde f(k) = \rho\int d^3r e^{\I \kvec\cdot\rvec}f(r)\label{eq:ft}\,.
\end{equation} 
The elementary diagram contributions $E(r)$ have to be included term
by term; they change the numerical values of the results, but not the
analytic structure of the equations. Three-body correlations also lead
only to a modification of that term \cite{EKthree}.

The pair correlation function can be eliminated entirely from the
theory by utilizing the Jackson-Feenberg identity
\begin{equation}F \nabla^2 F =
        \frac{1}{2} (\nabla^2 F^2+ F^2\nabla^2) + \frac{1}{2} F^2
        \left[\nabla,\left[\nabla ,\ln F\right]\right]
        - \frac{1}{4} \left[\nabla,\left[\nabla, F^2 \right]\right]\,.
\label{eq:JFIdentity}
\end{equation}
For a Jastrow wave function, the second term can be written as
\begin{equation}
F^2\left[\nabla,\left[\nabla ,\ln F\right]\right]
= \frac{1}{2}F^2\sum_{i<j}\nabla^2 u_2(r_{ij})\,.
\label{eq:JFint_kin}
\end{equation}
and, eliminating $u_2(r)$ via the HNC equations (\ref{eq:BoseHNC}), leads
after a few manipulations to
\begin{eqnarray}
  \frac{H_{\bf o}}{N}
  &=& \frac{\rho}{2}\int d^3r\left[
    g(r)v(r) + \frac{\hbar^2}{m}\left|\nabla\sqrt{g(r)}\right|^2\right]
  \label{eq:ERBose}\\
  &&- \frac{1}{4}
  \int \frac{d^3 k}{(2\pi)^3\rho} t(k)(S(k)-1)\tilde N(k)
  \label{eq:EQBose}\\
  &&- \frac{1}{4}
  \int \frac{d^3 k}{(2\pi)^3\rho} t(k)(S(k)-1)\tilde E(k)
  \label{eq:EeBose}\\
  &\equiv& \frac{E_{\rm R}}{N} + \frac{E_{\rm Q}}{N}
  + \frac{E_{\rm e}}{N}
\label{eq:EHNC}
\end{eqnarray}
where $t(k) = \hbar^2 k^2/2m$ is the kinetic energy.

It is then straightforward\cite{PPA1,FeenbergBook,LanttoSiemens}
to derive the Euler equation
\begin{equation}
  \frac{\delta}{\delta\sqrt{g(r)}} \frac{H_{\bf o}}{N} = 0\,.
\label{eq:optg}
\end{equation}
Skipping the technical details we display the resulting equations:
\begin{eqnarray}
  S(q) &=& \frac{1}{\sqrt{1+\displaystyle\frac{2\tilde V_{\rm p-h}(q)}{t(q)}}}
  \label{eq:BoseRPA}\\
  V_{\rm p-h}(r) &=& g(r)\left[v(r) + \Delta V_e(r)\right]
  + \frac{\hbar^2}{m}\left|\nabla\sqrt{g(r)}\right|^2
  + \left[g(r)-1\right]w_{\rm I}(r)\label{eq:BoseVph}\\
  \tilde w_{\rm I}(k) &=& -t(k)
\left[1-\frac{1}{ S(k)}\right]^2
\left[S(k)+\frac{1}{2}\right]\,.
  \label{eq:Bosewind}
\end{eqnarray}
Above,
\begin{equation}
  \Delta V_e(r) = \frac{\hbar^2}{4m}\nabla^2 E(r) + \rho\int d^3r'
  \frac{\delta E(r')}{\delta g(r)}\frac{\hbar^2}{4m}\nabla^2 g(r')
  \equiv \frac{\hbar^2}{4m}\nabla^2 E(r) + E'(r)
  \label{eq:Vedef}
\end{equation}
is the contribution from elementary diagrams and, if applicable, multiparticle
correlations.

A few algebraic manipulations show that the pair distribution function
satisfies the coordinate-space equation \cite{LanttoSiemens}
\begin{equation}
  \frac{\hbar^2}{m}\nabla^2\sqrt{g(r)} = \left[v(r) + \Delta V_e(r) +
    w_{\rm I}(r)\right]\sqrt{g(r)}
  \label{eq:BoseBG}\,.
  \end{equation}
Eq. (\ref{eq:BoseRPA}) recognized as a Bogoliubov equation in terms of
an effective ``particle-hole'' interaction $\tilde V_{\rm
  p-h}(k)$. Likewise, Eq. (\ref{eq:BoseBG}) is recognized as the
Bethe-Goldstone equation in terms of the interaction $v(r) + \Delta
V_e(r) + w_{\rm I}(r)$. This observation led Sim, Woo, and Buchler
\cite{Woo70} to the conclusion that ``it appears that the optimized
Jastrow function is capable of summing all rings and ladders, and
partially all other diagrams, to infinite order''.

The results (\ref{eq:BoseRPA}) and (\ref{eq:BoseBG}) also
substantiates the assertion made above that the HNC summation scheme
is the method of choice over alternatives because it facilitates the
optimization of the correlations: No matter which approximation we
choose for the elementary diagrams, and whether we include higher
order correlation functions, the only thing that changes is the
correction term $\Delta V_e(r)$, but the structure of the equation
remains the same.

\subsection{Pair density functional theory: The view from the top}
\label{ssec:PairDFT}

We have already commented in Section \ref{ssec:bhnc} that the pair
correlation function $f(r)$ can be eliminated entirely from energy expression
which is then formulated entirely in terms of the physical observable
$g(r)$. It is therefore natural to ask whether a general minimum
principle exists for the pair distribution function. Effectively, we
are looking for a two-body version of the Hohenberg-Kohn
\cite{KohnHohenberg,Levy} theorem.

Let us write the energy per particle as
\begin{equation}
	\frac{E}{N} = \frac{T}{N} + \frac{V}{N},
\label{eq:EHK}\end{equation}
where
\begin{equation}
	\frac{V}{N} = \frac{\rho}{2}\int d^3r v(r) g(r)
\label{eq:VHK}\end{equation}
is the potential energy, and $T$ the kinetic energy whose form
is yet unspecified. 

Following the line of arguments leading to the Hohenberg-Kohn theorem
for the one-body density, three statements can be made:
\begin{itemize}
\item [(1)] The kinetic energy $T$ depends only on $g(r)$ and not on
  $v(r)$.
\item [(2)] Assuming that the interaction goes to zero at large distances,
  there is a bijective mapping between $v(r)$ and $g(r)$.
\item [(3)] The total energy has a minimum equal to the ground state
  energy at the physical ground state distribution function, in other
  words the ground state distribution function can be obtained by
  functionally minimizing the energy (\ref{eq:EHK}) with respect to
  the distribution function $g(r)$.
\end{itemize}

The proof parallels exactly the proof of the original Kohn-Hohenberg
theorem and does not need to be repeated here.

Let us assume now that we have a variational problem of the form
(\ref{eq:optg}) with an energy functional (\ref{eq:EHK}),
(\ref{eq:VHK}). We then can calculate the pair distribution function
(or the static structure function) for {\em any\/} potential $\lambda
v(r)$ with $ 0 < \lambda < 1$.

Replacing, in Eq. (\ref{eq:VHK}) $v(r)$ by $\lambda v(r)$
and differentiating with respect to $\lambda$ gives
\begin{equation}
\frac{d}{d\lambda}\frac{E}{N} = \frac{\rho}{2}\int d^3r v(r)
g_\lambda(r) + \frac{1}{N}\int d^3 r \frac{\delta E}{\delta g_\lambda
  (r)}\frac{dg_\lambda (r)}{d\lambda}.
\label{eq:dEdlambda}
\end{equation}
The second term in Eq. (\ref{eq:dEdlambda}) vanishes, we can therefore
recover the energy by coupling constant integration,
\begin{equation}
  \frac{E}{N} = \frac{E_0}{N} +
  \frac{\rho}{2}\int d^3r v(r)\int_0^1 d\lambda g_\lambda(r),
  \label{eq:FeynmanHellman}
\end{equation}
where $g_\lambda(r)$ is the pair distribution function calculated for
a potential strength $\lambda v(r)$, and $E_0$ is the energy of the
non-interacting system which is zero for bosons, and equal to the
energy $T_{\rm F}$ of the non-interacting Fermi gas for fermions.

In Eq. (\ref{eq:FeynmanHellman}) we recover, of course, the
Hellmann-Feynman theorem \cite{hellmann1933,Feynman1939} which was
originally proven for the {\em exact\/} ground state. The above
derivation \cite{parquet3} shows that the theorem is true not only for
the exact ground state, but also for any {\em approximate\/} energy
functional, as long as the pair distribution function is obtained by
minimizing that functional.

Evidently, the above statement is much more general than the
optimization condition (\ref{eq:optg}) for the Jastrow--Feenberg wave
function because it defines a whole class of many--body theories which
can be characterized by the central role of the pair distribution
function.  To summarize, the above consideration shows that the
many-body theory of strongly interacting systems can indeed quite
generally be formulated in terms of a {\em local\/} two-body
quantity. This provides, similar to the tremendously successful
density functional theory of inhomogeneous electron systems, a
significant simplification compared to Greens's function theories.

The above analysis is evidently independent of the statistics,
it applies equally well for bosons and fermions. 

So far, our considerations were entirely parallel to conventional
density functional theory; the energy functional is still
unspecified. The next step is therefore the construction of a ``pair
density functional''.  Unlike the conventional density functional of
inhomogeneous electron systems, some exact features of the pair
distribution function are known that can be used for the construction
of a pair-density functional (\ref{eq:EHK}).

\begin{enumerate}
  \item{} The static structure function $S(q)$ is related to the
    density--density response function $\chi(q,\omega)$ through
 \begin{equation}
 S(q) = -\int_0^\infty \frac{d\omega}{\pi} \Im m\chi(q,\omega)
\label{eq:S}
\end{equation}

Assuming, for example, an RPA form of the density-density response
function {\em defines\/} a local ``particle--hole interaction''
$\tilde V_{\rm p-h}(q)$ by an RPA formula
 \begin{equation}
    \chi(q,\omega)= 
    \frac{\chi_0(q,\omega)} {1-\tilde V_{\rm
        p-h}(q)\chi_0(q,\omega)}\label{eq:chi}
  \end{equation}
where $\chi_0(q,\omega)$ is the Lindhard function.
 
\item{} The short-ranged structure of the pair distribution function
  $g(r)$ is determined by a Bethe-Goldstone equation in terms of a yet
  unspecified particle--particle interaction $V_{\rm p-p}(r)$ We can
  assume that, for short, distances, $V_{\rm p-p}(r)$ is dominated by
  the bare interaction $v(r)$, {\em i.e.\/} we can write
  \[ V_{\rm p-p}(r) = v(r) + w(r).\]
\item{} We require that $g(r)$ and $S(k)$ are consistent in the sense that
  they are related by Eq. (\ref{eq:sofk}).
\end{enumerate}

For bosons, all the calculations can be carried out analytically.
In the absence of Pauli operators, the Bethe Gladstone equation is
simply a zero-energy Schr\"odinger equation
  \begin{equation}
    \frac{\hbar^2}{m}\nabla^2\sqrt{g(r)} = V_{\rm p-p}(r) \sqrt{g(r)}\,.
    \label{eq:BGB}
  \end{equation}
Just as we can think of Eqs. (\ref{eq:S}) and (\ref{eq:chi}) as a {\em
  definition\/} of $\tilde V_{\rm p-h}(q)$ in terms of $S(q)$, we can
think of Eq. (\ref{eq:BGB}) as a {\em definition\/} of $V_{\rm
  p-p}(r)$ in terms of $g(r)$.

For bosons we have
 \begin{equation}
        \chi_0^{\rm Bose}(q,\omega) =
        \displaystyle \frac{2 t(q)}
        { (\hbar\omega+\I\eta)^2-
        t^2(q)} \,.
\label{eq:Chi0Bose}
 \end{equation}
Then, the frequency integration (\ref{eq:S}) can be carried out
analytically and leads to the familiar Bogoliubov formula
(\ref{eq:BoseRPA}).  Simply manipulating
Eqs. (\ref{eq:S})-(\ref{eq:Chi0Bose}) then leads to the HNC-EL
equations (\ref{eq:BoseRPA})-(\ref{eq:Bosewind}), (\ref{eq:BoseBG})
\cite{PairDFT}.  The only undetermined quantity is the correction
$V_e(r)$ which is, in HNC-EL, determined by the elementary diagrams
and multiparticle correlation functions retained; in local parquet
theory it is the set of diagrams that are both particle-particle and
particle-hole irreducible \cite{TripletParquet}.

To summarize, the HNC-EL theory supplements the variational
prescription following from the Hohenberg-Kohn theorem for the pair
distribution function by the requirement that $g(r)$ function
satisfies {\em both,\/} a Bethe-Goldstone equation and an RPA
equation.  Hence, we shall refer to the equations as {\em generic\/}
equations because they can be obtained without ever mentioning a
Jastrow-Feenberg function.

\subsection{Stability and Consistency}
\label{ssec:stability}

A condition for the existence of solutions of the Euler equation is
that the term under the square-root in Eq. (\ref{eq:BoseRPA}) is
positive. We must identify the long-wavelength limit with the
hydrodynamic speed of sound,
\begin{equation}
\tilde V_{\rm p-h}(0+) = mc^2\,,
\label{eq:BosemcfromVph}
\end{equation}
to obtain the correct long wavelength limit
\begin{equation}
  S(k) = \frac{\hbar k}{2mc}\qquad{\rm as}\qquad k\rightarrow 0+\,.
  \label{eq:Sklong}
\end{equation}
An immediate consequence is that the HNC-EL or local parquet equations
have no solution of the system is unstable against infinitesimal
density fluctuations. This is a very desirable feature and unique to
theories that have the diagrammatic completeness of the parquet
theory.

Alternatively we can calculate the hydrodynamic speed of sound
from the equation of state
\begin{equation}
mc^2 = \frac{d}{d\rho}\rho^2 \frac{d}{d\rho}\frac{E}{N}\,.
\label{eq:mcfromeos}
\end{equation}
The definitions (\ref{eq:BosemcfromVph}) and (\ref{eq:mcfromeos})
will, in any approximate theory, not be identical. In fact, it can be
shown in both Jastrow-Feenberg theory \cite{EKVar} and in
parquet-diagram theory \cite{parquet5} that they agree only when {\em
  all\/} diagrams and correlations {\em to all orders\/} are
included. Turning this feature in an advantage, the comparison between
the definitions (\ref{eq:BosemcfromVph}) and (\ref{eq:mcfromeos}) can
serve as a convergence test of approximate evaluations.  We will
utilize this feature in our numerical studies below.

\section{Variational and local parquet diagram theory for fermions}
\label{sec:variations}

The Jastrow-Feenberg method for fermions has been successfully applied
to the relatively simple electron liquid \cite{Lan80,Zab80,annals},
nuclear systems \cite{PAW79} as well as highly correlated Fermi
systems like $^4$He \cite{EKthree,lowdens} and $^3$He
\cite{OwenVar,EKVar,polish} at $T=0$ and finite temperatures
\cite{ChuckT,BradChuck,CKSC}.  The full fermion HNC equations
\cite{Mistig,MistigNP,Fantoni} are significantly more complicated than
the bosons equations (\ref{eq:BoseHNC}); instead of one set of nocal,
non-nodal, and elementary diagrams we have four sets. Moreover, very
specific truncation schemes of exchange diagrams are necessary to
permit an unconstrained optimization of the correlations \cite{EKVar},
the above-mentioned hierarchy of FHNC//n approximations.
 
We have shown in recent work \cite{ljium} that even the simplest
version of the FHNC-EL theory reproduces the equation of state within
better than one percent at densities less than 25\% of the saturation
density of liquid \he3. This statement applies to the energy, other
quantities are, as we shall see, more senesitive to level at which the
FHNC are implemented.  A similar statement applies for nuclear systems
\cite{ectpaper}. It is not much more complicated to solve the full set
of FHNC-EL equations, including elementary diagrams and triplet
correlations \cite{polish}. The version to be presented here permits,
however, a clearer identification of sets of FHNC diagrams with
Feynman diagrams.

\subsection{Generating functional and the generalized distribution functions}

We describe in this and the following sections the basic techniques of
cluster expansions and resummations for Fermi systems. The
manipulations of Section \ref{ssec:bhnc} relied on the simplicity of
the HNC equations (\ref{eq:BoseHNC}) for bosons. We must now be more
systematic, this is also necessary in view of the generalization to
superfluid systems to be described below.

The central quantity for all derivations is the ``generating
function'' $G(\beta)$ defined as follows: Substitute in the
variational wave function (\ref{eq:Jastrow})
\begin{equation}
  u_2(r) \rightarrow u_2(r,\beta) \equiv u_2(r) + \beta v_{\rm JF}(r)
\end{equation}
where
\begin{equation}
  v_{\rm JF}(r) = v(r) - \frac{\hbar^2}{4m}\nabla^2 u_2(r)
\end{equation}
is the ``Jackson-Feenberg effective interaction''. With this, all
quantities depend on the dummy parameter $\beta$. Define then the
generalized normalization integral
\begin{equation}
  I_{{\bf o}}(\beta) = \bra{\bf o}F^2(\rvec_1,\ldots\rvec_N;\beta)\ket{\bf o}
  \label{eq:Inormal}
\end{equation}
and the generating function
\begin{equation}
  G(\beta) \equiv \ln I_{{\bf o}}(\beta)\,.
  \label{eq:Gnormal}
\end{equation}
The value $G(0)$ is just the logarithm of the norm of the wave
function, but evidently it is technically no more complicated to
calculate $G(\beta)$ than it is to calculate $G(0)$.

The energy expectation value may be obtained from the generating
functional through
\begin{eqnarray}
  H_{{\bf o}} &=& T_{\rm F} + \frac{dG(\beta)}{d\beta}\Biggl|_{\beta=0}+T_{\rm JF}\nonumber\\
  &=& T_{\rm F} + N\frac{\rho}{2}\int d^3r g(r) v_{\rm JF}(r)
  +T_{\rm JF}\,.
\label{eq:dGbydbeta}
\end{eqnarray}
Here $T_{\rm F}$ is the kinetic energy of the non-interacting Fermi
gas, and
\begin{equation}
  T_{\rm JF} = \frac{\hbar^2}{8m}\frac{\bra{\Phi_0}\sum_i\left[\nabla_i,
      \left[\nabla_i,F^2\right]\right]\ket{\Phi_0}}{I_{{\bf o}}}
  \label{eq:TJF}
\end{equation}
is an energy correction in Fermi systems that will be dealt with below.
Further, it is convenient to define generalized densities
\begin{equation}
        \rho_2(\rvec,\rvec';\beta)= N(N-1)
        \int d^3r_3\ldots d^3r_N|\Psi_0(\rvec,\rvec',\rvec_3,\ldots,\rvec_N;\beta)|^2 \,.
\label{eq:rho2ofbeta}
\end{equation}
The (generalized) two-body density may be derived as a variational
derivative of the generating function with respect to the
(generalized) pair correlation function
\begin{eqnarray}
        \rho_2(\rvec,\rvec';\beta)&=& \rho^2 g(\rvec,\rvec';\beta)= 2
        \frac{\delta G(\beta)}{\delta u_2(\rvec,\rvec';\beta)}\label{eq:Gbyu2}
        \\
        &=& 2
        f^2(\rvec,\rvec')\frac{\delta G(\beta)}{\delta
          h(\rvec,\rvec';\beta)}\,,
\label{eq:gfactored}
\end{eqnarray}
a relation that will be particularly useful for the derivation of the
Euler equations. The last expression shows that the procedure
generates a $g(\rvec,\rvec')$ which has an overall factor $f^2(\rvec,\rvec') =
\exp(u_2(\rvec,\rvec')$.

For the case $\beta=0$, the definition (\ref{eq:rho2ofbeta}) reduces
to the two-body density introduced above (\cf\ Eq. (\ref{eq:rho2})),
\begin{equation}
  \rho_2(\rvec,\rvec';\beta=0) \equiv \rho_2(\rvec,\rvec')\,,
  \qquad g(\rvec,\rvec';\beta=0) \equiv g(\rvec_,\rvec')\,.
\end{equation}

The construction of both the energy and the distribution functions via
derivatives of one common generating functional is a welcome economy
since it is sufficient to develop an algorithm for the calculation of
$G(\beta)$; one does not need to start over for each individual
quantity of interest. For example, we can immediately write down the
exact Euler equation for the pair distribution function:
\begin{equation}
  \frac{\delta  H_{{\bf o},{\bf o}}}{\delta u_2(r)}
  = N\frac{\rho}{2}\left[-\frac{\hbar^2}{4m}\nabla^2 g(r) + g'(r)\right]=0
  \label{eq:Fermioptu2}
  \end{equation}
  where
  \begin{equation}
    g'(r) = \left.\frac{d g(r;\beta)}{d\beta}\right|_{\beta=0} + \frac{2}{N\rho}
    \frac{\delta T_{\rm JF}}{\delta u_2}(r)\,.
    \end{equation}

\subsection{Cluster expansions}
\label{ssec:cluster_expansions}

A very straightforward technique to derive cluster expansions for the
generating function and, or course, any other quantity of interest is
the power-series method \cite{Fantoni}. The method introduces a dummy
parameter $\alpha$ in the correlation operators
\begin{eqnarray}
  F_N^2(\rvec_1,\ldots,\rvec_N;\alpha)
  = \left.\prod_{i\le j}^N(1+\alpha h(\rvec_{ij}))\right|_{\alpha=1}\,,
  \label{eq:Fseries}
\end{eqnarray}
where $h(r) \equiv f^2(r)-1$ and we have suppressed the parameter
$\beta$. One then expands the quantity of interest in a power series
in $\alpha$, evaluated at $\alpha=1$.
For example,
\begin{eqnarray}
  G(\beta) &=&  \left. G(\alpha,\beta)\right|_{\alpha=1}\nonumber\\
  &=& \sum_{n}\frac{1}{n!}\frac{d^n}{d\alpha^n}\left. G(\alpha,\beta)\right|_{\alpha=1}\nonumber\\
  &=& \sum_{n}(\Delta G)^{(n)}(\beta)\,.
  \label{eq:pseries}
\end{eqnarray}
That way, a series of cluster contributions $(\Delta G)^{(n)}(\beta)$
of increasing number of $h(r_{ij})$ factors is generated.  These are
best represented diagrammatically \cite{Johnreview}:

\begin{enumerate}
\item Each point (open dot) represents a particle coordinate
$\rvec_i$.

\item Each filled point (solid dot) implies the integration over the
  coordinate $\rvec_i$, and multiplication with a density factor $\rho
  = N/\Omega $, where $\Omega$ is the normalization volume.

\item {\em Correlation line\/} connecting the points $\rvec_i$ and
  $\rvec_j$ represent a function $h(r_{ij})$. These are depicted as
  dashed lines connecting the two points.
\begin{eqnarray}
h(r_{ij}) \equiv \,\begin{tikzpicture}[baseline = {([yshift=-.5ex]current bounding box.center)},vertex/.style={anchor=base,
    circle,fill=black!25,minimum size=18pt,inner sep=2pt}]
\node[circle,inner sep=1.5pt,draw] at (0:0.5cm) (node1) {};
\node[circle,inner sep=1.5pt,draw] at (180:0.5cm) (node2) {};
\draw[thick,dashed] (node1) -- (node2);
\end{tikzpicture}\,
\end{eqnarray}

\item {\em Exchange lines\/} represent the function
\begin{equation}
    \ell(r_{ij}\KF) = \frac{3}{4\pi\KF^3}
    \int d^3k \theta(\KF-k) e^{\I\rvec_{ij}\cdot\kvec} =
    \frac{3}{r_{ij}\KF}j_1(r_{ij}\KF)
      \label{eq:ldef}\,.
      \end{equation}
These are depicted by {\em oriented solid line\/} connecting the point
$\rvec_i$ to $\rvec_j$,
\begin{eqnarray}
\ell(r_{ij}\KF)  = \,\begin{tikzpicture}[baseline = {([yshift=-.5ex]current bounding box.center)},vertex/.style={anchor=base,
    circle,fill=black!25,minimum size=18pt,inner sep=2pt}]
\node[circle,inner sep=1.5pt,draw] at (0:0.5cm) (node1) {};
\node[circle,inner sep=1.5pt,draw] at (180:0.5cm) (node2) {};
\draw[thick,->-] (node1) to (node2);
\end{tikzpicture}\,.
\end{eqnarray}
\end{enumerate}

We call a diagram {\em linked\/} if each point is connected to every
other point by at least one continuous path of graphical
elements. Linked diagram contributions to the generating functional
are proportional to the particle number. We call diagram {\em
  irreducible\/} if it cannot be calculated as a product of two or
more simpler diagrams.

The cluster expansion of the generating function is then represented
in terms of all topologically distinct irreducible diagrams without
external points constructed by the following rules:

\begin{enumerate}
\item{} Each point is attached by at least one correlation line.
\item{} Two points can be joined by at most one correlation line.
\item{} Any point of a contributing diagram is joined by at most one
  incoming exchange line which must be accompanied by a single
  outgoing exchange line. Hence, exchange lines come in closed loops.
  For each closed loop of $L$ exchange lines there is a factor
  $(-1/\nu)^{L-1}$ in the corresponding analytic contribution, as well
  as the exchange function. (Generally, the sign $(-)^{L-1}$ is
  displayed with the diagram, but the numerical factor $1/\nu^{L-1}$
  is left implicit.) Here, $\nu$ is the degree of degeneracy of the
  single particle states.
\end{enumerate}

\begin{figure}
  \centerline{\includegraphics[width=0.90\columnwidth]{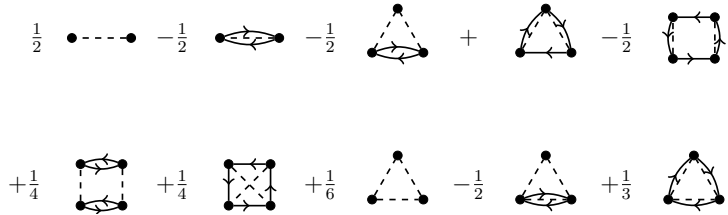}%
    \hspace{2.5cm}}
  \caption{The figure shows the diagrammatic representation of all
    two- and three- body contributions as well as the four-body
    contributions with two correlation lines to the generating
    function $G(\beta)$. The dashed line here is understood to be a
    generalized function $h(r_{ij},\beta) =
    \exp(u_2(r_{ij},\beta))-1$. Diagrams (d) and (e) have the same
    value and are normally drawn together; we spell them out here
    individually to clarify the topological factors, and also in view
    of the modifications necessary for the superfluid system.
    \label{fig:Gnormal}}
\end{figure}

The first terms in the diagrammatic expansion of the generating
function are shown in Fig. \ref{fig:Gnormal}.
The generalized pair distribution function can then be derived by the
variational prescription (\ref{eq:Gbyu2})
and the energy expectation is obtained from the general expression
(\ref{eq:dGbydbeta}) as follows
\begin{enumerate}
\item{} Replace, in turn, each correlation line $h(r_{ij})$ by a line
  $f^2(r_{ij}) v_{\rm JF}(r_{ij})$. This is what the
  $\beta$-derivative does. 
\item{} To calculate the kinetic energy corrections $T_{\rm JF}$
  replace any pair of exchange lines $\ell(r_{ij}\KF)\ell(r_{ik}\KF)$ by a
  pair $\frac{\hbar^2}{8m} \nabla_i^2\ell(r_{ij}\KF)\ell(r_{ik}\KF)$.
\end{enumerate}

\begin{figure}
  \centerline{\includegraphics[width=0.9\columnwidth]{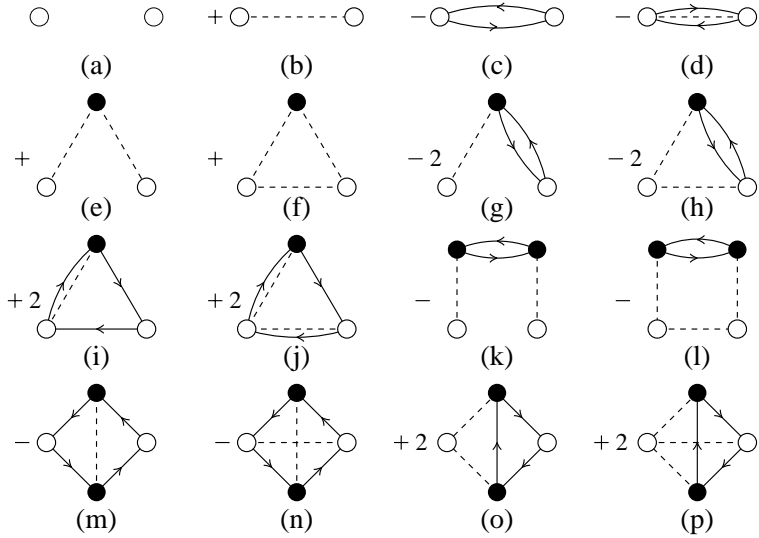}%
    }
  \caption{The figure shows the diagrammatic representation of a few
    low-order contributions to the pair distribution function. The
    combination of two adjacent diagrams always generates a common
    factor $f^2(r) = exp(u_2(r))$, see Eq. (\ref{eq:gfactored}). On the
    other hand, diagrams (c), (h), and (k) as well as diagrams (g) and
    (o) and diagrams (d), (i), and (m) must be combined to obtain the
    exact behavior of $S(k)$ as $k\rightarrow 0+$.
    \label{fig:g_normal}}
\end{figure}

Cluster contributions for the (generalized) pair distribution function
can then be generated by the prescription (\ref{eq:Gbyu2}).
Diagrammatically, the construction of $g(r,\beta)$
amounts to the following operation on the generating function
$G(\beta)$:
\begin{enumerate}
\item{} Remove, in turn, each correlation line and turn its endpoints
  into open (``reference'') points $\rvec$ and $\rvec'$
\item{} Multiply this by $\exp(u_2(\rvec,\rvec')) = 1 + h(\rvec
  ,\rvec')$.  This factor is needed to model the short-ranged
  structure of the wave function for hard-core potentials.
\end{enumerate}
 A few low-order
diagrams contributing to the pair distribution function are shown in
Fig. \ref{fig:g_normal}.
The figure shows the expansion of $g(r)$ in terms of correlation and
exchange lines. Two adjacent diagrams (a) and (b), (c) and (d) {\em
  etc.\/} can always be combined to obtain the form
(\ref{eq:gfactored}), in other words the function $C(r)$ is
represented by diagrams (a), (c), (e), {\em etc.}.  This is exactly
the intention of Jastrow-Feenberg theory
\cite{Jastrow55,FeenbergBook}, namely to have the correlation function
model the short-ranged structure of the wave function.

Close inspection of the individual contributions to $g(r)$ reveals,
however, a dilemma: For the stability of the system as well as for
meaningful solutions of the Euler equation, it is important long
ranged correlations are important, \cf Eq. (\ref{eq:Sklong}). To get
the correct behavior of $S(q)$ for $q\rightarrow 0+$, the diagrams
must be grouped differently. For example, diagrams (b), (h), and (k)
combine to
\[\SF^2(q)\tilde h(q)\]
where
\begin{equation}
  \SF(q) = \begin{cases}
     \displaystyle \frac{3q}{4\KF}-\frac{q^3}{16\KF^3},
                & q < 2\KF ;\\
      1,      & q \ge 2\KF.
      \end{cases}
\end{equation}
is the static structure function of the non-interacting Fermi gas.
Likewise, it is straightforward to show \cite{EKVar} that the sum of
diagrams (g) and (o) as well as the sum of diagrams (d), (i), and (m)
go as $q^2$ in the limit $q\rightarrow 0+$. The analysis can be easily
extended to other, more complicated cases.

To summarize, {\em there is no finite truncation of the expansion of
  the pair distribution function that is exact in both, the
  short-distance and the long-wavelength limit.} One can deal with
this situation in three ways:

\begin{itemize}
\item{} One can ignore the problem entirely and use, consistent with the
  original idea of Jastrow-Feenberg theory, an approximation for
  $g(r)$ of the form (\ref{eq:gfactored}). This is, among others, the idea
  of the FHNC summations of Ref. \citenum{Fantoni}. One must live with the fact
  that the static structure function has the incorrect long-wavelength limit
  and the correlation functions are limited to simple parameterized forms.
\item{} One can use {\em different approximations\/} for $g(r)$ and
  $S(q)$ depending on which quantity is of interest. Choosing a form
  for $S(q)$ that has the exact long-wavelength behavior permits the
  unconstrained optimization of the pair correlations. Of course, one
  must then construct a pair distribution $g(r)$ of the form
  (\ref{eq:gfactored}).
  \item{} One must sum infinite sets of exchange diagrams or
    approximations thereof as was done in Ref. \citenum{polish}.
\end{itemize}
For the purpose of this work, we shall use the second approach because
it is sufficiently accurate for all of our purposes
\cite{ljium,ectpaper} and allows the most direct identification of JF
diagrams with parquet diagrams.

\begin{figure}
  \centerline{\includegraphics[width=0.9\columnwidth]{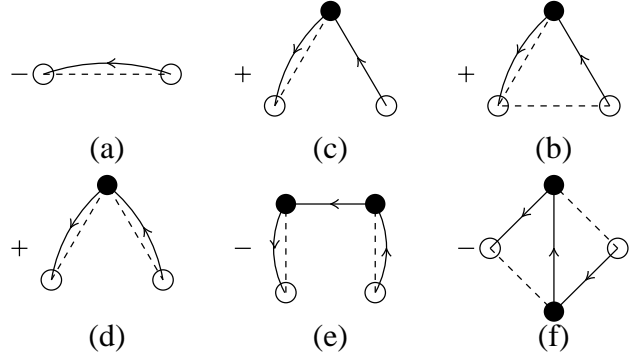}%
    }
  \caption{The figure shows a few ``cyclic chain'' diagrams.
    \label{fig:ccdiagrams}}
\end{figure}

To conclude this section, we mention another specific set of diagrams,
the so--called ``cyclic chain'' (cc-) diagrams. These are two body
diagrams that have a continuous exchange path connecting the external
points. A few examples are shown in Fig. \ref{fig:ccdiagrams}. We can
again identify ``chain'' diagrams (diagrams (c), (d), and (e)),
``parallel connections'' (diagrams (a) and (b)) and one ``elementary''
diagram (diagram (f)).
The summation of these diagrams suggests the introduction of a
``dressed'' exchange line $L(r)$ which is the sum of all 2-point
diagrams having an exchange path connecting the two external points.
The ``cc'' diagrams are related to self-energy corrections in
perturbation theory, see Sections \ref{ssec:CBF} and \ref{ssec:selfen}
below.

\subsection{FHNC and Euler equations}
\label{ssec:FHNC-EL}

We discuss here the simplest implementation of the FHNC theory that is
compatible with the variational problem, called FHNC//0 approximation.
The approximation keeps in $g(r)$ only those diagrams that contain
exchange loops of the form $\ell^2(r_{ij}\KF)$. These are, for
example, the diagrams (a-g), (k) and (l) shown in
Fig. \ref{fig:g_normal}. We must also keep in mind that a different
approximation has to be used in $S(q)$.  The implementation and
relevance of higher order exchange corrections will be discussed below
in Section \ref{ssec:exchanges}.

In the FHNC//0 approximation, the FHNC equations are no more complicated
than the Bose HNC equations (\ref{eq:BoseHNC}):
\begin{eqnarray}
  \Gamma_{\!\rm dd}(r) &=& \exp(u_2(r) + N_{\rm dd}(r)) -1\\
  X_{\rm dd}(r) &=&  \Gamma_{\!\rm dd}(r) -   N_{\rm dd}(r)\label{eq:FHNC0}\\
  \tilde N_{\rm dd}(k) &=& \frac{\tilde X_{\rm dd}^2(k) \SF(k)}
         {1-\tilde X_{\rm dd}^2(k) \SF(k)}\\
         S(k) &=& \SF(k)(1+\Gamma_{\!\rm dd}(k)\SF(k))\nonumber\\
         &=& \frac{\SF(k)}{1-\tilde X_{\rm dd}(k)\SF(k)}\label{eq:SofkFermi}\,.
\end{eqnarray}
Above, $\Gamma_{\!\rm dd}(r)$ is the set of all diagrams contributing
to $g(r)$ that have no exchange lines attached to the external points.
Examples are diagrams (b), (e), (f), (k) and (l) shown in Fig.
\ref{fig:g_normal}.  We need these equations to eliminate $u_2(r)$
from the $v_{\rm JF}(r)$ and to formulate the theory entirely in terms
of the physically observable static structure function $S(k)$ and
derived quantities.

In this approximation, the energy correction
$T_{\rm JF}$ has the form
\begin{equation}
  \frac{T_{\rm JF}}{N}\approx \frac{T^{(2)}_{\rm JF}}{N}
    = -\frac{\hbar^2 \rho}{8 m \nu}
    \int d^3r \Gamma_{\!\rm dd}(r)\nabla^2\ell^2(r\KF)\,.
\label{eq:TJF2}
\end{equation}
Similar to the Bose case, the Euler equation is best formulated in
momentum space. Formally, the optimization condition` can be written
as \cite{FeenbergBook}
\begin{equation}
  \frac{1}{2}t(k)\left[S(k)-1\right] + S'(k)= 0\,.\label{eq:Euler}
\end{equation}

The $S'(k)$ is generated by the $\beta$-derivative technique outlined
above.  This amounts to
\begin{enumerate}
\item{} Replace, in turn, each correlation line by $f^2(r) v_{\rm JF}(r)$.
  This can be done with the $\beta$-derivative trick.
\item{} To add the correction from $T_{\rm JF}$, replace, in turn, each
  $\SF(k)$ by $-\frac{t(k)}{2}\left[\SF(k)-1\right]$.
\end{enumerate}

Using the version (\ref{eq:FHNC0}), (\ref{eq:SofkFermi}) of the FHNC00
equations as well as the ``priming'' operation outlined above we
obtain
\begin{equation}
S'(k) = S^2(k)\left[\tilde  V_{\rm p-h}(k) + \frac{t(k)}{2}
\left(\frac{1}{\SF(k)}-\frac{1}{S(k)}\right)
+\frac{\SF'(k)}{\SF^2(k)}\right]\,,\label{eq:Sprime}
\end{equation}
and, inserting this in the Euler
equation (\ref{eq:Euler}) and solving for $S(k)$:
\begin{equation}
  S(k) = \frac{\SF(k)}{\sqrt{1 + 2 \displaystyle\frac{\SF^2(k)}{t(k)}
      \tilde  V_{\rm p-h}(k)}}\,.
\label{eq:PPA}
\end{equation}
In the FHNC//0 approximation, the effective interaction $\tilde V_{\rm
  p-h}(k)$ is approximated by the ``direct'' part of the
particle--hole interaction, $\tilde V_{\rm p-h}(k)\approx \tilde
V_{\rm dd}(k)$; we will discuss the importance to exchange corrections
further below. The quantity is structurally identical to the one for
bosons:
\begin{equation}
  V_{\rm p-h}(r) = v_{\rm CW}(r)+
  \Gamma_{\!\rm dd}(r)w_{\rm I}(r)\label{eq:Vph}
\end{equation}
where
\begin{equation}
v_{\rm CW}(r) = (1+\Gamma_{\!\rm dd}(r))v(r)+
\frac{\hbar^2}{m}\left|\nabla\sqrt{1+\Gamma_{\!\rm dd}(r)}\right|^2
\label{eq:vcw}
\end{equation}
is the ``Clark-Westhaus'' effective interaction \cite{Johnreview} and
$w_{\rm I}(r)$ is the ``induced interaction''.
\begin{eqnarray}
  \tilde w_{\rm I}(k) &=&\left[(1+\SF(k)\tilde\Gamma_{\rm dd}(k))^2-1\right]\tilde V_{\rm p-h}(k)
  + \frac{t(k)}{2}\tilde\Gamma_{\rm dd}^2(k)\label{eq:wind1}\\
& =&-t(k)
\left[\frac{1}{\SF(k)}-\frac{1}{ S(k)}\right]^2
\left[\frac{S(k)}{\SF(k)}+\frac{1}{2}\right]\,.
  \label{eq:wind}
\end{eqnarray}
The second line is obtained by using Eq. (\ref{eq:PPA}) to eliminate
$\tilde V_{\rm p-h}(k)$. The Bose limit is obtained by replacing
$\SF(k) \rightarrow 1$.
 
To derive the equation determining the short-ranged structure of the
correlations, begin with Eq. (\ref{eq:wind}) which we can write, using
the Euler equation (\ref{eq:PPA}), as (\cf Eq. (2.62) of
Ref. \citenum{mixmonster})
\begin{eqnarray}
V_{\rm p-h}(r) +   w_{\rm I}(r) &=&
(1+\Gamma_{\!\rm dd}(r))\left[v(r)+w_{\rm I}(r)\right]+
\frac{\hbar^2}{m}\left|\nabla\sqrt{1+\Gamma_{\!\rm dd}(r)}\right|^2\nonumber\\
&=& -\left[\frac{t(k)}{S_F(k)}\tilde\Gamma_{\!\rm dd}(k)\right]^{\cal F}(r)
  \,.
  \label{eq:windBG}
\end{eqnarray}
Using the identity
\begin{equation}
  \left|\nabla\sqrt{1+\Gamma_{\!\rm dd}(r)}\right|^2
  = \frac{1}{2}\nabla^2\Gamma_{\!\rm dd}(r) -
  \sqrt{1+\Gamma_{\!\rm dd}(r)}\nabla^2\sqrt{1+\Gamma_{\!\rm dd}(r)}\,,
\end{equation}
Eq. (\ref{eq:windBG}) becomes
\begin{equation}
\sqrt{1+\Gamma_{\!\rm dd}(r)}\left[-\frac{\hbar^2}{2m}\nabla^2 + v(r) + w_{\rm I}(r)\right]
  \sqrt{1+\Gamma_{\!\rm dd}(r)}%\nonumber\\
  =\left[t(k)(1-\SF^{-1}(k))\tilde\Gamma_{\!\rm dd}(k)\right]^{\cal F}(r)
  \label{eq:ELSchr}
\end{equation}
The right-hand side is evidently zero for bosons, and the Euler
equation is a simple zero-energy Schr\"odinger equation where the bare
interaction is supplemented by the induced potential which guarantees
that the scattering length of the effective interaction $v(r)+w_{\rm
  I}(r)$ is zero. This is the well-known result of
Refs. \citenum{LanttoSiemens} and \citenum{parquet2}. For fermions,
the right hand side changes the short-ranged behavior of the
correlation function $\Gamma_{\!\rm dd}(r)$ and, hence, the
short-ranged behavior of the pair distribution function $g(r)$.  This
is consistent with the prediction of the Bethe-Goldstone equation that
the Pauli principle changes the short-ranged behavior of the wave
function \cite{GWW58}, see Eq. (\ref{eq:BGlocal}) below.

\subsection{Energy}
\label{ssec:energy}

\begin{figure}[H]
    \centerline{\includegraphics[width=0.6\columnwidth]{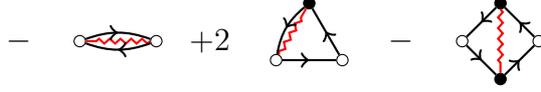}}
    \caption{The figure shows the first order exchange corrections
      $(\Delta \tilde X_{\rm ee})_1(r)$ to the pair distribution
      function Here, the wavy red line represents the function
      $\Gamma_{\rm dd}(r)$. For corrections to the effective
      interactions, we have to re-interpret that line as exchange
      interaction, see below.
       \label{fig:eelink}}
  \end{figure}

For $E_{\rm FHNC}$ we use Eq. (2.14) of Ref. \citenum{cbcs} which implies
the form (\ref{eq:TJF2}) for the $T_{\rm JF}$.
Summarizing, we use
\begin{eqnarray}
  \frac{E}{N}
  &=&\frac{T_{\rm F}}{N} + \frac{E_{\rm R}}{N} +
  \frac{E_{\rm Q}}{N}
\,,\nonumber \\
\frac{E_{\rm R}}{N} &=& \frac{\rho }{2}\int\! d^3r\>\bigl[g_{\rm F}(r) + C(r)\bigr]
v_{\rm CW}(r)\,,
\label{eq:EJF}\\
\frac{E_{\rm Q}}{N} &=& \frac{1}{4}\int\!\frac{d^3k}{(2\pi)^3\rho}\>
t(k)\tilde\Gamma_{\!\rm dd}^2(k)\left[S^2_{\rm F}(k)/S(k)-1\,\right]\nonumber
\end{eqnarray}
where $T_{\rm F}$ is the kinetic energy of the non-interacting Fermi gas,
theory, and the pair distribution function is given
by \cite{cbcs}
\begin{equation}
g(r) = \left[1 + \Gamma_{\!\rm dd}(r)\right]\left[g_{\rm F}(r)+C(r)\right]\,.
\label{eq:gFHNC0}
\end{equation}
Above, 
\begin{equation}
  \tilde C(k) = (\SF^2(k)-1)\tilde\Gamma_{\!\rm dd}(k) + (\Delta
  \tilde X_{\rm ee})_1^{(3)}(k) + (\Delta \tilde X_{\rm
    ee})_1^{(4)}(k)\,.
\label{eq:C0ofk}
\end{equation}
The two terms $(\Delta\tilde\tilde X_{\rm ee})_1^{(3)}(k)$ and
$(\Delta \tilde X_{\rm ee})_1^{(4)}(k)$ are diagrammatically
represented bu the 3-body and 4-body diagram shown in
Fig. \ref{fig:eelink}
That term is omitted the FHNC//0 approximation, it is for energy
calculations in our case very small and only needed to establish the
exact low density expansion of the energy in powers of $\a0\KF$ to
second order \cite{HuangYang57}. We found that non-universal
contributions to the equation of state are overwhelming well below the
densities where the second-order terms become visible. However, we
will see that the correction to the particle-hole interaction
originating from these diagrams is substantial even in the low-density
limit.

\subsection{Uniform limit approximation}
\label{ssec:UL}

The so-called ``uniform limit'' approximation \cite{FeenbergBook} is
obtained by assuming a {\em weak, but possibly long--ranged\/}
interaction. We study this because it will permit direct contact to be
made to the random phase approximation. It will also highlight a
special feature for the superfluid system, see \ref{ssec:ULBCS} below.

Specifically, the uniform limit approximation is valid
when
\begin{eqnarray}
  v(r)\Gamma_{\!\rm dd}(r) &\ll& v(r)\nonumber\\
  \Gamma^2_{\!\rm dd}(r)  &\ll&\Gamma_{\!\rm dd}(r)\,,\nonumber
\end{eqnarray}
note that we do {\em not\/} assume that $\tilde v(k)\tilde
\Gamma_{\!\rm dd}(k)$ or $\Gamma^2_{\!\rm dd}(k)$ are negligible.

In the energy term $E_{\rm R}$ in Eq. (\ref{eq:EJF}) we can use
\[\left|\nabla\sqrt{1+\Gamma_{\!\rm dd}(r)}\right|^2 \approx 
\frac{1}{4}\left|\nabla\Gamma_{\!\rm dd}(r)\right|^2\,.\]
This term is {\em not\/} negligible because
  we can write the energy contribution as
  \[\frac{\rho}{2}\int d^3r \frac{\hbar^2}{m}\left|\nabla\sqrt{1+\Gamma_{\!\rm dd}(r)}\right|^2 \approx \frac{1}{8}\int\frac{ d^3 k}{(2\pi)^3\rho} t(k)\tilde\Gamma_{\!\rm dd}^2(k)\,.
  \]
Then, the energy becomes
\begin{eqnarray}
  \frac{E}{N} &=& \frac{T_{\rm F}}{N}  + \frac{1}{2}\tilde v(0+) + \frac{1}{2}\int \frac{d^3k}{(2\pi)^3\rho}
  \left[S(k)-1\right]\tilde v(k)\nonumber\\
  &&\qquad+\frac{1}{2}\int \frac{d^3k}{(2\pi)^3\rho}
  \frac{t(k)}{2}\frac{(S(k)-\SF(k))^2}{\SF^2(k)S(k)}\nonumber\\
   &=& \frac{E_{\rm HF}}{N} - \frac{1}{2}\int \frac{d^3k}{(2\pi)^3\rho}
  \frac{t(k)}{2}\frac{(S(k)-\SF(k))^2}{\SF(k)S^2(k)}
\label{eq:EULnormal}
\end{eqnarray}
where $E_{\rm HF}$ is the energy expectation value in
Hartree-Fock approximation.  The Euler equation (\ref{eq:PPAUL})
becomes
\begin{equation}
  S(k) = \frac{\SF(k)}{\sqrt{1 + 2 \displaystyle\frac{\SF^2(k)}{t(k)}
      \tilde  v(k)}}\,.
\label{eq:PPAUL}
\end{equation}
follows from this expression by minimization with respect to $S(k)$.
We will see below that both (\ref{eq:EULnormal}) and (\ref{eq:PPAUL})
are is indeed approximations for of the RPA energy and structure
function. Since the $S(k)$ is obtained from a variational principle,
it follows from our analysis of Section \ref{ssec:PairDFT} that the
energy (\ref{eq:EULnormal}) can be obtained by coupling constant
integration which is now, of course, best performed in momentum space.

\subsection{The low-density limit}
\label{ssec:lowdens}

Cold gas applications focus on very low density systems where
information about the interaction can be reduced to a single
parameter, the $S$-wave scattering length $\a0$. We have studied this
area very carefully in Ref. \citenum{cbcs}, see also
sec. \ref{ssec:limits}. We have identified three areas where the
correlations are determined by different effects:
\begin{enumerate}
\item{} The short-distance regime of the order of the interaction
  range is, of course, dominated by the shape of the interaction.
\item{} The intermediate regime is the range between the interaction
  regime and the average particle distance. If this regime is large,
  which is the case for very low--density systems, the correlations are
  determined by the vacuum scattering Knight $\a0$,
  \[\sqrt{1+\Gamma_{\rm dd}(r)} \sim 1- \frac{\a0}{r}\,.\]
\item{} For interparticle distances larger that $1/\KF$, the
  correlations fall off as
\begin{equation}
  \Gamma_{\!\rm dd}(r) \sim -\frac{3}{4}\frac{V_{\rm p-h}(0+)}{\mcf}
\frac{1}{ r^2\KF^2}\,,
\label{eq:Gddlong}
\end{equation}
where $\mcf = \hbar^2\KF^2/3m$ is the incompressibility of a
non-interacting Fermi gas. Evidently this is a many-body effect, the
fall-off of the long-range correlations is needed to have the wave
function normalized.
\end{enumerate}

A manifestly microscopic calculation begins, of course, with the bare
interaction which determines the vacuum scattering length. Since all
diagrammatic calculations imply some approximations, we need to make
sure that any approximations we are making will not destroy this
property. This will turn out to be an important consideration in the
next section.

To derive the low-density limit of the Euler equation, begin with the
diagrammatic expansion of the pair distribution function shown in
Fig. \ref{fig:g_normal}. In that limit, only the first line of
diagrams contributes; moreover we can identify $\Gamma_{\rm dd}(r) =
f^2(r)-1$.  Then the Euler equation reads
 \begin{equation}
   \frac{\hbar^2}{4m}\nabla^2 \left[g_{\rm F}(r)(1+\Gamma_{\rm dd}(r))\right]
   =  g_{\rm F}(r)\Gamma_{\!\rm dd}'(r) + (1+\Gamma_{\rm dd}(r))
   \frac{\hbar^2}{4m}\nabla^2 g_{\rm F}(r)\,.
 \end{equation}
where $g_{\rm F}(r) = 1-\frac{1}{\nu}\ell^2(r\KF)$ is the pair
distribution function of the non-interacting system, and
$\Gamma_{\!\rm dd}'(r)$ is constructed in analogy to $S'(k)$ above
which yields
\begin{equation}
\Gamma_{\!\rm dd}'(r)
  = v_{\rm CW}(r)+
  (1+\Gamma_{\rm dd}(r))w_{\rm I}(r) -\frac{\hbar^2}{4m}\nabla^2\Gamma_{\!\rm dd}(r)\,.
  \end{equation}
We can cancel the
term $\frac{\hbar^2}{4m}\nabla^2 g_{\rm F}(r)$ on the left and the
right; the term $\nabla g_{\rm F}(r)\cdot\nabla\Gamma_{\rm dd}(r)$
goes as $\KF^2$ and can be ignored; hence we end up with
  \begin{equation}
   \frac{\hbar^2}{4m}\nabla^2(1+\Gamma_{\rm dd}(r))
   = \Gamma_{\rm dd}'(r)\,.
  \end{equation}
  which is exactly the bosons form.
  Going through the same manipulations as in Section
  \ref{ssec:FHNC-EL} we finally can write the low-density limit of the
  Euler equation as
\begin{equation}
\left[-\frac{\hbar^2}{2m}\nabla^2 + v(r) + w_{\rm I}(r)\right]
  \sqrt{1+\Gamma_{\!\rm dd}(r)}
  =0\,.
\end{equation}
Evidently, this is identical to the low-density limit of
Eq. (\ref{eq:ELSchr}) saying that the FHNC//0 approximation as spelled
out in the previous section describes both the short-- and the long--
ranged correlations consistently. Of course, the simplicity of this
equation is caused by the fact that the Pauli principle plays no role
in the low-density limit, see Eq. (\ref{eq:ELSchr}). 

\subsection{Exchange corrections}
\label{ssec:exchanges}

We have above formulated a version of FHNC-EL that contains the
simplest versions of the RPA and the Bethe-Goldstone equation. These
describe the {\em qualitatively\/} correct physics, but have, even at
very low densities, some {\em quantitative\/} inconsistencies which we
have to address and handle.

Let us go back to the energy expression (\ref{eq:EJF}).
The dominating term at low densities is
\begin{equation}
\frac{E_{\rm R}}{N} \approx \frac{\rho }{2}\int\! d^3r\>g_{\rm F}(r)
v_{\rm CW}(r)\,.\label{eq:EHFnormal}
\end{equation}

Taking this Hartree-Fock like expression and ignoring the density
dependence of the correlation functions, we get
\begin{equation}
  \frac{d}{d\rho}\rho^2 \frac{d}{d\rho}\frac{E_{\rm R}}{N}
  = \rho\int d^3r V_{\rm CW}(r)\left[1-\frac{1}{\nu}j_0^2(r\KF )
    +\frac{1}{\nu}j_1^2(r\KF )\right]\label{eq:mc2HF}
\end{equation}
which evidently differs from 
\begin{equation}
\tilde V_{\rm p-h}(0+) \approx \rho\int d^3r V_{\rm CW}(r)
\label{eq:mcfromVph2}
\end{equation}
by approximately a factor of $1-1/\nu$.

That means, exchange diagrams must be included to obtain consistency
between the hydrodynamic derivative (\ref{eq:mcfromeos}) and $\tilde
V_{\rm p-h}(0+)$ in the low-density limit. The simplest version of the
FHNC hierarchy that corrects for this deficiency is FHNC//1 which
includes the exchange diagrams shown in Fig. \ref{fig:eelink}.
 We can extract the relevant modification from the full
FHNC-EL equations as formulated in Ref. \citenum{polish} by keeping
only the exchange term $V_{\rm ee}(k)$. The Euler equation remains
practically the same, except that the static structure function of the
non-interacting system becomes
\begin{equation}
  \SF(k) \rightarrow \SF(k) + (\Delta \tilde X_{\rm ee})^{(1)}(k)
  \approx S_\sigma(k)\,.
  \label{eq:Ssigma}
\end{equation}
where $S_\sigma(k)$ is the spin-structure function obtained from the
wave function (\ref{eq:Jastrow}), and the combination $\SF(k) +
(\Delta\tilde X_{\rm ee})^{(1)}(k)$ is the two-body approximation for
that function. Note that $(\Delta\tilde X_{\rm ee})^{(1)}(k)\propto
k^2$ as $k\rightarrow 0+$ whereas $\SF(k)\propto k$.

The particle-hole interaction is modified by
\begin{equation}
  \tilde V_{\rm p-h}(k) \rightarrow \tilde V_{\rm p-h}(k) +
  \tilde V_{\rm ex}(k)\,,\qquad \tilde V_{\rm ex}(k)\equiv
  \frac{\tilde V_{\rm ee}(k)}{S_\sigma^2(k)}
  \label{eq:Vphexc}
\end{equation}
where $\tilde V_{\rm ee}(k)$ is represented by the sum of the three
diagrams shown in Fig. \ref{fig:eelink}.
The red wavy line must then be identified with
\begin{equation}
  W(r) = V_{\!\rm p-h}(r) + w_I(r)\label{eq:Wdef}
\end{equation}
which is, of course, in the low density limit equal to $V_{\rm
  CW}(r)$. 
The Euler equation becomes
   \begin{equation}
     S(k) = \frac{\SF(k) + \tilde X_{\rm ee}(k)}
     {\sqrt{1+\displaystyle\frac{2\SF^2(k)}{t(k)}\tilde V_{\rm p-h}(k)}}\,.
       \label{eq:Stemp}
   \end{equation}
   The induced interaction is also modified and has, in the form
   of Eq. (\ref{eq:wind1}) an additional term
\begin{equation}
  \tilde w_{\rm I}(k) =\left((1+S_\sigma(k)\tilde\Gamma_{\rm dd}(k))^2-1\right)
  \tilde V_{\!\rm p-h}(k)
  + \frac{t(k)}{2}\tilde\Gamma_{\!\rm dd}^2(k) +
  \tilde\Gamma^2_{\rm dd}(k)
  \tilde V_{\rm ee}(k)\,.\label{eq:windex}
\end{equation}

Skipping the technical details (see also \ref{app:exchanges}), the
long-wavelength expansion of $\tilde V_{\rm ee}(k)$ is found to be
\begin{equation}
  \lim_{k\rightarrow 0}\tilde V_{\rm ex}(k)
  = \lim_{k\rightarrow 0}\frac{\tilde V_{\rm ee}(k)}{S_{\sigma}^2(k)}
  = -\frac{\rho}{\nu}\int d^3r W(r)\left[
    j_0^2(r\KF )-\frac{4}{3}j_1^2(r\KF )\right]\,.
  \label{eq:Vexlong}
\end{equation}

The factor $4/3$ compared to Eq. (\ref{eq:mc2HF}) is an incorrect
consequence of local correlation functions, but that term vanishes in
the low-density limit. The {\em leading\/} term in the density
expansion comes out correctly if the first order exchange diagrams are
included. Moreover, $W(r)\approx V_{\rm p-h}(r)$, and we see
that the first order exchange corrections lead to the desired factor
$1-1/\nu$ as they should.

However, the na\"\i ve addition of exchange diagrams is problematic in
the limit of low densities. In that limit, the three and four-body
diagram in $\tilde V_{\rm ex}(k)$ can be neglected, and we have
\begin{equation} V_{\rm ex}(r) = (g_{\rm F}(r)-1) W(r)\,.\label{eq:Veelowdens}
\end{equation}
Following the derivations of Section \ref{ssec:FHNC-EL} we and up
with a coordinate space equation of the form
   \begin{eqnarray}
   &&g_{\rm F}(r)\sqrt{1+\Gamma_{\rm dd}(r)}\left[v(r)+w_{\rm I}(r) - \frac{\hbar^2}{m}
    \nabla^2\right]\sqrt{1+\Gamma_{\rm dd}(r)}\nonumber\\
      &=&-(g_{\rm F}(r)-1)\frac{\hbar^2}{2m}\nabla^2\Gamma_{\rm dd}(r)\nonumber\\
      &&
      +\left[t(k)(1-S_\sigma^{-1}(k))\tilde\Gamma_{\rm dd}(k) - 2 S_\sigma(k)\Gamma_{\rm dd}(k)\tilde V_{\rm ee}(k)\right]^{\cal F}\!\!\!\!(r)\,.
      \label{eq:BGFHNC1}
   \end{eqnarray}
The last line in Eq. (\ref{eq:BGFHNC1}) goes to zero in the low
density limit, but the term in the second line does not. This leads to
solutions that are very different from the vacuum solution derived in
the previous section. The only way to rectify this situation (short of
solving the full FHNC-EL) equations is to use a slightly modified
relationship between $S(k)$ and the effective interactions, namely

\begin{equation}
 S(k) =
 \SF(k)\sqrt{\frac{1+\displaystyle\frac{2\SF^2(k)}{t(k)}\tilde V_{\rm
       ex}(k)} {1+\displaystyle\frac{2\SF^2(k)}{t(k)}\tilde
       V_{\rm p-h}(k)}}\,.
   \end{equation}
This relationship has a number of very interesting and very desirable
features: First, the square-root term in the numerator may be
identified with a ``collective RPA'' expression for the spin-structure
function,
   \begin{equation}
     S_\sigma(k) = \frac{\SF(k)}{\sqrt{1+\displaystyle\frac{2\SF^2(k)}{t(k)}
         \tilde V_{\rm ex}(k)}}\,,\label{eq:SsigmaColl}
   \end{equation}
we refer the reader to Section \ref{ssec:rings} for a
justification. The expression (\ref{eq:Stemp}) is then obtained by
expanding $S_\sigma(k)$ to first order in the interactions and
identifying
   \[\tilde X_{\rm ee}(k) \approx -\frac{\SF^3(k)}{t(k)} \tilde V_{\rm ee}(k)\,.\]

As mentioned above, the FHNC//1 approximation only leads to the
two-body approximation which is unsatisfactory for another reason: As
stated above, the spin-static structure function in that approximation
goes, for small $k$, as $\SF(k)$ which disagrees with experiments. The
expression (\ref{eq:SsigmaColl}) does not have this problem. In other
words, the thorough examination of the variational problem and the
demand for consistent treatment of short-- and long ranged
correlations as well as the proper low-density limit provides crucial
information on adequate approximation schemes for the Euler equation.

We have commented above on the fact that the positivity of the term
under the square root in the denominator is, with the
qualification that the Jastrow-Feenberg wave function is not exact,
related to the stability against density fluctuations. Likewise, the
positivity of the numerator is connected with the stability against
spin-density fluctuations.

In perturbation theory, the diagrams shown in Fig. \ref{fig:eelink}
correspond to the particle-hole ladder diagrams, driven by the
{\em exchange\/} term of the particle-hole interaction
\begin{equation}
  W_{\rm ex}(\hvec,\hvec';\qvec)
  = \Omega\bra{\hvec+\qvec,\hvec'-\qvec} W \ket{\hvec',\hvec}\,.
  \label{eq:Wex}
\end{equation}
This non-local term leads to a rather complicated addition to the
summation of the ring diagrams in the sense that it would supplement
the RPA sum by the RPA-exchange (or particle-hole ladder) summation.
We can again make the connection to the (local) FHNC expression
(\ref{eq:Vphexc}) by realizing that this expression is obtained from
the exact expression (\ref{eq:Wex}) by exactly the same hole-state
averaging process that was discussed in Section \ref{ssec:rings},
Eq. (\ref{eq:favg}):
\begin{equation}
  V_{\rm ex}(q) = \frac{\tilde V_{\rm ee}(q)}{S_{\rm F}^2(q)}
  = \left\langle  W_{\rm ex}(\hvec,\hvec';\qvec)\right\rangle(q)\,.
\end{equation}
The discussion of the preceding section shows that this localization
procedure of the exchange term maintains the dominant part of the
long-wavelength limit.

\subsection{Limitations of local correlation functions}
\label{ssec:limits}

One often learns most about a theory by examining situations where it
fails.  The wave function (\ref{eq:Jastrow}) is in principle exact for
bosons when correlation functions to all orders are included. It is
{\em not\/} exact for fermions since the nodes of the wave function
are identical to those of the non-interacting system.

A first consequence of the limitations of local correlation functions
was pointed out by Zabolitzky \cite{Zab80}: The exact high-density
expansion of the equation of state of a homogeneous electron gas is,
in units of the Wigner-Seitz radius $r_s$
\cite{Mac50,GellMannBrueckner}.
\begin{equation}
  \frac{E}{N} \approx \frac{2.21}{r_s^2} - \frac{0.916}{r_s} + 0.0622
  \ln r_s + C\quad\mathrm{Ry}\,.
  \label{eq:eegas}
  \end{equation}
A wave function of the form (\ref{eq:Jastrow}) leads, in this case, to
0.05690$\,\ln r_s\,$Ry for the logarithmic term. This deficiency is
corrected \cite{LanttoKroSmithOaxtepec} by second order correlated
basis functions (CBF) theory which will be reviewed in the next
Section \ref{ssec:CBF}

More recently \cite{cbcs}, we have examined the low-density limit of
equation of state of the weakly interacting Fermi gas. The exact
limit is expressed as a power series expansion in terms of the
parameter $\a0\KF$ \cite{HuangYang57,Landau5}
\begin{equation}
  \frac{E}{N} =\frac{\hbar^2 \KF^2}{2m}\left[\frac{3}{5}
    + \frac{2}{3}\frac{\a0 \KF }{\pi} + \frac{4(11-2\ln 2)}{35}
    \left(\frac{\a0\KF }{\pi}\right)^2 +\ldots\right]\,.
  \label{eq:elowdens}
\end{equation}
With the wave function (\ref{eq:Jastrow}) one obtains for the
coefficient of the third term the result of $1.5415$ instead of the
exact value $4(11-2\ln 2)/35 = 1.098$. Again, second order CBF theory
corrects this limit \cite{cbcs}.

A third issue is the stability of the system against infinitesimal
density fluctuations. We have commented about the connection between
the long-wavelength limit $V_{\rm p-h}(0+)$ and the hydrodynamic
compressibility for bosons. In a Fermi fluid, we also have the
Pauli repulsion, {\em i.e.\/} we should identify
\begin{equation}
  mc^2 = mc_{\rm F}^{*2} + \tilde V_{\rm p-h}(0+) \equiv  mc_{\rm F}^{*2}(1+F_0^S)
\label{eq:FermimcfromVph}
\end{equation} 
where $c_{\rm F}^{*2} = \frac{\hbar^2\KF^2}{3mm^*}$ is the speed of
sound of the non-interacting Fermi gas with the effective mass $m^*$,
and $F_0^s$ is Landau's Fermi liquid parameter.  Requiring a positive
compressibility leads to Landau's stability condition $F_0^s > -1$.

Solutions of the FHNC-EL equation exist if the expression under the
square root of Eq. (\ref{eq:PPA}) is positive, which leads to the
stability condition $F_0^s > -4/3$. This result is again a
manifestation of the fact that the wave function (\ref{eq:Jastrow}) is
not exact.  We shall return to this issue when we discuss the
generalization of the Euler equation to superfluid systems.

A final word is in order on the identification of $\tilde V_{\!\rm
  p-h}(0+)$ with the Fermi-liquid parameter $F_0^s$. We have already
seen above that exchange corrections are important and improve the
agreement in leading order in the exchange corrections. More
generally, the identification is, of course, only approximate and,
just like for bosons, exact agreement can be achieved only in an exact
theory \cite{parquet5}. A diagrammatic analysis that makes no
assumptions on the level of FHNC theory used is found in
Refs. \citenum{EKVar,KrKundt}. The FHNC//0 and FHNC//1 versions makes
more approximations, we have discussed these in Section
\ref{ssec:exchanges}.

\subsection{Elements of correlated basis functions}
\label{ssec:CBF}

We have seen above that a locally correlated wave function
(\ref{eq:Jastrow}) fails to reproduce several known exact features of
many-particle systems.  A way to cure the problem is provided by
perturbation theory with correlated basis functions (CBF theory)
\cite{FeenbergBook,ClW66}. Some of the basic ingredients of CBF theory
are also required for the formulating a theory for weakly coupled
superfluid systems \cite{HNCBCS,CBFPairing}. We review CBF theory here
only very briefly for the purpose of defining the essential
ingredients, details may be found in pedagogical material
\cite{KroTrieste} and review articles \cite{Johnreview,polish}. The
diagrammatic construction of the relevant ingredients has been derived
in Ref. \citenum{CBF2}.

CBF theory uses the correlation operator $F$ to generate a complete
set of correlated and normalized $N$-particle basis states through
where the correlated states $N$-body $\ket {\Psi_{\bf m}^{(N)}}$ are
\begin{eqnarray}
  \ket{ \Psi_{\bf m}^{(N)} } \equiv
  \left[I_{{\bf m}}^{(N)}\right]^{-1/2}F_N\ket{{\bf m}^{(N)}}\,,
  \qquad I_{{\bf m}}^{(N)}\equiv 
         {\bra{{\bf m}^{(N)} } F_{\!N}^{\dagger} F^{\phantom{\dagger}}_{\!N}
           \ket{{\bf m}^{(N)} }}
 \,,\label{eq:Psim}
\end{eqnarray}
where the $\{\ket{{\bf m}^{(N)}}\}$ form complete sets of $N$-particle
states, normally Slater determinants of single particle orbitals. When
unambiguous, we will omit the superscript $N$ referring to the
particle number. Although the $\ket{\Psi_{\bf m}^{(N)}}$ are not
orthogonal, perturbation theory can be formulated in terms of these
states \cite{MF1,FeenbergBook}.

In general, we label ``hole'' states which are occupied in $
\ket{{\bf o}}$ by $h$, $h'$, $h_i\;, \ldots\,$, and unoccupied
``particle'' states by $p$, $p'$, $p_i\;,$ \textit{etc.}.  To display
the particle-hole pairs explicitly, we will alternatively to the
notation $\ket{\Psi_{\bf m}}$ use
$\ket{\Psi_{p_1 \ldots p_d\, h_1\ldots h_d}} $.  A basis
state with $d$ particle-hole pairs is then
\begin{equation}
\ket{\Psi_{p_1 \ldots p_d\, h_1\ldots h_d}} 
=\left[I_{p_1,\dots h_1}^{(N)}\right]^{-1/2}
\creat{p_1}\cdots\creat{p_d}\annil{h_d}\cdots\annil{h_1}\ket{\bf o}
\,.
\label{eq:psimph}
\end{equation}

For the off-diagonal elements $O_{\bf m,n}$ of an operator $O$, we
sort the quantum numbers $m_i$ and $n_i$ such that $\ket{{\bf m}}$
  is mapped onto $\ket{{\bf n}}$ by
\begin{equation}
\label{eq:defwave}
\ket{{\bf m}} = \creat{m_1}\creat{m_2}
\cdots 
\creat{m_d} \; \annil{n_d} \cdots \annil{n_2}\annil{n_1}  
\ket{{\bf n}} \;.
\end{equation}
From this we recognize that, to leading order in the particle number
$N$, any matrix element of an operator $\hat O$
\begin{equation}
  O_{\bf m,n} = \bra{\Psi_{\bf m}} \hat O \ket{\Psi_{\bf n}}
\end{equation}
depends only on the {\em difference\/} between the states
$\ket{{\bf m}}$ and $\ket{{\bf n}}$, and {\em
not\/} on the states as a whole.  Consequently, $O_{\bf m,n}$ can be
written as matrix element of a $d$-body operator
\begin{equation}
\label{eq:defmatrix}
O_{\bf m,n} \equiv \bra{ m_1\, m_2 \, \ldots m_d \,} 
{\cal O}(1,2,\ldots d) \,\ket{n_1\,
n_2 \, \ldots n_d}_a \;.
\end{equation}
(The index $a$ indicates antisymmetrization.) 

The key quantities for the execution of the theory are diagonal and
off-diagonal matrix elements of unity and $H\!-\!H_{\bf o}$,
\begin{eqnarray}
M_{\bf m,n} &=& \ovlp{\Psi_{\bf m}}{\Psi_{\bf n}}
\equiv \delta_{\bf m,n} +  N_{\bf m,n}\;,
\label{eq:defineNM}
\\
W_{\bf m,n} &=& \bra{\Psi_{\bf m}}H-\frac{1}{2}\left(H_{\bf m}+H_{\bf n}\right)\ket{\Psi_{\bf m}} \,.
\label{eq:defineW}
\end{eqnarray}
Eq. (\ref{eq:defineW}) defines a natural decomposition
\cite{CBF2,KroTrieste} of the matrix elements of $H$ into the
off-diagonal quantities $W_{\bf m,n}$ and $N_{\bf m,n}$ and diagonal
quantities $H_{\bf m}$.

To leading order in the particle number, the {\em diagonal\/} matrix
elements of $ H\!-\!H_{\bf o}$ become additive, so
that for the above $d$-pair state we can define the CBF single
particle energies
\begin{equation}
\bra{\Psi_{\bf m}} H\!-\!H_{\bf o}  \ket{\Psi_{\bf m}} \>\equiv\>
\sum_{i=1}^d e_{p_ih_i}  + {\cal O}(N^{-1}) \;,
\label{eq:CBFph}
\end{equation}
with $e_{ph} = e_p - e_h$ where
\begin{align}
e_p &=\phantom{-}\bra{\Psi_p}\,\ H\!-\!H_{\bf o}
\ket{\Psi_p} =& t(p) + u(p)\nonumber\\
e_h &=-\bra{\Psi_h}\,H\!-\!H_{\bf o}\ket{\Psi_h}\ =& t(h) + u(h)\,
\label{eq:spectrum}
\end{align}
and $u(p)$ is an average field that can be expressed in terms of the
compound diagrammatic quantities of FHNC theory \cite{CBF2}

According to (\ref{eq:defmatrix}),
$W_{{\bf m},{\bf n}}$  and $N_{{\bf m},{\bf n}}$ define 
$d-$particle operators ${\cal N}$ and ${\cal W}$, {\em e.g.\/}
\begin{eqnarray}
N_{{\bf m},{\bf o}} &\equiv& N_{p_1p_2\ldots p_d \,h_1h_2\ldots h_d,0} \nonumber\\
&\equiv& \bra{ p_1p_2\ldots p_d \,}\, {\cal N}(1,2,\ldots,d)\,
\ket{\,h_1h_2\ldots h_d }_a  \;,\nonumber\\
W_{{\bf m},{\bf o}} &\equiv& W_{p_1p_2\ldots p_d \,h_1h_2\ldots h_d,0}\nonumber\\
&\equiv&  \bra{ p_1p_2\ldots p_d \,}\, {\cal W}(1,2,\ldots,d)\,
\ket{\,h_1h_2\ldots h_d }_a  \;.\qquad\;
\label{eq:NWop}
\end{eqnarray}
Diagrammatic representations of ${\cal N}(1,2,\ldots,d)$ and ${\cal
W}(1,2,\ldots,d)$ have the same topology \cite{CBF2}.  In
homogeneous systems, the continuous parts of the $p_i,h_i$ are wave
numbers ${\bf p}_i,{\bf h}_i$; we abbreviate their difference as ${\bf
q}_i$.

In principle, the ${\cal N}(1,2,\ldots,d)$ and ${\cal
  W}(1,2,\ldots,d)$ are non-local $d$-body operators. In the next
section, we will show that we need, for examining pairing phenomena,
only the two-body operators. Moreover, the low density of the systems
we are examining permits the same simplifications of the FHNC theory
that we have spelled out in Sec. \ref{ssec:FHNC}. In the same
approximation, the operators ${\cal N}(1,2)$ and ${\cal W}(1,2)$ are
local, and we have \cite{polish}
\begin{eqnarray}
{\cal N}(1,2) &=& {\cal N}(r_{12}) = \Gamma_{\!\rm dd}(r_{12})\nonumber\\
{\cal W}(1,2) &=& {\cal W}(r_{12})\,,\quad \tilde {\cal W}(k) =
\tilde W(k) = - \frac{t(k)}{\SF(k)}\tilde \Gamma_{\!\rm dd}(k)\,.
\label{eq:NWloc}
\end{eqnarray}

Explicit formulas for the single-particle spectrum $e_k$ may be found
in Ref. \citenum{polish}. Since we are only interested in relatively
weakly correlated systems, and also want to make the connection to
perturbation theory as transparent as possible, we spell out the
simplest form:
\begin{equation}
  e_k = t(k) + \frac{\tilde X_{\rm cc}'(k)}{1-\tilde X_{\rm cc}(k)} + \mathrm{const.}\,.
  \label{eq:uhncdef}
\end{equation}
where
\begin{eqnarray}
  \tilde X'_{\rm cc}(k) &=& -\frac{\rho}{\nu}
  \int d^3r\, e^{\I\kvec\cdot\rvec}W(r)\ell(r\KF)\,,\label{eq:Xccpdef}\\
  \tilde X_{\rm cc}(k) &=& -\frac{\rho}{\nu}
  \int d^3r\, e^{\I\kvec\cdot\rvec}\Gamma_{\rm dd}(r)\ell(r\KF)\,,
\label{eq:Xccdef}
\end{eqnarray}
Note that we have above approximated, among others, $L(r)\approx \ell(r\KF)$.

One of the most straightforward application of CBF theory is to
calculate corrections to the ground state energy. In second order we
have, for example,
\begin{equation}
\delta E_2 = -\frac{1}{4}\sum_{pp'hh'}
\frac
    {\left|\bra{ pp'}{\cal W}\ket{hh'}_a
+ \frac{1}{2}\left[e_p + e_{p'}-e_h -e_{h'}\right]
\bra{ pp'}{\cal N}\ket{hh'}_a
\right|^2}
{e_p + e_{p'}-e_h -e_{h'}}\,.
\label{eq:E2CBF}
\end{equation}
The magnitude of the CBF correction is normally comparable to the
correction from three-body correlations \cite{polish}. It is also
important to note that there are significant cancellations between the
two terms in the numerator. We have shown in previous work
that including the second order CBF correction (\ref{eq:E2CBF})
leads in the electron gas \cite{LanttoKroSmithOaxtepec} to the correct expansion
(\ref{eq:eegas}) as well to Huang-Yang expansion (\ref{eq:elowdens})
for a low-density Fermi gas \cite{cbcs}.

\section{Connections between FHNC and parquet diagrams}
\label{sec:parquet}

\subsection{Rings}
\label{ssec:rings}

The expression (\ref{eq:PPA}) reduces to the Bogoliubov equation for
the case of bosons, $\SF(k)=1$. It is therefore expected that it can
also be derived, for fermions, from the random phase approximation for
the dynamic structure function
 \begin{eqnarray}
    \chi(q,\omega) &=& 
    \frac{\chi_0(q,\omega)} {1-\tilde V_{\rm
        p-h}(q)\chi_0(q,\omega)}\label{eq:chiRPA}\,,\\
 S(q) &=& -\int_0^\infty \frac{d\omega}{\pi} \Im m\chi(q,\omega)
\label{eq:SRPA}
\end{eqnarray}
where $\chi_0(k,\omega)$ is the Lindhard function, and $\tilde V_{\rm
  p-h}(k)$ is a local quasiparticle interaction or ``pseudopotential''
\cite{Aldrich,ALP78}. Consistent with the convention (\ref{eq:ft})
according to which $\tilde V_{\rm p-h}(k)$ has the dimension of an
energy, we have defined the density-density response function slightly
different than usual \cite{FetterWalecka}, namely such that has the
dimension of an inverse energy.

Eq. (\ref{eq:PPA}) can be obtained by approximating the Lindhard
function $\chi_0(k,\omega)$ by a ``collective'' Lindhard function
(occasionally also referred to as ``one-pole approximation'' or
``mean spherical approximation'') $\chi_0^{\rm coll}(k,\omega)$.

This approximation may be justified in several ways. For further
reference, define for any function $f(\pvec,\hvec)$ depending on a
``hole momentum'' $|\hvec|<\KF$ and a ``particle momentum''
$\pvec=\hvec+\qvec$ with $|\pvec|>\KF$ the Fermi-sea average
\begin{equation}
\left\langle f(\pvec,\hvec) \right\rangle(q)
= \frac{\sum_{\hvec} \bar n(\hvec+\qvec) n(\hvec) f(\hvec+\qvec,\hvec)}
{\sum_{\hvec} \bar n(\hvec+\qvec) n(\hvec)}
= \frac{1}{S_{\rm F}(q)}\int \displaystyle\frac{d^3h}{V_{\rm F}}\bar n(\hvec+\qvec) n(\hvec)  f(\hvec+\qvec,\hvec)\,.
\label{eq:favg}
\end{equation}
where $V_{\rm F}$ is the volume of the Fermi sphere, $n(k) =
\theta(\KF-k)$ is the Fermi distribution, and $\bar n(k) = 1-n(k)$.
Replacing the particle-hole energies in the Lindhard function
$\chi_0(q,\omega)$ by the above ``collective'' or ``Fermi-sea
averaged'' energies
\begin{equation}
\left\langle t(\hvec+\qvec)-t(\hvec)\right\rangle(q)
= \frac{t(q)}{\SF(q)}\label{eq:tqcoll}\,.
\end{equation}
leads to a ``collective approximation'' of
the Lindhard function
\begin{equation}
        \chi_0^{\rm coll}(q,\omega) =
        \displaystyle \frac{2 t(k)}
        { (\hbar\omega+\I\eta)^2-
        \left(\displaystyle{\frac{t(q)}{\SF(q)}}\right)^2} \,.
\label{eq:Chi0Coll}
\end{equation}
and the RPA response function
\begin{equation}
        \chi^{\rm coll}(q,\omega) =
        \displaystyle \frac{2 t(q)}
        { (\hbar\omega+\I\eta)^2-
          \displaystyle{\left(\frac{t(q)}{\SF(q)}\right)^2-
            2t(q)\tilde V_{\rm p-h}(q)}} \,.
\label{eq:ChiColl}
\end{equation}

The frequency integration (\ref{eq:SRPA}) can then be carried out
analytically and leads to equation (\ref{eq:PPA}).

Another way to justify the collective approximation
(\ref{eq:Chi0Coll}) is to approximate the particle-hole band by an
effective single pole such that the $m_0$ and $m_1$ sum rules are
satisfied \cite{Rip79,polish}
\begin{eqnarray}
        -\Im m\int_0^\infty\frac{ d\omega}{\pi}\chi_0^{\rm coll}(q,\omega) &&=
        -\Im m\int_0^\infty\frac{ d\omega}{\pi}\chi_0(q,\omega) = \SF(q)
        \label{eq:m0MSA}\\
        -\Im m\int_0^\infty\frac{ d\omega}{\pi}\omega\chi_0^{\rm coll}(q,\omega) &&=
        -\Im m\int_0^\infty\frac{ d\omega}{\pi}\omega\chi_0(k,\omega) = t(q)\,.
\label{eq:m1MSA}
\end{eqnarray}
This leads to the same form (\ref{eq:Chi0Coll}).  A third way to
understand the approximation (\ref{eq:Chi0Coll}) is to realize that
the Lindhard function can be thought of as the Fermi-sea average of
the particle-hole excitations.
\begin{equation}
  \chi_0(q,\omega) =
  \SF(q)\left\langle\frac{1}{\hbar\omega+\I\eta+t(\hvec+\qvec)-t(h)}+
\frac{1}{\hbar\omega+\I\eta -t(\hvec+\qvec)+t(h)}\right\rangle(q)\,.
\end{equation}
Approximating this average value of the inverse of the particle-hole
spectrum by the inverse of the average value,
\begin{equation}
  \left\langle\frac{1}
{\hbar\omega+\I\eta\pm t(\hvec+\qvec)\mp t(h)}\right\rangle(q)
\approx\frac{1}
{\left\langle\hbar\omega+\I\eta\pm t(\hvec+\qvec)\mp t(h)\right\rangle(q)}
\label{eq:invav}
\end{equation}
also leads to the collective Lindhard function
(\ref{eq:Chi0Coll}).

The uniform limit approximation discussed in Section \ref{ssec:UL}
amounts to replacing the particle--hole interaction
$\tilde V_{\!\rm p-h}(k)$ by $\tilde v(k)$. We can then use the
familiar coupling constant integration (\ref{eq:FeynmanHellman}),

\begin{equation}
        \frac{E_{\rm RPA}}{ N} = \frac{E_{\rm HF}}{N}
         -\frac{1}{2\rho}\int\frac{ d^3 k d\omega}{(2\pi)^4}
        \int_0^1 d\lambda
        \tilde v(k)\Im m\left[\chi_\lambda (k,\omega)-\chi_0(k,\omega)\right]\,,
\label{eq:DERPA}
\end{equation}
where $\chi_\lambda (k,\omega)$ is the RPA density--density response
function (\ref{eq:chiRPA}) for an interaction $\lambda v(r)$.  Using
the collective approximation (\ref{eq:Chi0Coll}) in the RPA energy
correction, we can carry out both the coupling constant and the
frequency integration exactly and end up with the ``uniform limit''
expression (\ref{eq:EULnormal}) \cite{mixmonster}.

The connection between the chain diagrams of FHNC-EL theory and the
RPA ring diagrams can also be made more technical by summing all ring
diagrams in CBF perturbation theory. This is a rather tedious
procedure which involves calculating all contributions to both the
energy and the effective interaction ${\cal W}$ and ${\cal N}$
introduced in Section \ref{ssec:CBF} with the momentum flux of a ring
diagram in FHNC-EL {\em and\/} infinite order CBF perturbation theory.
The plausible result of the calculation is that the contribution of
all FHNC ring diagrams which involve the ``collective'' particle-hole
propagator (\ref{eq:Chi0Coll}) are canceled and replaced by the RPA
ring diagrams involving the exact Lindhard function.  For details, see
Refs.  \citenum{CBF2,rings} and also Ref. \citenum{KroTrieste} for
pedagogical material.

\subsection{Ladders}

We now turn to discuss the connection between the Euler equations of
FHNC-EL, specifically Eq. (\ref{eq:ELSchr}), and the Bethe-Goldstone
equation.  A treatment that is similarly rigorous as the one of the
ring diagrams is not available for the ladder diagrams that describe
short-ranged correlations
\cite{BRU55a,BRU55,BetheGoldstone57,Goldstone57,BruecknerLesHouches}.
We can nevertheless highlight the relationship between the
coordinate-space formulation (\ref{eq:ELSchr}) of the Euler equation
and the Bethe Goldstone equation.

We begin with the Bethe-Goldstone equation as formulated in
Eqs. (2.1), (2.2) of Ref. \citenum{BetheGoldstone57}. As above, it is
understood that $\pvec, \pvec'$ are particle states and $\hvec,
\hvec'$ are hole states; it is then unnecessary to spell out the
projection operators.

For the present purpose it is not convenient to go to the center of
mass frame. Then, the Bethe Goldstone equation reads
\begin{eqnarray}
  &&\bra{\kvec,\kvec'}G\ket{\hvec,\hvec'}=
\bra{\kvec,\kvec'}v\ket{\hvec,\hvec'}\label{eq:BG}
  \\
&-&
 \sum_{\pvec,\pvec'}
\frac{\bra{\kvec,\kvec'}v\ket{\pvec,\pvec'}
    \bra{\pvec,\pvec'}G\ket{\hvec,\hvec'}}
  { e(\pvec) + e(\pvec')
-e(\hvec)-e(\hvec')\,.}\,.\nonumber
\end{eqnarray}
Following Ref. \citenum{BetheGoldstone57}, introduce the pair wave
function
\begin{eqnarray}
  \bra{\kvec,\kvec'}\psi\ket{\hvec,\hvec'}
  &=&   \ovlp{\kvec,\kvec'}{\hvec,\hvec'}
  - \bar n(k)\bar n(k')\frac{\bra{\kvec,\kvec'}G\ket{\hvec,\hvec'}}
  {e(\kvec) + e(\kvec')
-e(\hvec)-e(\hvec')}\label{eq:fullpsi}\,.
\end{eqnarray}

Comparing this with Eq. (\ref{eq:BG}) we see that
\begin{eqnarray}
  \bra{\kvec\kvec'}G\ket{\hvec,\kvec'}
  &=& \sum_{\kvec_1,\kvec_1'}
  \bra{\kvec,\kvec'}v\ket{\kvec_1,\kvec_1'}
    \bra{\kvec_1,\kvec_1'}\psi\ket{\hvec,\hvec'}\nonumber\\
&=&\bra{\kvec,\kvec'}v\psi \ket{\hvec,\hvec'}\,.\label{eq:BGpsi}
\end{eqnarray}

This gives the relationship
\begin{eqnarray}
  \bra{\kvec,\kvec'}\psi\ket{\hvec,\hvec'}
  &=&   \ovlp{\kvec,\kvec'}{\hvec,\hvec'}
  - \bar n(k)\bar n(k')\frac{\bra{\kvec,\kvec'}v\psi
    \ket{\hvec,\hvec'}}
  {e(\kvec) + e(\kvec')
-e(\hvec)-e(\hvec')}\label{eq:fullpsi2}\,.
\end{eqnarray}
for the pair wave function $\psi$.  This is obviously still an
quantity that depends on three meomenta.

Bethe and Goldstone set the center of mass momentum zero which leaves
us still with a function of two variables. This is customary in
nuclear physics applications, what follows is what local parquet
diagram theory or FHNC-EL would suggest.

Making the connection to FHNC-EL we can proceed again in two different
ways. One is to approximate the energy denominator in
Eq. (\ref{eq:fullpsi2}) by its Fermi-sea average, note that the
$\bar n(k_i)$ factors in Eq. (\ref{eq:fullpsi2}) restrict the
integration range to unoccupied states.
\begin{equation}
\frac{1}{E_1(q)}\equiv
\left\langle \frac{1}{ e(\hvec+\qvec) + e(\hvec'-\qvec)
-e(\hvec)-e(\hvec'))}\right\rangle(q)\,.
\label{eq:E1ofq}
\end{equation}
Since the bare interaction is, per assumption, a function of momentum
transfer, the pair wave function $\psi$ also becomes a function of
momentum transfer only.  This gives a local equation
\begin{equation}
\psi(q)-\delta(q) = -\frac{[v\psi](q)}{E_1(q)}\,.
\end{equation}

The Fermi-sea integrals of the energy denominator defining $E_1(q)$
can be carried out analytically, see problem 1.5 in
Ref. \citenum{FetterWalecka}.

Another way to deal with this, which is
more in the spirit of Bethe and Goldstone, is to write
Eq. (\ref{eq:fullpsi2}) as
\begin{equation}
   \left[e(\kvec) + e(\kvec')-e(\hvec)-e(\hvec')\right]
\left[\bra{\kvec,\kvec'}\psi\ket{\hvec,\hvec'}-
\ovlp{\kvec,\kvec'}{\hvec,\hvec'}\right]\nonumber\\
  =-\bra{\kvec,\kvec'}v\psi\ket{\hvec,\hvec'}
  \label{eq:BGpsi3}\,.
\end{equation}
Approximating now
\begin{eqnarray}
  &&\left[e(\hvec+\qvec) + e(\hvec'-\qvec)-e(\hvec)-e(\hvec')\right]\nonumber\\
  &\approx& \left\langle e(\hvec+\qvec) + e(\hvec'-\qvec)-e(\hvec)-e(\hvec')
  \right\rangle(q) = \frac{2 t(q)}{\SF(q)}\,.
\end{eqnarray}
gives
\begin{equation}
2 \frac{t(q)}{\SF(q)}\left[\psi(q)-\delta(q)\right] = -[v\psi](q)
  \label{eq:BGSchq}
\end{equation}
or, in coordinate space
\begin{equation}
  \left[-\frac{\hbar^2}{m}\nabla^2 + v(r)\right]\psi(r)
  = \left[2t(q)(1-\SF^{-1}(q))(\tilde\psi(q)-\delta(q))\right]^{\cal F}(r)\,.
  \label{eq:BGSchr}
\end{equation}
Making the connection to the collective response functions of Section
\ref{ssec:rings} we note that we can write (\ref{eq:E1ofq}) as
\begin{equation}
  \frac{1}{E_1(q)}=
  -\displaystyle\int_0^\infty \frac{d\omega}{2\pi} \Im m\chi_0^2(q,\omega)\,.
\end{equation}
Replacing here $\chi_0(q,\omega)$ by  $\chi_0^{\rm coll}(q,\omega)$
leads to the same expression,
\[E_1^{\rm coll}(q) = 2\frac{t(q)}{\SF(q)}\,.\]
Thus, the different localization prescriptions discussed in Section
\ref{ssec:rings} all lead to the localized Bethe-Goldstone equation
(\ref{eq:BGSchq}). The direct calculation of the localized energy
denominator (\ref{eq:E1ofq}) offers an alternative which we have not
further pursued.

Both procedures are, of course, approximations; in particular the
second form permits to rewrite the Bethe-Goldstone equation in the
form of a non-local Schr\"odinger equation. The legitimacy can be
tested by comparing $\left\langle E(\qvec,\kvec,\kvec')
\right\rangle(q)$ with $E_1(q)$ as defined in Eq. (\ref{eq:E1ofq}).
The maximum deviation of the ratio $\left\langle E(\qvec,\kvec,\kvec')
\right\rangle(q)/E_1(q)$ from 1 is about 18 percent around $k\approx
2\KF$, in the relevant regime $0 \le k \le \KF$ the agreement is
better than 10 percent.

Comparing Eq. (\ref{eq:BGSchr}) with Eq. (\ref{eq:ELSchr}), it makes
sense to identify $\psi(r) \approx \sqrt{1+\Gamma_{\!\rm
    dd}(r)}$. Eq. (\ref{eq:BGSchr}) is then obtained by the further
assumption $\psi^2(r)-1 \ll 1$. More importantly, the bare interaction
of the Bethe Goldstone equation is supplemented by the induced
interaction.  This has the very important feature that the scattering
length of the effective interaction $v(r) + w_{\rm I}(r)$ is zero and
that, hence, the pair wave function falls off faster than
$1/r$. Historically, a rapid ''healing'' of the wave function was often
accomplished by introducing a gap in the single particle spectrum
\cite{BetheBrandowPetschek}. The literature on this subject is vast;
the reader is directed to review articles \cite{MahauxReview,MahauxSartor}
for details.

The identification between the two expressions (\ref{eq:BGSchr}) and
(\ref{eq:ELSchr}) is not as precise as in the case of ring diagrams,
but note that FHNC-EL//0 contains more than just particle-particle
ladders but also particle-hole and hole-hole ladders \cite{Rip79}.

One can also apply the same procedure to define a localized $G$ matrix
which is then
\begin{equation}
  G(q) = v(q) - \int \frac{d^3q'}{(2\pi)^3}v(\qvec-\qvec')
  \frac{\SF(q')G(q')}{2t(q')}\,.
  \label{eq:BGlocal}
  \end{equation}

The answer is, for $\SF(q)=1$, the well-known Bose limit which
should come out. The right-hand side of Eqs. (\ref{eq:BGSchr}) and
(\ref{eq:ELSchr}) manifests the fact that the {\em short--ranged\/} behavior
of the pair wave function is modified by the Pauli principle.
The effect has already been noted by Gomes, Walecka, and Weisskopf
\cite{GWW58}.

\subsection{Rungs}
\label{ssec:rungs}

Let us now turn to the localization procedure of parquet diagram
theory. There are two issues to clarify: One is to determine the
approximations that are made such that the procedure leads, similar to
the Bose case, to the Euler equations for the
Jastrow correlations. The second is to generalize the procedure to
fermions {\em without\/} such approximations.

We have commented above about the importance of the induced
interaction correction $w_{\rm I}(r)$ in the Bethe-Goldstone equation
or the coordinate-space Euler equation. From diagrammatic analysis, we
should identify $w_{\rm I}(r)$ with a local, {\em energy
  independent\/} approximation for the {\em energy-dependent\/}
particle-hole reducible vertex.  Assuming a local particle-hole
interaction $\tilde V_{\rm p-h}(q)$ we can write the chain
approximation for the full, energy dependent vertex as
\begin{equation}
  \widetilde W(q,\omega) =
  \frac{\tilde V_{\rm p-h}(q)}{1-\tilde V_{\rm p-h}(q)\chi_0(q,\omega)}
\label{eq:Vchain}
\end{equation}
and the particle-hole reducible part of that as
\begin{equation}
  \tilde w_{\rm I}(q,\omega) = \widetilde W(q,\omega) - \tilde V_{\rm p-h}(q)
  = 
  \frac{\tilde V^2_{\rm p-h}(q)\chi_0(q,\omega)}{1-\tilde V_{\rm p-h}(q)\chi_0(q,\omega)}\,.
\label{eq:Wchain}
\end{equation}

Following Refs. \citenum{parquet1,parquet2}, we now define
an energy-independent vertex by taking $W(q,\omega)$
at an average frequency $\bar\omega(q)$ defined by
 \begin{eqnarray}
 S(q) &=& -\int_0^\infty \frac{d\omega}{\pi} \Im m
\left[\chi_0(q,\omega) + \chi_0(q,\omega)\widetilde W(q,\bar \omega(q))
  \chi_0(q,\omega)\right]\nonumber\\
&=&\SF(q) - \widetilde W(q,\bar \omega(q))\int_0^\infty \frac{d\omega}{\pi}\,
\Im m \chi_0^2(q,\omega) \,.
\label{eq:Scond}
 \end{eqnarray}
The frequency integral in the second term can be carried out
analytically \cite{FetterWalecka}.  In the ``collective
approximation'' (\ref{eq:Chi0Coll}) for $\chi_0(q,\omega)$, we obtain
\begin{equation}
  \tilde W(q,\bar \omega(q)) = -\frac{t(k)}{\SF(k)}\tilde\Gamma_{\!\rm dd}(k)
  = \tilde W(q)\label{eq:Wlocal}
\end{equation}
where we recover the $\tilde W(q)$ of Eq. (\ref{eq:Wdef}) and the
induced interaction (\ref{eq:wind}). We can also calculate the
averaged frequency
\begin{equation}
  \hbar^2\bar\omega^2(q) = -\frac{t^2(q)}{\SF(q)(\SF(q)+2S(q))}\,.
  \label{eq:wbar}
\end{equation}
This is the same as Eq. (8) of Ref. \citenum{parquet2} for bosons
when we replace $\SF(q) \rightarrow 1$.

In fact, the collective approximation is not necessary; the frequency
integral can be carried out exactly for the full Lindhard function.
Of course, {\em both\/} versions are approximations for the fully
energy-dependent induced interaction, their comparison gives
information on the robustness of this approximation. We have commented
about this already in the section on the ladder diagrams. The above
analysis clarifies the relationships between the ring diagrams in
parquet theory and those of FHNC-EL.

The above-mentioned experience with the accuracy of the collective
approximation for energy calculations must of course be qualified:
Approximating the response function by a ``collective'' response
function can, of course, be expected to be a good approximation only
if the system indeed has a strong collective mode. This is indeed the
case for the density channel in \he3 which has, at zero pressure, a
Fermi-liquid parameter $F_0^s = 9.15$\cite{GRE83} or for electrons
where the plasmon contains all the strength.  If $F_0^s$ is small or
even negative, the collective mode is Landau damped. The collective
approximation (\ref{eq:Chi0Coll}) in the RPA expression
(\ref{eq:SRPA}) is still reasonably accurate, but not at the percent
level.

Another issue that needs to be addressed when moving from the
Jastrow-Feenberg description to parquet diagrams is the definition of
$\tilde\Gamma_{\!\rm dd}(k)$. In FHNC//0 we can obtain this quantity
from $S(k)$ via Eq. (\ref{eq:SofkFermi})
\begin{equation}
  \tilde\Gamma_{\!\rm dd}^{\rm FHNC}(q) = \frac{S(q)-\SF(q)}{\SF^2(q)}\,.
  \end{equation}
To construct the equivalent of this relationship in parquet theory, we
go back to Eq. (\ref{eq:Scond}). We should identify
\begin{equation}
  \tilde\Gamma_{\!\rm dd}(q)\SF^2(q) = -\tilde W(q,\bar \omega)
  \int_0^\infty \frac{d\omega}{\pi} \Im m\chi_0^2(q,\omega)
\end{equation}
Here, we encounter again the integral (\ref{eq:Scond}) which can
be carried out either exactly of in the collective approximation.
The latter leads to the connection (\ref{eq:wbar}) as a definition
for $\tilde\Gamma_{\!\rm dd}(q)$. The FHNC approximation
$\tilde\Gamma_{\!\rm dd}^{\rm FHNC}(q)$ is then obtained by replacing
$\chi_0(q,\bar\omega(q))$ by $\chi_0^{\rm coll}(q,\bar\omega(q))$.

\subsection{Self-energy}
\label{ssec:selfen}

Single-particle properties are in perturbation theory normally
discussed in terms of the Dyson-Schwinger equation
\cite{Dyson49,Schwinger51a,Schwinger51b}. While in principle exact,
approximations are necessary to make the theory useful. A popular
approximation is the so-called G0W approximation
\cite{Hedin65,Rice65,FetterWalecka,FrimanBlaizot} for the
self--energy,
\begin{equation}
\Sigma(k,E) = U(k) + \I\int \frac{d^3 q \, d(\hbar\omega)}{
  (2\pi)^4} G_0({\bf k}-{\bf q},E-\hbar\omega)
\tilde V_{\rm p-h}^2(q)\chi(q,\omega)
\label{eq:G0W}
\end{equation}
where
\begin{equation}
G^{(0)}(k,\omega) = \frac{\bar n(k)}{\hbar\omega - t(k) + \I\eta}
+ \frac{n(k)}{\hbar\omega - t(k) - \I\eta}
\label{eq:green}
\end{equation}
is the free single-particle Green's function. $U(k)$ is a static
field, the Fock term in Hartree-Fock approximation, possibly
the Brueckner-Hartree-Fock term in Brueckner theory \cite{MahauxReview}.

In the term $\tilde V_{\rm p-h}^2(q)\chi(q,\omega)$ we recover the RPA
for the energy dependent effective interactions $\tilde w_{\rm
  I}(q,\omega)$ (\ref{eq:Wchain}). Setting, as above, $\tilde w_{\rm
  I}(q,\omega) \approx \tilde w_{\rm I}(q,\bar\omega(q))$ we can carry
out the frequency integration and obtain a {\em static\/} self-energy
simply in the form of a Fock term in terms of the induced interaction
$w_I(q)$.  Since this sums all chain (or particle-hole reducible)
diagrams, the static field $U(k)$ should be the Fock term generated by
the particle--hole {\em irreducible\/} interaction, {\em i.e.\/}
$\tilde V_{\rm p-h}(q)$. To summarize, we obtain in this simplest
approximation of the FHNC-EL equations that the numerator $\tilde
X_{\rm cc}'(k)$ of Eq. (\ref{eq:uhncdef}) is the same as static
approximation for the G0W self-energy when the effective interaction
is taken at the average frequency $\bar\omega(q)$.  Of course, the
summation of the full set of the ``cc'' equations and the associated
linear equation for $\tilde X_{\rm cc}'(k)$ as well as retaining the
denominator $1-\tilde X_{\rm cc}(k)$ implies that much larger classes
of diagrams are routinely calculated in FHNC.

The study of the self--energy highlights, at a somewhat more technical
level, another limitation of locally correlated wave functions:
Strictly speaking, $\Sigma(k,\omega)$ depends on energy and
momentum. Approximating this function by an ``average''
energy-independent function misses the important non-analytic
structure of the self-energy around the Fermi surface. This has the
well-known consequence of an enhancement of the effective mass in
nuclei around the Fermi surface \cite{BrownGunnGould}. The effect is
quite dramatic in $^3$He \cite{PethickMass,ZaringhalamMass} where a
Jastrow-Feenberg wave function predicts an effective mass ratio $m^*/m
< 1$ in massive contrast to the experimental value around $m^*/m
\approx 3$ \cite{GRE83,GRE86} which are well reproduced when the full
self-energy is calculated \cite{Bengt,he3mass}.

To conclude this discussion, a few more remarks are in order:
\begin{itemize}
\item{} The strong enhancement of the effective mass in \he3 is due
  to an interplay between density-fluctuations describing basically
  hydrodynamic backflow, and spin-fluctuations describing the emission
  and re-absorption of a low-energy magnon \cite{he3mass}. We have
  spelled out in Eq. \ref{eq:G0W} only the density channel which
  is the most important one for out purposes.
\item{} One can, of course, also use the collective density--density
  response function (\ref{eq:ChiColl}). This would describe the
  coupling to density fluctuations but treat the coupling to
  particle-hole excitations only very approximately; in particular
  it would wipe out the structure of the self-energy around the Fermi
  momentum. The approximation can be useful for describing the motion
  of impurities in a Fermi liquid.
\end{itemize}

\section{BCS theory for local correlations}
\label{sec:hncbcs}

Let us now turn to the generalization of the correlated wave functions
method to superfluid systems. Since we have reviewed the FHNC-EL
theory and its relation to parquet diagrams above, we can restrict
ourselves to the discussion of what changes for a superfluid system.
Previous work has either assumed that the superfluid state deviates
little from the normal state
\cite{ectpaper,HNCBCS,CBFPairing,shores,CCKS86,cbcs} and/or adopted
low-order cluster expansions
\cite{YangClarkBCS,Fabrocinipairing,Pavlou2017,Benhar}. The
Jastrow-Feenberg variational approach has never been developed to a
level comparable to the normal system which made the identification
with parquet-diagrams possible. This is one of the tasks of our work.

We construct a correlated wave function for a superfluid system by
combining the BCS wave function of a weakly interacting system,
\begin{equation}
\ket{\rm BCS} =
{\prod_{\kvec}}
\left[ u_{\kvec} +  v_{\kvec} a_{ {\bf  k} \uparrow }^\dagger
 a_{-{\bf  k} \downarrow}^\dagger  \right] \ket{{\bf o}}\,
\label{eq:BCS}
\end{equation}
where $u_\kvec$, $v_\kvec$ are Bogoliubov amplitudes satisfying
$u_\kvec^2 + v_\kvec^2 = 1$, with the Jastrow-Feenberg wave function
(\ref{eq:wavefunction}), (\ref{eq:Jastrow}), to the form
\begin{equation}
\ket{\rm CBCS} =  \sum_{{\bf m},N} \ket {\Psi_{\bf m}^{(N)}}
\ovlp{{\bf m}^{(N)}}{\rm BCS}\,.
\label{eq:CBCS}
\end{equation}
We have commented on alternative choices
\cite{Fantonipairing,Fabrocinipairing} of the correlated BCS wave
function in Ref. \citenum{cbcs}.

In what follows, we will refer to expectation values with respect to
the {\em uncorrelated\/} state (\ref{eq:BCS}) as
$\left\langle\ldots\right\rangle_0$ and those with respect to the {\em
  correlated state\/} (\ref{eq:CBCS}) as
$\left\langle\ldots\right\rangle_c$.  Physically interesting
quantities like (zero temperature) Landau potential of the superfluid
system
\begin{equation}
  \left\langle H'\right\rangle_c = \frac{\bra{\mathrm{CBCS}} \hat H'
    \ket{\mathrm{CBCS}}}
  {\bigl\langle\mathrm{CBCS}\big|\mathrm{CBCS}\bigr\rangle}\,,\qquad
  \hat H' \equiv \hat H - \mu\hat N\,.
  \label{eq:EBCS}
\end{equation}
are then calculated by cluster expansion and resummation techniques;
the correlation functions are determined by the variational principle
\begin{equation}
\frac{\delta  \left\langle H'\right\rangle_c }
{\delta u_n}({\bf r}_1,\ldots,{\bf r}_n) = 0\,.
\label{eq:euler}
\end{equation}

\subsection{Weakly coupled systems}
\label{ssec:weak}

We have simplified in Refs. \citenum{cbcs} and \citenum{ectpaper} the
problem by expanding $\left\langle H'\right\rangle_c$ (\ref{eq:EBCS})
in the {\em deviation\/} of the Bogoliubov amplitudes $u_{\kvec}$,
$v_{\kvec}$ from their normal state values $u^{(0)}_{\kvec}= \bar
n(k)$, $v^{(0)}_{\kvec}=n(k)$. This approach adopts a rather different
concept than the original BCS theory: A wave function of the form
(\ref{eq:BCS}) begins by creating Cooper pairs out of the
vacuum. Instead, the approach (\ref{eq:CBCS}) begins with the {\em
  normal, correlated\/} ground state and generates one Cooper pair at
a time out of the normal system as suggested recently by Leggett
\cite{LeggettQFS2018}. Adopting an expansion in the number of Cooper
pairs, the correlation functions $u_n(\rvec_1,\ldots\rvec_n)$ can be
optimized for the normal system.

Carrying out this expansion in the number of Cooper pairs, we have
arrived at the energy expression of the superfluid state
\begin{eqnarray}
\langle \hat H' \rangle_c &=& H_{\bf o}^{(N)} - \mu N + 2 \sum_{\kvec
  ,\,|\, \kvec \,|\,>\KF} v_{\kvec}^2 (e_{\kvec} - \mu ) - 2
\sum_{\kvec , \,|\, \kvec \,|\,<\KF} u_{\kvec}^2 (e_{\kvec} - \mu )
\nonumber \\ &\quad& + \sum_{\kvec,\kvec'}u_\kvec v_\kvec u_{\kvec'}
v_{\kvec'} {\cal P}_{\kvec\kvec'}\,.
\label{eq:Ebcs}
\end{eqnarray}
Above, $H_{\bf o}^{(N)}$ is the energy expectation value of the normal
$N$-particle system, and $\mu$ is the chemical potential. The
$e_{\kvec}$ are the single particle energies derived in correlated
basis function (CBF) theory \cite{CBF2}, see also Section
\ref{ssec:selfen}. The paring interaction has the form
\begin{eqnarray}
{\cal P}_{\kvec\kvec'} &=& {\cal W}_{\kvec\kvec'}+(|e_{\kvec}- \mu | 
+ |e_{\kvec'}- \mu |)
{\cal N}_{\kvec\kvec'}\label{eq:Pdef}\,,\\
{\cal W}_{\kvec\kvec'} &=& \bra{\kvec \uparrow ,-\kvec\downarrow}
{\cal W}(1,2)\ket{\kvec'\uparrow ,-\kvec'\downarrow}_a\,,\label{eq:Wnldef}\\
{\cal N}_{\kvec\kvec'}&=&
\bra{\kvec \uparrow ,-\kvec\downarrow}
{\cal N}(1,2)\ket{\kvec'\uparrow , - \kvec'\downarrow}_a\,.
\label{eq:Ndef}\end{eqnarray}
The effective interaction ${\cal W}(1,2)$ and the correlation
corrections ${\cal N}(1,2)$ are given by the compound-diagrammatic
ingredients of the FHNC-EL method for off-diagonal quantities in CBF
theory \cite{CBF2}, see Section \ref{ssec:CBF}.

The Bogoliubov amplitudes $u_\kvec $, $v_\kvec $ are obtained in the
standard way by variation of the energy expectation (\ref{eq:Ebcs}).
This leads to the familiar gap equation
\begin{equation}
\Delta_\kvec = -\frac{1}{2}\sum_{\kvec'} {\cal P}_{\kvec\kvec'}
\frac{\Delta_{\kvec'}}{\sqrt{(e_{\kvec'}-\mu)^2 + \Delta_{\kvec'}^2}}\,.
\label{eq:gap}
\end{equation}

The conventional ({\em i.e.\/} ``uncorrelated'' or ``mean-field'') BCS
gap equation \cite{FetterWalecka} is retrieved by replacing the
effective interaction ${\cal P}_{\kvec\kvec'}$ by the pairing matrix
of the bare interaction.

\subsection{Cluster expansions for a superfluid system}
\label{ssec:cbcs}

The expansion in the number of Cooper pairs created out of the normal
ground state described above is legitimate as long as the superfluid
gap function $\Delta_\kvec$ is small. When that assumption is not
satisfied, one must evaluate all physical quantities of interest for
the fully correlated BCS state (\ref{eq:CBCS}). This is the purpose of
this section.

The central quantity in the development of our method is the
zero-temperature grand (or Landau) potential (\ref{eq:EBCS}) of the
superfluid system which we can write as
\begin{equation}
  \left\langle H'\right\rangle_c
  =  \frac{\sum_{N,{\bf m},{\bf n}} \ovlp{\rm BCS}{{\bf m}^{(N)}}
  \bra{\Psi_{\bf m}^{(N)}}\hat H'\ket{\Psi_{\bf n}^{(N)}}
\ovlp{{\bf n}^{(N)}}{\rm BCS}}{\ovlp{\rm CBCS}{\rm CBCS}}\,.
\label{eq:EBCSsum}
\end{equation}
For the development of the formal theory, we utilize the methods
developed for the cluster expansions of the normal system \cite{CBF2}
and outlined in Section \ref{ssec:CBF} and modify them for the present
case.  We can write $\left\langle H\right\rangle_c$ in terms of these
quantities as

\begin{eqnarray}
  \left\langle H'\right\rangle_c &=&  \sum_{N,{\bf m}}
  \ovlp{\rm BCS}{{\bf m}^{(N)}}
  {H'}_{\mathbf{m},\mathbf{m}}^{(N)}\ovlp{{\bf m}^{(N)}}{\rm BCS}\nonumber\\
  &+&
  \sum_{N,{\bf m},{\bf n}}\frac{
    \ovlp{\rm BCS}{{\bf m}^{(N)}}W_{\mathbf{m},\mathbf{n}}^{(N)}\ovlp{{\bf n}^{(N)}}{\rm BCS}}{\ovlp{\rm CBCS}{\rm CBCS}}
  \nonumber\\
  &+&\frac{1}{2}\frac{\sum_{N,{\bf m},{\bf n}}
    \ovlp{\rm BCS}{{\bf m}^{(N)}}\left({H'}_{\mathbf{m},\mathbf{m}}^{(N)}
    + {H'}_{\mathbf{n},\mathbf{n}}^{(N)}-2E_{\rm diag}\right)
    N_{{\bf m},{\bf n}}^{(N)}\ovlp{{\bf n}^{(N)}}{\rm BCS}}{\ovlp{\rm CBCS}{\rm CBCS}}\nonumber\\
  &\equiv& E_{\rm G} + E_{\rm enum} \,.
  \label{eq:FCBCS}
  \end{eqnarray}
where $E_{\rm diag}$ is the term in the first line.

The decomposition (\ref{eq:FCBCS}) has the following purpose: In what
follows, we will see that one can obtain the combination $E_{\rm G}$
by appropriate generalization of the diagrammatic expansions for the
{\em normal\/} system, in particular the cluster expansion of this
term is {\em irreducible} and one can utilize the technique of a
``generating function'' to obtain it. The ``energy numerator'' term
$E_{\rm enum}$ requires separate treatment; it leads to the second
term in the definition (\ref{eq:Pdef}) of the pairing matrix elements.

Carrying out the cluster expansions is a rather tedious matter
\cite{FanThesis}; their main purpose is to verify, to a convincingly
high order, the diagrammatic rules according to which the basic
quantities of the theory are constructed. We display here only the
characteristic steps and the most relevant result.  A few simple
examples on how the cluster expansions are carried out will be shown
in \ref{app:AppA} and \ref{app:AppB}.

We utilize the ``generating function'' technique by replacing
$u_2(r)$ by $u_2(r;\beta)$ in the correlated state (\ref{eq:CBCS}).
Apart from the ``Jackson-Feenberg'' kinetic energy terms which
originate from the last term in Eq. (\ref{eq:JFIdentity}) and will be
spelled out below, the diagonal and off-diagonal energy terms are
\begin{eqnarray}
  G_{\rm diag}(\beta) &=& \sum_{N,{\bf m}}
  \ovlp{\rm BCS}{{\bf m}^{(N)}} \ln
  I_{\mathbf{m},\mathbf{m}}^{(N)}(\beta)\ovlp{{\bf m}^{(N)}}{\rm
    BCS}\label{eq:Gdiag}\\
  G_{\rm offd}(\beta) &=&
  \ln\ovlp{\mathrm{CBCS(\beta)}}{\mathrm{CBCS(\beta)}}\nonumber\\ &=&
  \ln\left[1+\sum_{N,{\bf m} \neq {\bf n}} \ovlp{\rm BCS}{{\bf
        m}^{(N)}}N_{\mathbf{m},\mathbf{n}}^{(N)}(\beta)
    \ovlp{\mathbf{n}^{(N)}}{\mathrm{BCS}}\right]\label{eq:Goffd}\,.\\
  G(\beta) &\equiv& G_{\rm diag}(\beta) + G_{\rm offd}(\beta)\,.\label{eq:Gdef}
\end{eqnarray}

\begin{figure}[H]
\centerline{\includegraphics[width=0.9\columnwidth]{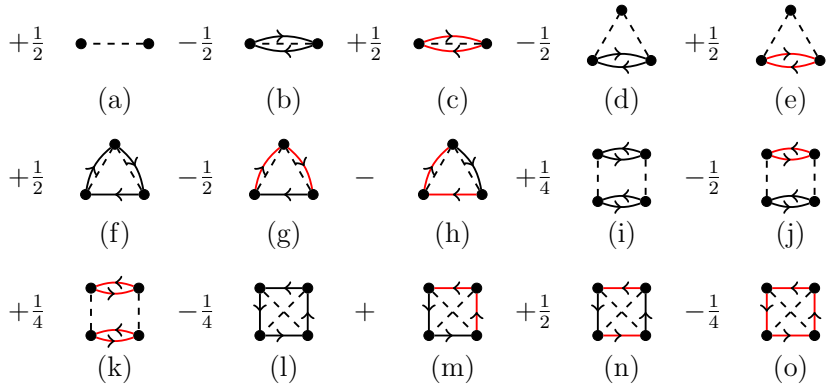}}
\caption{(color online) The figure shows the leading diagrams
  contributing to the generating function $G(\beta)$.  The black
  and red exchange lines depict the exchange functions
  $\ell_v(r_{ij})$ and $\ell_u(r_{ij})$, {\em cf.}  Eqs.
  (\ref{eq:lvdef}) and (\ref{eq:ludef}), respectively. The dashed
  lines denote, as usual, correlation functions $h(r_{ij},\beta)$. The
  term $G_{\rm diag}(\beta)$ is represented by the subset of diagrams
  containing only black exchange lines.
\label{fig:G}}
\end{figure}

The purpose of the representation (\ref{eq:Gdef}) is that cluster
expansion and resummation techniques for $G$ can be
constructed directly from the corresponding expression for the normal
system. We have discussed the rules for the normal system in Section
\ref{ssec:cluster_expansions} above.  Our diagrammatic analysis for
the {\em superfluid\/} system has resulted in the assertion that the
corresponding expansion is derived from that by the following rule:

\begin{enumerate}
\item{} Interpret the density factor as
\begin{equation}
  \rho = \frac{\nu}{\Omega}\sum_{\kvec} v_{\kvec}^2\,,
  \label{eq:density_factor}
\end{equation}
where $\Omega$ is the normalization volume. Note that this is the
density of the model system described by the uncorrelated BCS state
(\ref{eq:BCS}) and not the density corresponding to the correlate
state (\ref{eq:CBCS}).  See Eq. (\ref{eq:density_correction}) for the
correlation correction to the density.

  \item{} Re-interpret all exchange lines $\ell(r\KF)$ as
\begin{equation}
    \ell_v(r) \equiv \frac{\nu}{\rho\Omega}
    \sum_{\kvec} v_{\kvec}^2e^{\I\kvec\cdot\rvec} = \frac{\nu}{\rho}
\int \frac{d^3 k}{(2\pi)^3} v_{\kvec}^2e^{\I\kvec\cdot\rvec}  \,.  
  \label{eq:lvdef}
\end{equation}
\item{} In each exchange loop, replace, in turn, each pair of exchange
  lines $\ell_v(r_{ij})\ell_v(r_{kl})$ by a pair
  $-\ell_u(r_{ij})\ell_u(r_{kl})$, where
    \begin{equation}
   \ell_u(r) \equiv \frac{\nu}{\rho\Omega}
   \sum_{\kvec} u_{\kvec}v_{\kvec}e^{\I\kvec\cdot\rvec} =  \frac{\nu}{\rho}
\int \frac{d^3 k}{(2\pi)^3} u_{\kvec}v_{\kvec}e^{\I\kvec\cdot\rvec}\,.
   \label{eq:ludef}
\end{equation}
\end{enumerate}
The terms containing only $\ell_v(r_{ij})$ lines come from the
``diagonal'' term $G_{\rm diag}$ and those containing at least one
pair of $\ell_u(r)$ lines originate from $G_{\rm offd}(\beta)$.  The
diagrammatic expansion of the generating function $G(\beta)$ can
therefore be read off from the diagrammatic expansions for the normal
system discussed in Section \ref{sec:parquet}. As an example, we show
in Fig. \ref{fig:G} the diagrammatic representation of some leading
cluster contributions to $G(\beta)$.

From the cluster expansion for the generating function $G(\beta)$
we can construct the expansion for $E_{\rm G}$
\begin{equation}
E_{\rm G} =  \left\langle \hat T-\mu \hat N\right\rangle_0
  + \left. \frac{d}{d\beta}
   G(\beta)\right|_{\beta=0} + T_{{\rm JF},{\rm G}}\,.
\end{equation}
Note, in particular, that the expansion of $E_{\rm G}$ is {\em
  irreducible\/}. Diagrammatically, the corresponding expansion of the
energy $E_{\rm G}$ can be generated from the generating functional by
generalizing the construction rules spelled out in Section
\ref{ssec:cluster_expansions} by
\begin{enumerate}
  \setcounter{enumi}{1}
\item{} To calculate the kinetic energy corrections $T_{{\rm JF},{\rm
    G}}$ replace any pair of exchange lines with one common point,
  $\ell_{\{u,v\}}(r_{ij})\ell_{\{u,v\}}(r_{ik})$, by a pair
  $\frac{\hbar^2}{8m}
  \nabla_i^2\ell_{\{u,v\}}(r_{ij})\ell_{\{u,v\}}(r_{ik})$.
\end{enumerate}

\begin{figure}[H]
\centerline{\includegraphics[width=0.9\columnwidth]{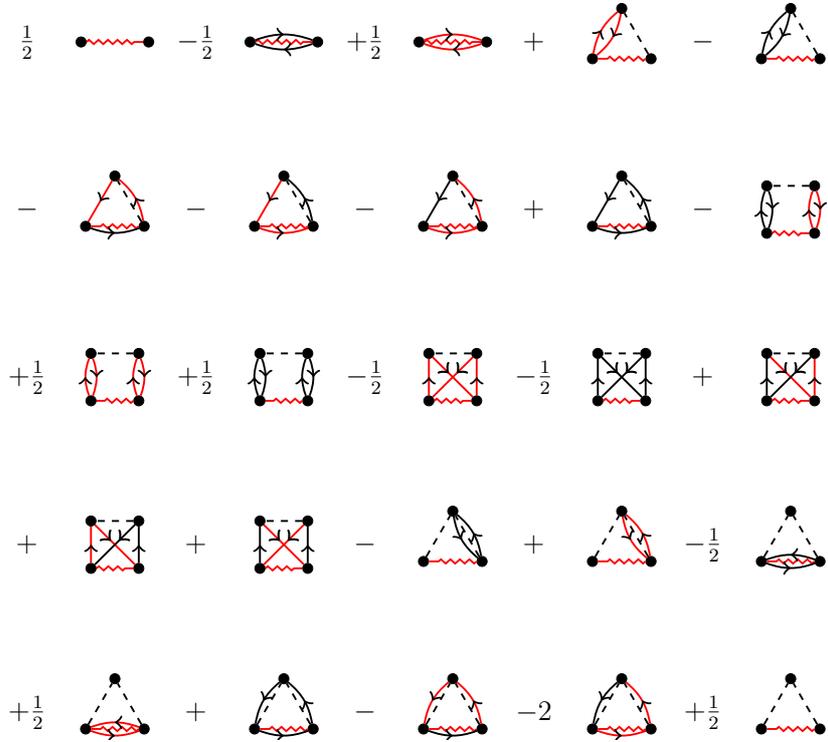}}
\caption{(color online) The figure shows the leading diagrams
  contributing to the potential energy. The black and red exchange
  lines depict the exchange functions $\ell_v(r_{ij})$ and
  $\ell_u(r_{ij})$, \cf  Eqs.  (\ref{eq:lvdef}) and
  (\ref{eq:ludef}), respectively, and the red wavy line denotes an interaction
  $f^2(r_{ij})v_{\rm JF}(r_{ij})$. Otherwise we follow the usual
  diagrammatic conventions \cite{Johnreview}.
\label{fig:epot}}
\end{figure}

The last term, $E_{\rm enum}$ gives rise to the ``energy numerator
corrections'' shown in (\ref{eq:Pdef}). A discussion of the
significance of these terms is found in Refs.  \citenum{cbcs} and
\citenum{ectpaper}. Basically, the formulation
(\ref{eq:Pdef})-(\ref{eq:gap}) amounts to a reformulation of the gap
equation in terms of the $T$-matrix as carried out, for example, in
Ref. \citenum{PethickSmith}.

Cluster expansions for the energy numerator terms must be treated
separately. These corrections consist of a series of
products of {\em diagonal\/} matrix elements of the Hamiltonian, and
{\em off-diagonal\/} matrix elements of the correlation operator. We
classify the individual terms according to the number of common states
contained in these two sets of matrix elements. The simplest, and
dominating, energy numerator terms are those that have only one common
state; in particular these are the only ones that survive in the
weakly coupled limit discussed in Section \ref{ssec:weak}. The
derivation of these terms will be outlined in \ref{app:AppC}. To
express the construction rules, we need the following two generalized
exchange lines
\begin{eqnarray}
  \ell'_u(r) &\equiv&
  \frac{\nu}{\rho}\int \frac{d^3k}{(2\pi)^3} (e_{\kvec}^{(v)}-\mu) (u_\kvec^2-v_\kvec^2)
  u_\kvec v_\kvec
  e^{\I\kvec\cdot\rvec} \label{eq:lupdef}\\
  \ell_v'(r) &\equiv&
  \frac{2\nu}{\rho}\int \frac{d^3k}{(2\pi)^3}
  (e_{\kvec}^{(v)}-\mu) u_\kvec^2v_\kvec^2
   e^{\I\kvec\cdot\rvec} \label{eq:lvpdef}
\end{eqnarray}
where the $e_{\kvec}^{(v)}$ are single particle energies generated
from the ordinary CBF single particle energies $e_{\kvec}$ by
replacing all exchange lines by $\ell_v(r)$. The leading term in the
$e_{\kvec}$ is simply the kinetic energy $t(k)$.

The construction rule to obtain these diagrams from the generating
functional is then:
\begin{enumerate}
\item Replace, in turn, each $\ell_u(r_{ij})$ (red-arrow line) by
  $\ell_u'(r_{ij})$ (double red-arrow line) as defined in
  Eq. (\ref{eq:lupdef}).
\item Replace, in turn, $\ell_v(r_{ij})$ (black-arrow line) by
  $\ell_v'(r_{ij})$ (double black-arrow line) as defined in
  Eq. (\ref{eq:lvpdef}). For this, we must interpret the single dot as
  an $\ell_v(r_{ii})$ line returning into itself.
\end{enumerate}

\subsection{FHNC and Euler equations}
\label{ssec:FHNC}

We formulate in this section the simplest version of the FHNC-EL
equations, or FHNC//0 approximation. It is straightforward to
formulate and implement the full FHNC equations \cite{polish}; in fact
we have used the FHNC//1 approximation in our numerical
applications. As above, the FHNC//0 form displays the physical
content of the theory more clearly. Moreover, we shall show that
corrections {\em beyond\/} the local correlation function are much
more important than adding more complicated FHNC diagrams.

\begin{figure}[H]
  \centerline{\includegraphics[width=0.90\textwidth]{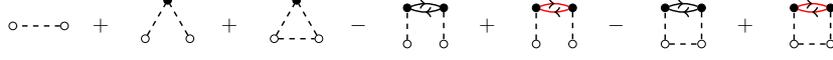}}
  \caption{Diagrammatic expansion of the ``dressed'' correlation line
    $\Gamma_{\!\rm dd}(r)$ in terms of the ``bare'' correlation
    function $h(r) = f^2(r)-1 \equiv \exp(u_2(r))-1$ and exchange
    lines. The first diagram is the bare line, diagrams \#2, \#4, and
    \#5 are the simplest chaining operations containing two
    correlation lines, and diagrams \#3, \#6, and \#7 are parallel
    connections of a correlation line and the chains diagrams \#2,
    \#4, and \#5.  \label{fig:GddFHNC0} }
\end{figure}

For further reference, we need the following quantities
\begin{eqnarray}
  \sigma_v(r) &=& \frac{1}{\rho}\ell_v(0)\delta(r)-\frac{1}{\nu}\ell_v^2(r)
  \label{eq:sigmavdef}\\
  \sigma_u(r) &=& \frac{1}{\nu}\ell_u^2(r)
  \label{eq:sigmaudef}\,.\\
  \sigma'_v(r) &=& \frac{1}{\rho}\ell_v'(0)\delta(r)-\frac{2}{\nu}\ell_v(r)\ell_v'(r)
  \label{eq:sigmavpdef}\\
  \sigma'_u(r) &=& \frac{2}{\nu}\ell_u(r)\ell_u'(r)\,.\label{eq:sigmaupdef}
  \end{eqnarray}
as well as their Fourier transforms $\tilde \sigma_v(k)$, $\tilde
\sigma_u(k)$, $\tilde \sigma_v'(k)$, and $\tilde \sigma_u'(k)$.  We
begin with the expansion of the ``dressed'' correlation line
$\Gamma_{\!\rm dd}(r)$. In FHNC//0, one keeps only the diagrams shown
in Fig. \ref{fig:GddFHNC0} {\em plus,\/} of course, all longer chain
diagrams and all parallel connections.

The FHNC//0 equations for a superfluid system are then identical to
those for the normal system, except that the structure function of the
normal system is replaced by the one of the BCS state,
\begin{equation}
  \SF(k) = \frac{\left\langle\hat\rho_{\kvec}\hat\rho_{-\kvec}\right\rangle_0}
     {\NO}\,,
\label{eq:SFdef0}
\end{equation}
where $\hat\rho_{\kvec}$ is the density operator. It has the form
\begin{equation}
  \SF(k) = 1 - \frac{\rho}{\nu}
  \int d^3 r e^{\I \kvec\cdot\rvec}\left[
    \ell_v^2(r) - \ell_u^2(r)\right]= \tilde\sigma_v(k)+\tilde\sigma_u(k)
\,.\label{eq:SFdef}
\end{equation}
From the definitions (\ref{eq:SFdef}), (\ref{eq:sigmavdef}), and
\ref{eq:sigmaudef}) it is seen that the long-wavelength limit of
$\SF(k)$ is
\begin{equation}
  \SF (0+) = 2\frac{\sum_{\kvec} u_{\kvec}^2 v_{\kvec}^2}{\sum_{\kvec}v_{\kvec}^2} > 0\,.
\label{eq:SF0}
\end{equation}
Hence, $\SF(0+)\ne 0$ for the superfluid system. The FHNC equations
are, in this approximation, the same as
Eqs. (\ref{eq:FHNC0})-(\ref{eq:SofkFermi}).  Note, however, that we
can in general not directly construct the pair distribution function
$g(r)$ from Eqs. (\ref{eq:gFHNC0}), (\ref{eq:C0ofk}) if energy
numerator terms beyond the kinetic energy are included.

The derivation of the Euler equation also is very similar to that for
the normal system; the only change is that we must take the energy
numerator terms into account.  The $S'(k)$ then consists of three
instead of two terms, the rule spelled out in Section
\ref{ssec:FHNC-EL} has to be augmented by
\begin{enumerate}\setcounter{enumi}{2}
\item{} Replace, in turn, each $\SF(k)$ by
  $\tilde \sigma_u'(k)$.
  This may be thought of a correction to the second item $\SF'(k)$.
\end{enumerate}

The remaining derivation is then identical to that for the normal system if we
re-interpret
\begin{equation}
  \SF'(k)= -\frac{t(k)}{2}\left[\SF(k)-1\right] + \tilde \sigma_u'(k)
\,.\label{eq:SFprime}
  \end{equation}
Inserting this in Eq. (\ref{eq:Sprime}) and the Euler equation (\ref{eq:Euler})
and solving for $S(k)$:
\begin{equation}
  S(k) = \frac{\SF(k)}{\sqrt{1 + 2 \displaystyle\frac{\SF^2(k)}{t(k)}
      \left[\tilde  V_{\rm p-h}(k)+\tilde R(k)\right]}}\,,
\label{eq:PPAs}
\end{equation}
where we have defined
\begin{equation}
 \tilde R(k) \equiv \frac{\tilde\sigma_u'(k)}{\SF^2(k)}\,.\label{eq:Rdef}
  \end{equation}

\subsection{Energy}
\label{ssec:BCSenergy}

Since the $E_{\rm G}$ is, with replacement of the exchange lines by
$\ell_v(r_{ij})$ and $\ell_u(r_{ij})$ described above, identical to
that of the normal system, we only need to discuss the correction
$E_{\rm enum}$.  We include in the energy numerator all those terms
that have the same topology as those retained in the above energy
expressions (\ref{eq:EJF}).  There are the ones derived in
\ref{app:AppB} containing $\tilde\sigma_u'(k)$ and
$\tilde\sigma_v'(k)$ as defined in Eqs. (\ref{eq:sigmaupdef}) and
(\ref{eq:sigmavpdef}).
  \begin{eqnarray}
    \frac{E_{\rm enum}}{\NO}
    &=&  \frac{1}{2}\int\!\frac{d^3k}{(2\pi)^3\rho}\>
      \tilde\sigma_u'(k)\tilde\Gamma_{\!\rm dd}(k)\nonumber\\
     &+& \frac{1}{2}\int\!\frac{d^3k}{(2\pi)^3\rho}\>
        \tilde\sigma_v'(k)\left[\tilde\Gamma_{\!\rm dd}(k)-
          \tilde\Gamma_{\!\rm dd}^{(v)}(k)\right]\nonumber\\
        \label{eq:Eenum}\,,
  \end{eqnarray}
  where $\tilde\Gamma_{\!\rm dd}^{(v)}(k)$ is represented by the
  subset of diagrams containing only $\ell_v(r_{ij})$ exchange
  lines. In our numerical calculations we found that the term
  $\left[\tilde\Gamma_{\!\rm dd}(k)-\tilde\Gamma_{\!\rm
      dd}^{(v)}(k)\right]$ is tiny, the term is henceforth omitted.

\subsection{Uniform limit approximation}
\label{ssec:ULBCS}

Eq. (\ref{eq:PPAs}) has an interesting consequence: Note that the
correction term $R(k)$ does not contain correlation contributions. In
other words it as also present {\em even if the interaction is set to
  zero.\/} The consequence of this is most easily seen in the
``uniform limit approximation'' \cite{FeenbergBook} discussed in
Section \ref{ssec:UL} which amounts to identifying $V_{\rm p-h}(r) =
v(r)$ in Eq. (\ref{eq:PPAs}). The energy is, in this approximation
\cite{bcsmbt}
  
\begin{eqnarray}
  &&\frac{E_{\rm G}+E_{\rm enum}}{\NO}
  = \frac{\left\langle \hat T-\mu\hat N\right\rangle_0}{\NO} + \frac{1}{2}\tilde v(0+) + \frac{1}{2}\int \frac{d^3k}{(2\pi)^3\rho}
  \left[S(k)-1\right]\tilde v(k)\nonumber\\
  &&\qquad+\frac{1}{2}\int \frac{d^3k}{(2\pi)^3\rho} \frac{t(k)}{2}\frac{(S(k)-\SF(k))^2}{\SF^2(k)S(k)} +  \frac{1}{2}\int \frac{d^3k}{(2\pi)^3\rho}\tilde R(k)
 \left(S(k)-\SF(k)\right)\nonumber\\
 &&= \frac{\left\langle \hat H-\mu\hat N\right\rangle_0}{\NO}
 - \frac{1}{4}
\int \frac{d^3k}{(2\pi)^3\rho}t(k)\frac{(S(k)-\SF(k))^2}{\SF(k)S^2(k)}\,.
\label{eq:EpairsUL}
\end{eqnarray}
The first terms in this expression is simply the (Hartree-Fock) energy
expectation value with respect to the uncorrelated BCS state
\begin{equation}
\frac{\left\langle \hat H-\mu \hat N \right\rangle_0}{\NO} = 
\nu\int\frac{d^3k}{(2\pi)^3\rho}v_\kvec^2(t(k)-\mu) +
\frac{\rho}{2}\int d^3r v(r)\left(1-\frac{1}{\nu}\ell_v^2(r)
+ \frac{1}{\nu}\ell_u^2(r)\right)\label{eq:EHF}\,,
\end{equation}
and the remaining terms have been manipulated, using the Euler
equation (\ref{eq:PPAs}) in the uniform limit to eliminate $\tilde
v(k)$.

Our result (\ref{eq:EpairsUL}) demonstrates that the energy correction
due to correlations is always negative.  This result is not entirely
surprising, it simply says that adding correlations will lower the
energy expectation value. The more interesting statement is that the
second term is negative {\em even in a non-interacting system.\/} The
reason for this is that the BCS wave function (\ref{eq:BCS}), when
projected to a fixed particle number, can be written in the form of an
independent pair wave function \cite{Schrieffer1999}
\begin{equation}
\Phi_{\rm BCS}^{(N)} = {\cal N}^{-1}{\cal A}
  \{ \phi(12)  \phi(34) \cdot \cdot \cdot  \phi (N-1,N) \} \, ,
\label{eq:BCSN}
\end{equation}
where ${\cal N}$ is the normalization integral, ${\cal A}$ stands for
antisymmetrization, and $\phi(ij)$ is a pair wave function given by
the Fourier transform of $v_\kvec / u_\kvec$. Thus, if we begin with
an {\em incorrect\/} assumption about the Bogoliubov amplitudes (they
are, of course, equal to normal state values
$u^{(0)}_{\kvec}=\bar n(k)$, and $v^{(0)}_{\kvec}=n(k)$
for a non-interacting system) the Jastrow-Feenberg correlations try
to compensate for that and lower the energy expectation value.

\subsection{Euler equation for the Bogoliubov amplitudes}
\label{ssec:gapeq}

The study of the Euler equation has so far kept the Bogoliubov
amplitudes $u_{\kvec}, v_{\kvec}$ fixed and only dealt with the
optimization of pair correlations for such a fixed model state. In
particular the analysis of the preceding section showed that the
Jastrow correlations and the correlations implicit to the BCS wave
function are not completely independent. It is clear that, in an
independent step, the Bogoliubov amplitudes must also be optimized.

The derivation of the Euler-equation of the Bogoliubov amplitudes
$u_\kvec$, $v_\kvec$ follows closely the derivation of the
conventional gap equation by minimization of the Hartree-Fock
approximation \cite{BeliaevLesHouches}. One difference is that one
has, in Hartree-Fock-Bogoliubov approximation, only one pair of
$\ell_v(r)$ and one pair of $\ell_u(r)$ lines, see Eq. (\ref{eq:EHF}).
The full variational energy expression $\left\langle
H'\right\rangle_c$ in Eq. (\ref{eq:FCBCS}) can, on the other hand,
have any number of these exchange lines. This means, in practice, that
the effective pairing interaction and the single-particle energies
depend implicitly on the $u_\kvec$, $v_\kvec$.  The second difference
is the appearance of the ``energy numerator'' terms $E_{\rm enum}$.

As usual, we guarantee the normalization condition $u_\kvec^2 + v_\kvec^2 = 1$
by setting
\begin{equation}
  u_\kvec = \sin\chi_\kvec\,,\qquad v_\kvec = \cos\chi_\kvec\,.
\end{equation}

The minimization of the energy with respect to the $\chi_\kvec$ can be done
in both momentum and coordinate space. In coordinate space we have
\begin{equation}
  \frac{\delta(\rho\ell_v(r))}{\delta\chi_\kvec} = -\frac{\nu}{(2\pi)^3}u_\kvec v_\kvec e^{\I\kvec\cdot\rvec}\,,\qquad
  \frac{\delta(\rho\ell_v(r))}{\delta\chi_\kvec} = \frac{\nu}{(2\pi)^3}
    (v_\kvec^2-u_\kvec^2)e^{\I\kvec\cdot\rvec}\,.
\end{equation}
or, in momentum space with
\begin{equation}
  \tilde{\ell}_v(k) = \nu v_\kvec^2\qquad \tilde{\ell}_u(k) = \nu u_\kvec v_\kvec
  \end{equation}
we have
\begin{equation}
  \frac{\delta(\tilde{\ell}_v(k))}{\delta\chi_\kvec} = - \nu \sin 2 \chi_\kvec \,,\qquad
  \frac{\delta(\tilde{\ell}_u(k))}{\delta\chi_\kvec} = \nu 
    \cos 2 \chi_\kvec \,.
\end{equation}

Then, the optimization of $\left\langle H'\right\rangle_c$ with
respect to $\chi_\kvec$ results in
\begin{eqnarray}
  &&\frac{\delta\left\langle H'\right\rangle_c} {\delta\chi_\kvec}
  =
 -\frac{\delta\left\langle
      H'\right\rangle_c} {\delta \tilde{\ell}_v(k)} \sin 2 \chi_\kvec
    +\frac{\delta\left\langle H'\right\rangle_c}
    {\delta \tilde{\ell}_u(k)}\cos 2 \chi_\kvec\label{eq:BogOpt}\\
  &&=
  \frac{\nu}{(2\pi)^3} \int d^3r_2d^3r_2\left[-\frac{\delta\left\langle
      H'\right\rangle_c} {\delta(\rho\ell_v(r_{12}))}u_\kvec v_\kvec
    +\frac{\delta\left\langle H'\right\rangle_c}
    {\delta(\rho\ell_u(r_{12}))}(v_\kvec^2-u_\kvec^2)\right]e^{\I\kvec\cdot\rvec_{12}}
  =0 \,. \nonumber
\end{eqnarray}

There are two contributions to the variational derivatives
appearing in Eq. (\ref{eq:BogOpt}): One comes from $E_{\rm G}$ and
the other from the energy numerator terms $E_{\rm enum}$.  The
variations of $E_{\rm G}$ and $E_{\rm enum}$ with respect to
$\ell_v(r_{ij})$ or $\ell_u(r_{ij})$ are best done diagrammatically by
removing, in turn, from the expansion shown in Fig. \ref{fig:epot} one
exchange line and opening its external points.

\begin{itemize}
\item{}
  The variation of $E_{\rm G}$ with respect to the $\tilde{\ell}_v(k)$ gives
\begin{equation}
\frac{E_{\rm G}}
     {\delta \tilde{\ell}_v(k)} = \left[ \frac{1}{\rho^2}
       \frac{E_{\rm G}}
       {\delta \ell_v(r_{12})} \right]^{\cal F}(k) \equiv   e_\kvec-\mu\,.
\end{equation}
where the $e_\kvec$ are the generalization of the single particle
energies of CBF theory \cite{CBF2}, performed according to the rules
formulated in Section \ref{ssec:cbcs}.
\item{} The variation of $E_{\rm G}$ with respect to the
  $\tilde{\ell}_u(k)$ is constructed according to the same rules. It
  generally leads to a non-local pairing interaction of the general
  CBF form \cite{CBF2}.  Since we keep, in the FHNC//0 version, only
  diagrams with $\ell_v^2(r)$ or $\ell_u^2(r)$ loops, we can write the
  result as
\begin{equation}
  \frac{1}{\Omega}\frac{E_{\rm G}}
       {\delta(\rho\ell_u(r))} = \frac{1}{\nu}{\cal W}(r)\ell_u(r)\,.
\end{equation}
where ${\cal W}(r)$ is diagrammatically obtained by (a) taking all
diagrams in $E_{\rm G}$ than contain only $\ell_v^2(r)$ or
$\ell_u^2(r)$ loops, and (b) removing, in turn, each $\ell_u^2(r)$
loop and opening its external points.

\item{}

  We have restricted the discussion of the energy numerator terms to the
  simple approximation (\ref{eq:Eenum}), omitting the second term.
  Then the variation with respect to $\chi_\kvec$ yields two terms:

  \begin{eqnarray}
  \frac{\delta}{\delta\chi_\kvec}\frac{E_{\rm enum}}{\NO}
  &=&
  \int \frac{d^3k'}{(2\pi)^6\rho^2}\tilde \Gamma_{\!\rm dd}(|\kvec-\kvec'|)
  (v_\kvec^2-u_\kvec^2)  u_{\kvec'}v_{\kvec'}\times\nonumber\\
  &&\qquad\times\left[
    (e_\kvec^{(v)}-\mu)(v_\kvec^2-u_\kvec^2) + (e_{\kvec'}^{(v)}-\mu)(v_{\kvec'}^2-u_{\kvec'}^2)\right]
  \nonumber\\
  &-&
  \int \frac{d^3k'}{(2\pi)^6\rho^2}\tilde \Gamma_{\!\rm dd}(|\kvec-\kvec'|)
  u_\kvec^2 v_\kvec^2(e_{\kvec}^{(v)}-\mu) u_{\kvec'}v_{\kvec'}
  \,.
\label{eq:Ered2App}
\end{eqnarray}
  The second term, which is proportional to $u_\kvec^2 v_\kvec^2$,
  contributes a correction $\delta e_k$ to the single particle
  spectrum whereas the first term gives a correction to the pairing
  matrix element.  There is a third term, which we have not spelled
  out in Eq. (\ref{eq:Ered2App}), which originates from the implicit
  dependence of the single-particle energies $e_{\kvec}^{(v)}$ on the
  $\ell_v(r)$. This term also contribute, if kept, to the shift
  $\delta e_k$ of the single particle spectrum.
\end{itemize}
We can now write the minimization condition for the
Bogoliubov amplitudes as
\begin{equation}
  0 = -\left(e_\kvec-\mu + \delta e_\kvec\right)u_\kvec v_\kvec + \int
  d^3 k' {\cal
    P}(\kvec,\kvec')u_{\kvec'}v_{\kvec'}(v_{\kvec}^2-u_{\kvec}^2)
\end{equation}
where
\begin{equation}
  {\cal P}_{\kvec,\kvec'} = \tilde {\cal W}(|\kvec-\kvec'|)
+  \tilde\Gamma_{\!\rm dd}(|\kvec-\kvec'|)
\left[(e_\kvec^{(v)}-\mu) (u_{\kvec}^2-v_{\kvec}^2)
  +(e_{\kvec'}^{(v)}-\mu) (u_{\kvec'}^2-v_{\kvec'}^2)\right]\,.\label{eq:PPair}
\end{equation}
With that, we have brought the minimization condition for the
Bogoliubov amplitudes in exactly the same form as the one for the
weakly interacting system \cite{BeliaevLesHouches,Tinkham2004}, the
remaining manipulations can therefore be skipped. The only change
compared to the weakly coupled limit (\ref{eq:gap}) is that all
ingredients depend implicitly on the $u_{\kvec}, v_{\kvec}$ and the
equation must be solved iteratively. Of course, in a more complete
evaluation of the energy numerator terms as outlined in
\ref{app:AppB}, the kinetic energy terms in Eq. (\ref{eq:PPair}) are
also replaced by the CBF single particle energies {\rm plus\/}
corrections from higher order diagrams. Also, note that the single
particle energies and the pairing matrix elements depend implicitly on
the Bogoliubov amplitudes. We also recover the weak pairing limit
(\ref{eq:Pdef}) by setting $u^{(0)}_{\kvec}= \theta(k-\KF)$,
$v^{(0)}_{\kvec}=\theta(\KF-k)$, then $(t(k)-\mu)
(u_{\kvec}^2-v_{\kvec}^2)\rightarrow |t(k)-\mu|$.

\subsection{Long wavelength analysis}

The result (\ref{eq:PPAs}) points to one of the major problems of
local correlation functions and is, as such, one of the key messages
of our paper. At the first glance, the result looks innocuous. Recall
that we have discussed in Section \ref{ssec:limits} how Landau's
stability condition $F_0^s > -1$ is a condition for existence of the
parquet equations.  In particular, within that limit, a negative value
of $\tilde V_{\rm p-h}(0+)$ is permitted.

If, on the other hand, the system is superfluid, we have $\SF(0+) >
0$, and therefore $\tilde V_{\rm p-h}(0+)+R(0+)$ must be positive to
have a solution. $R(0+)$ vanishes in the limit of a normal system
whereas $\tilde V_{\rm p-h}(0+)$ remains finite. {\em Thus, if\,
  $\tilde V_{\rm p-h}(0+)<0 $, the Euler equation ceases to have a
  solution even if the gap is infinitesimally small.\/} We have shown
this here only for the case of the FHNC//0 approximation, but it is
also true in the general case that all FHNC diagrams, and possibly
also higher-order correlation functions, are included since the Euler
equation remains structurally the same, see Section
\ref{ssec:exchanges}. Our observation applies, of course, equally to
``fixed-node'' Monte Carlo calculations which may see this instability
only in large stochastic fluctuations.

On a less drastic level, assume that $\tilde V_{\!\rm p-h}(0+) +
\tilde R(0+) > 0$. Eq. (\ref{eq:PPAs}) then predicts the long-wavelength
limit
\[S(k) = \frac{\hbar k}{\sqrt{4m(\tilde V_{\rm p-h}(0+) + \tilde R(0+))}}\]
  which is obviously the wrong behavior since one should have
  $S(k)\sim \hbar k/2mc$. In other words, even if the interaction is
  repulsive, $\tilde V_{\rm p-h}(0+) > 0$, Eq. (\ref{eq:PPAs}) predicts
  the incorrect behavior of the static structure function at long wave
  lengths since the contribution of the free kinetic energy is missing.

This is evidently not a statement about the physics, but rather a
statement on the approximations implicit to the wave function
(\ref{eq:CBCS}), specifically the collective approximation
(\ref{eq:Chi0Coll}) for the Lindhard function.  One can expect that
the correct Lindhard function removes this spurious instability.

There have been several suggestions for a Lindhard function for a
superfluid system
\cite{PhysRevB.61.9095,PhysRevA.74.042717,Steiner2009,Vitali2017}, the
most frequently used form for $T=0$ is given below.  In view of the
need for the spin-spin response function in Section
\ref{ssec:NeutronMatter} we cite here both the spin and the density
channels. In terms of the usual relationships of BCS theory,
\begin{eqnarray}
  u_k^2 &=& \frac{1}{2}\left(1+\frac{\xi_\kvec}{E_\kvec}\right)
  \nonumber\\
  v_k^2 &=& \frac{1}{2}\left(1-\frac{\xi_\kvec}{E_\kvec}\right)\,.
\end{eqnarray}
with $\xi_{\kvec} = t(k)-\mu$ and $E_{\kvec} =
\sqrt{\xi_{\kvec}^2+\Delta_{\kvec}^2}$
we have \cite{Schrieffer1999,Kee1998,Kee1999,PhysRevB.61.9095}

\begin{equation}
  \chi_0^{(\rho,\sigma)}(\kvec,\omega)
  = \frac{\nu}{\NO}\sum_{\pvec}
  b_{\pvec,\kvec}^{(\rho,\sigma)}\Biggl[\frac{1}
    {\omega-E_{\kvec+\pvec}-E_{\pvec}+\I\eta}
    - 
    \frac{1}{\omega+E_{\kvec+\pvec}+E_{\pvec}+\I\eta}\Biggr]
    \label{eq:BCSLindha}
\end{equation}
where
\begin{eqnarray}
  b_{\pvec,\kvec}^{(\rho,\sigma)}
   &=&\frac{1}{4}\left[\left(1-
    \frac{\xi_{\pvec}}{E_{\pvec}}\right)
    \left(1+\frac{\xi_{\kvec+\pvec}}{E_{\kvec+\pvec}}
     \right)
     \pm\frac{\Delta_{\pvec}}{E_{\pvec}}
     \frac{\Delta_{\kvec+\pvec}}{E_{\kvec+\pvec}}\right]
   \nonumber\\
   &=&v_{\pvec}^2u_{\kvec+\pvec}^2 \pm
   u_{\pvec} v_{\pvec} u_{\kvec+\pvec} v_{\kvec+\pvec}\,,
\end{eqnarray}
where the upper sign applies to the density channel, and the lower
to the spin channel, respectively.

In the limit of a normal system, the coefficient
$b_{\kvec,\qvec}^{\rho,\sigma}$ become
\begin{equation}
  b_{\pvec,\kvec}^{(\rho,\sigma)} \rightarrow
  n_{\pvec}(1-n_{\lvec+\qvec})\,,
  \end{equation}
as it should come out.

This superfluid Lindhard function is consistent with the $\SF(k)$ as
defined in Eq. (\ref{eq:SFdef0}).

  \begin{equation}
    \SF(k) = -\int_0^\infty \frac{d\omega}{\pi} \Im m \chi_0^{(\rho)}(k,\omega)
    =  \frac{\nu}{\NO}\sum_{\qvec}b_{\pvec,\kvec}^\rho
    = \tilde\sigma_u(k) + \tilde\sigma_v(k)\,.
\label{eq:m0bcs}
  \end{equation}

We can now return to the frequency integration (\ref{eq:SRPA}). All we
need to show is that this expression exists, for small gaps, for $-1 <
\frac{\tilde V_{\rm p-h}(0+)}{mc_{\rm F}^2}$.  For that, it is
sufficient to look at the limit $k\rightarrow 0$ of the {\em static\/}
response function and the {\em static\/} Lindhard function.

For $\omega=0$ we get for the static Lindhard function
\begin{equation}
  \lim_{p\rightarrow 0}\chi_0^{(\rho)}
  (k,0) = -\frac{\nu}{2\rho}\int\frac{ d^3p}{(2\pi)^3}
  \frac{\Delta_\pvec^2}{E_{\pvec}^3} \rightarrow
  - \frac{1}{\mcf}\quad\mathrm{as}\quad\Delta\rightarrow 0\,.
\end{equation}
This is identical to the same limit of the Lindhard function of the
normal system and leads to the correct stability condition.

Thus, the conclusion of our analysis is a more precise version of the
statement made in Sections \ref{ssec:limits} and \ref{sec:parquet}
about the validity of locally correlated wave functions for physical
phenomena that involve mostly particles close to the Fermi
surface. Previous work \cite{shores,CCKS86} pointed out {\em
  quantitative\/} deficiencies; we have demonstrated here a much more
profound problem, namely that there are indeed very serious {\em
  qualitative\/} difficulties in the sense that the minimization
problem (\ref{eq:optu2}) has no solution. We stress again that this
feature is {\em not\/} a consequence of the specific level of FHNC
approximations. It is a general problem of the Jastrow-Feenberg form
of the wave function. More elaborate versions of the FHNC summations and/or
the inclusion of triplet correlations can change the numerical values
but not the general features.

We conclude this section by recalling that the
$\chi_0^{(\rho)}(\kvec,\omega)$ does not satisfy the $f$-sumrule.
Improved versions have been suggested
\cite{PhysRevA.74.042717,Vitali2017}. It would be very interesting to
examine the wave function that corresponds to that work. Recall
\cite{PinesNoz} that, to satisfy the $f$-sumrule, the wave function
must be an eigenfunction of the real-world Hamiltonian
(\ref{eq:Hamiltonian}), which would be simply the Hamiltonian of a
non-interacting Fermi system. Since that Hamiltonian commutes with the
particle number operator, the wave function would also be, unlike the
state (\ref{eq:BCS}), an eigenstate of the particle number operator.

\section{Applicatios}

Most of the calculations reported in this paper use the ``weak
coupling approximation'' spelled out in Section \ref{ssec:weak} with
the additional qualification that the ``collective'' Lindhard function
(\ref{eq:Chi0Coll}) that was used throughout our earlier work was
replaced by the exact Lindhard function in the rings and rungs.  Only
very few of the ground state properties changes visibly by including
the superfluid Lindhard functions (\ref{eq:BCSLindha}), we will
mention that where appropriate. Only for the calculations of the gap
function $\Delta(k)$ we have used also used the superfluid Lindhard
function.

We have also used the simple RPA formulas (\ref{eq:SRPA}) and
(\ref{eq:Wchain}) to sum the rings and rungs, for that it is
appropriate to set $\tilde R(k)=0$.  A more elaborate version of
time-dependent Hartree-Fock is available \cite{TohyamaSchuck2015}
which should provide an interpretation of the correction $\tilde
R(k)$; we have not included this improvement because the corrections
due to including the functions (\ref{eq:BCSLindha}) are throughout
very small, but computationally very time consuming.

\subsection{Neutron Matter}
\label{ssec:NeutronMatter}

Neutron matter is the first natural application of our method.  It is
of astrophysical interest because the magnitude of the superfluid gap
is of critical importance for the cooling rate of neutron stars
\cite{PAB90,PAG94,PethickSchaeferSchwenk}. The system has been the
subject of two extensive recent reviews
\cite{SchuckBCS2018,SedrakianClarkBCSReview} which give a very
complete account of the current literature. We can therefore restrict
ourselves here to those aspects that are specific to the high-level
many-body treatment of our paper.

From a quantitative point of view one should question the validity of
the ``weakly paired'' approximation described in Section
\ref{ssec:weak} because the gap is, at low densities, of the order of
$0.5\EF$ \cite{GC2008,GC2010,ectpaper}. This concern is one of the
reasons for developing the methods described in Section
\ref{sec:hncbcs}.

A few additions are necessary to deal with the dependence of the
neutron-neutron interaction on the relative spin of the interacting
particles. The Jastrow-Feenberg correlation function
(\ref{eq:Jastrow}) does not contain spin-dependent correlations. It
is, of course, straightforward to calculate the energy expectation
value for an interaction of the type (\ref{eq:Vop}). Indeed, the
spin-channel of the interaction gives an important correction to the
energy; we have included it in Ref. \citenum{ectpaper}. In the Euler
equation, the spin-channel operator only contributes to the exchange
terms discussed in Section \ref{ssec:exchanges}; this correction
is omitted in the FHNC//0 approximation.

Within CBF, it is also straightforward to include the simplest
spin-correlations, either in low order perturbation theory
\cite{shores}, or by summing the CBF ring diagrams \cite{rings}. The
latter provides an RPA-type energy correction from spin-correlations;
the procedure then corresponds to an optimized ``single operator
chain'' approximation \cite{IndianSpins}.

\subsubsection{Energetics}
\label{ssec:nuclenergy}

Let us now go through the relevant computations step-by-step. All
intermediate results that are shown are for the Reid $V_4$ soft-core
potential which we write in the operator basis \cite{Day81}
\begin{equation}
  \hat v(r_{ij}) = v_c(r_{ij}) + v_\sigma(r_{ij}){\mathbf{\sigma}_i}
  \cdot{\mathbf{\sigma}_j}\,.
\label{eq:Vop}
\end{equation}
We have carried out the identical calculations for the Argonne
$V_{4}'$ nucleon-nucleon interaction \cite{AV18}. The results are
sufficiently similar to those of the Reid potential; we will therefore
display only the final results for that case.

The essential difference compared with what we have discussed in
previous work \cite{cbcs,qfs2018} is the operator form
(\ref{eq:Vop}) of the interaction.  Instead of a local ${\cal W}(r)$,
we have the spin--dependent interaction
\begin{equation}
 \tilde W(k) + \tilde V_{\rm p-h}^{(\sigma)}(k)\sigma_1\cdot\sigma_2\,.
\label{eq:WexcFHNC}
\end{equation}
where
\begin{equation}
    V_{\rm p-h}^{(\sigma)}(r) = \left[1 + \Gamma_{\!\rm dd}(r)\right]v_\sigma(r)\,.
    \label{eq:Vphspin}
\end{equation}
in $V_{\rm ee}(r)$. Since the wave function (\ref{eq:Jastrow})
contains no spin correlations, there are no chain diagrams
contributions to $\tilde W(k)$ in the spin channel. In the
CBF or parquet calculation, the appropriate interaction is then
\begin{equation}
  \widetilde W^{(\rho)}(k,\bar\omega(k)) + \widetilde
  W^{(\sigma)}(k,\bar\omega(k)) \sigma_1\cdot\sigma_2\,,
  \label{eq:WexcParquet}
\end{equation}
where
  \begin{equation}
    \tilde W^{(\rho,\sigma)}(k,\omega)
    = \frac{\tilde V_{\rm p-h}^{(\rho,\sigma)}
      (k)}{1-\chi_0^{(\rho,\sigma)}(k,\omega)
      \tilde V_{\rm p-h}^{(\rho,\sigma)}(k)}\,.\label{eq:Weff}
\end{equation}

\begin{figure}[H]
\centerline{\includegraphics[width=0.65\columnwidth,angle=-90]{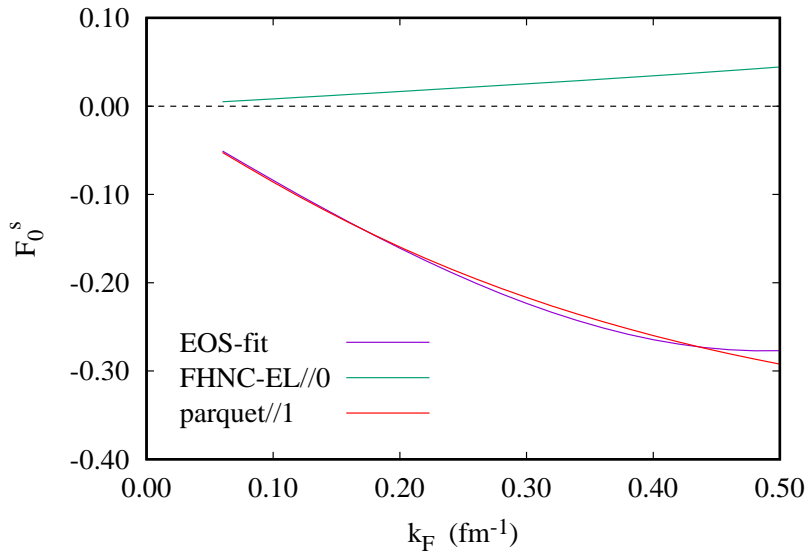}}
\caption{(color online) The figure shows the Fermi-liquid parameter
  $F_0^s$ as obtained from the equation of state via a polynomial fit
  (\ref{eq:eospoly}) to the equation of state Eq. (\ref{eq:mcfromeos})
  (purple, solid line), and from the FHNC-EL//0 and parquet//1
  approximation (green and blue solid lines). The FHNC//1 results are
  indistinguishable from the parquet//1 results and not shown.  The
  horizontal dashed line separates the area of positive and negative
  $F_0^s$ as a guide to the eye.
  \label{fig:re_fitplot}}
\end{figure}

We have commented above about the fact that the two ways to calculate
$F_0^s$, Eqs. (\ref{eq:FermimcfromVph}) and (\ref{eq:mcfromeos}) agree only
in an exact theory. To assess the consistency between the two
definitions, we have fitted, in the regime $0\le\KF\le 0.5\,{\rm fm}^{-1}$
the equation of state by a function
\begin{equation}
  \frac{E}{N} = \frac{3 \hbar^3\KF^2}{10m} + a\KF^3 + b\KF^5
  \label{eq:eospoly}
  \end{equation}
and calculated $F_0^s$ from both the fit (\ref{eq:mcfromeos}) and the
long-wavelength limit (\ref{eq:FermimcfromVph}). Here, and throughout
the rest of this paper, we will set the effective mass ratio
$m^*/m=1$. Justification for
this is derived from calculations of the self-energy as described in
Sections \ref{ssec:selfenergy}, \ref{ssec:ljmass} and
\ref{ssec:PTenergy}.  Fig. \ref{fig:re_fitplot} shows the situation
for the low-density equation of state of neutron matter, interacting
via the Reid $V_4$ potential.

We find in particular the FHNC//0 approximation for the
particle--hole interaction leads to an incorrect positive value of
$F_0^s$. The FHNC-EL//1 approximation improves the agreements by
including the exchange term in the particle-hole interaction, but it
still disregards the chain-diagrams in the spin-channel who seem to be
insignificant. These are included in the parquet calculations.  The
agreement between $F_0^s$ calculated in these two ways is, for low
densities, evidently quite satisfactory.

\subsubsection{Effective interactions}
\label{ssec:nuclinteractions}

The energetics of neutron matter is relatively insensitive to the
quality of the many-body wave function.  We have already seen that
this is not the case for the quasiparticle interaction. Similarly, one
should expect visible corrections to the pairing interaction where
medium polarization effects, especially due to spin fluctuations, are
expected to have an impact on the superfluid gap
\cite{JWCgap,CKY76,Wam93,Sch96}.

A very thorough examination of polarization effects has been performed
by Schulze \etal \cite{SPR2001}; our calculations go beyond that work
in the sense that we determine the effective interactions by summing
the parquet-diagrams, and take the superfluid particle-hole
propagators.  We have included the exchange terms that have been
included in Ref. \citenum{SPR2001} in the localized form described in
Section \ref{ssec:exchanges}. Following the work of
Ref. \citenum{SPR2001}, we have also taken the dynamic interactions
$\tilde W^{(\rho,\sigma)}(q,\omega)$ and $\tilde
W^{(S)}(q,\omega)$ at $\omega=0$.  The interactions derived from the
FHNC approximation directly are, of course, energy independent.

\begin{figure}[H]
    \centerline{\includegraphics[width=0.65\columnwidth,angle=-90]{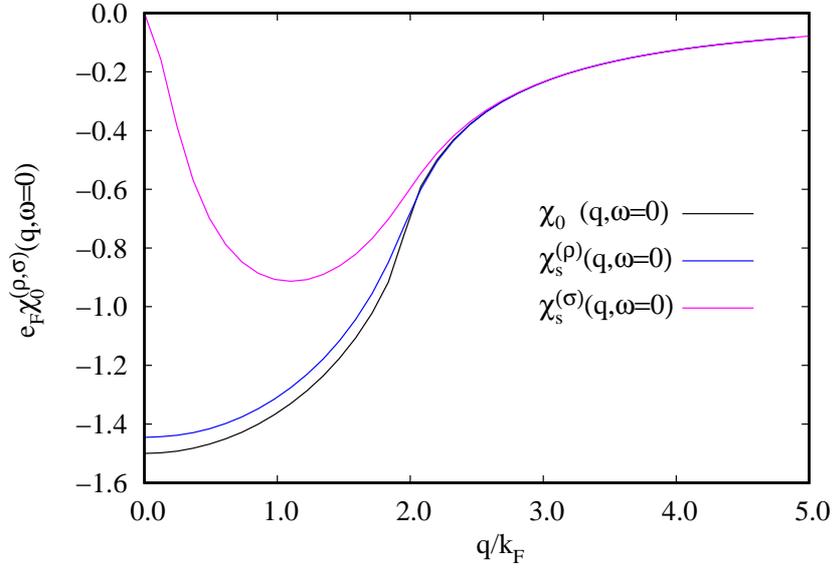}}
      \caption{(color online) The figure shows the static Lindhard
        function $\chi_0(q,\omega=0)$ of the normal system, and the
        superfluid Lindhard function in both the density and the spin
        channel as indicated in the legend. The calculation is done at
        $\KF = 1.0\,\mathrm{fm}^{-1} $ for a gap function obtained for
        the Reid $V_4$ potential.\label{fig:chiplot}}
  \end{figure}

A calculation for the superfluid phase that goes beyond the ``weakly
coupled'' approximation requires the implementation of the full
parquet-level theory because the Euler equations have, due to $F_0^s <
0$, no solution. Among others, we need the rather time-consuming
computation of the superfluid Lindhard function which is an essential
input for the theory. The fact that this is indeed necessary is
documented in Fig. \ref{fig:chiplot}. There we show, for $\omega=0$,
the Lindhard function of the normal system, as well as the density-
and the spin-channel of the Lindhard function of the superfluid
system, Eqs. (\ref{eq:BCSLindha}).

The discrepancy between the density and the spin-channel superfluid
Lindhard functions is quite interesting and has, to our knowledge, not
been noted in neutron matter calculations. It is caused by the fact
that the superfluid Lindhard function in the spin-channel,
Eq. (\ref{eq:BCSLindha}) goes to zero as $q\rightarrow 0$ for all
values of the gap. Of course, the regime where the spin-channel
function deviates from the density-channel function becomes smaller
with decreasing gap. The density-channel function is expectedly close
to the one of the normal system even when the gap is relatively large.

\begin{figure}[H]
    \centerline{\includegraphics[width=0.35\textwidth,angle=-90]{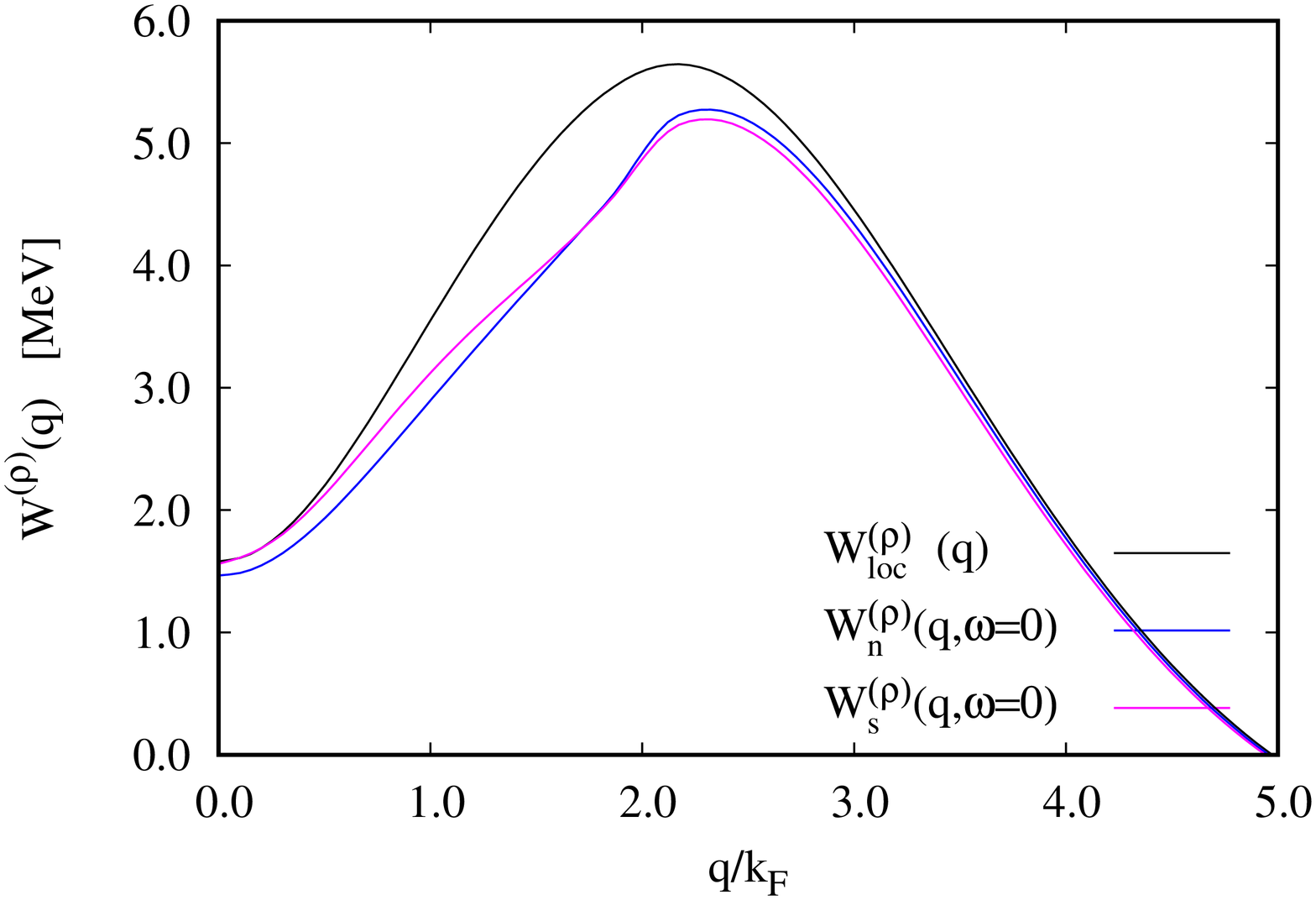}
      \hspace{0.05\textwidth}%
      \includegraphics[width=0.35\textwidth,angle=-90]{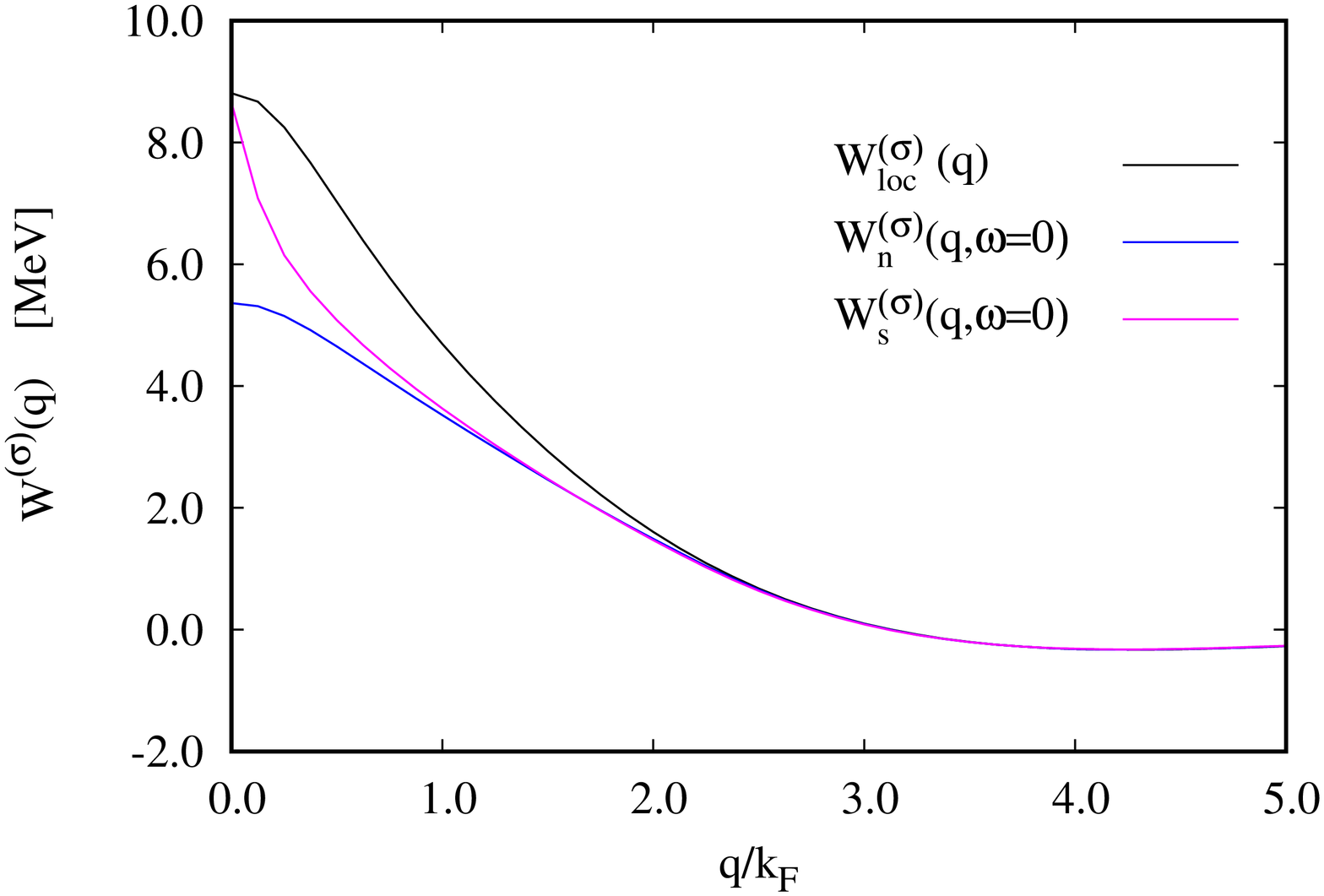}
    }
      \caption{The left figure shows the density channel of the static
        effective interaction (\ref{eq:Wchain}), as well as the
        energy-dependent effective interaction
        $W^{(\rho)}(q,\omega=0)$ defined in Eq. (\ref{eq:Weff}). The
        version of $W_n^{(\rho)}(q,\omega=0)$ uses the normal-system
        Lindhard function and the density-channel function
        (\ref{eq:BCSLindha}), whereas $W_s^{(rho)}(q,\omega=0)$ uses
        the superfluid Lindhard function. The right figure shows the
        same interactions in the spin-channel.
        \label{fig:Vrhosigma}
}
  \end{figure}

\begin{figure}[H]
  \centerline{\includegraphics[width=0.65\columnwidth,angle=-90]{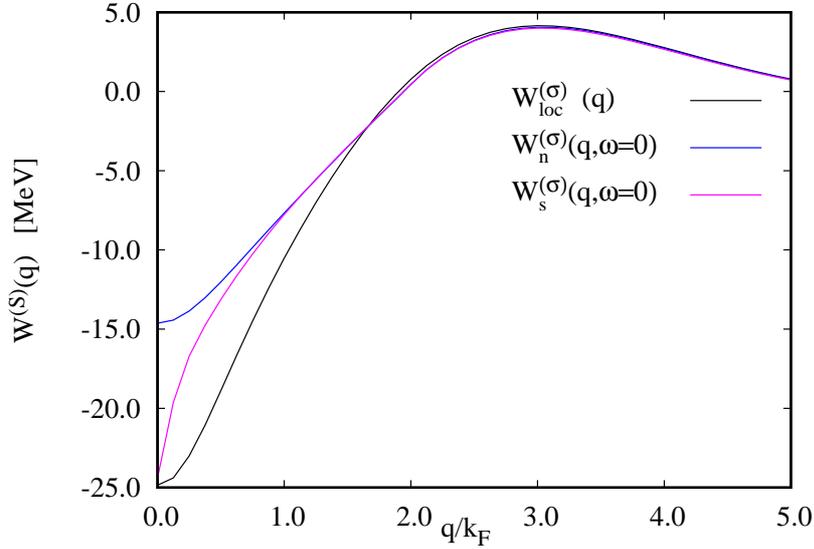}}
      \caption{The figure shows the particle-hole interaction $\tilde
        V_{\rm p-h}(q)^{S} = V_{\rm p-h}(q)^{(\rho)}(q) - 3 V_{\rm
          p-h}(q)^{(\sigma)}(q)$ as well as the energy-dependent
        effective interaction $W^{(S)}(q,\omega=0)$ defined in
        Eq. (\ref{eq:Wsinglet}). The versions of $W^{(S=0)}(q,\omega=0)$
        using the normal-system Lindhard function is labeled by
        $W_n^{(S)}(q,\omega=0)$\,.\label{fig:Vsing}
}
  \end{figure}

This result has significant consequences for the effective
interactions (\ref{eq:Weff}). These interactions are shown in figures
\ref{fig:Vrhosigma} and \ref{fig:Vsing}

The conclusion to be drawn from Fig. \ref{fig:Vrhosigma} is that the
use of the dynamic effective interaction in the {\em density channel
  potential\/} at $\omega=0$ makes the interaction somewhat less
attractive, but there is practically no difference between using the
normal system Lindhard function or the one for the superfluid one.
The situation is quite different in the spin-channel: Using the
superfluid Lindhard function changes the effective interaction in
that channel significantly, note that the gap at the density
considered here is only about 20 percent of the Fermi energy, see
Fig. \ref{fig:gaps} below. This is to a large extent due to the fact
that superfluid $\chi_0^{(\sigma)}(q,0)$ goes to zero at zero momentum
transfer.  The particle-hole interaction and the paring interaction
are therefore identical in that limit. This is not the case if one
uses the normal system Lindhard function to calculate
$W^{(\sigma)}(q)$.

In the pairing matrix elements, we must couple the interactions
(\ref{eq:Weff}) into the singlet and triplet channel:
  \begin{eqnarray}
    \widetilde W^{(S)}(q,\omega) &=&
    \widetilde W^{(\rho)}(q,\omega)-3\widetilde W^{(\sigma)}(q,\omega)
    \label{eq:Wsinglet}\\
    \widetilde W^{(T)}(q,\omega) &=&
    \widetilde W^{(\rho)}(q,\omega)+\widetilde W^{(\sigma)}(q,\omega)
    \label{eq:Wtriplet}\,.
  \end{eqnarray}
The most relevant quantity for our purposes is the singlet pairing
interaction $W^{(S)}(q,\omega=0)$ which is shown in
Fig. \ref{fig:Vsing}. Similar to the spin-channel, using the
superfluid Lindhard function changes the interaction visibly, in
particular for long wavelengths $0\le q \le 2\KF$ which is the
important regime for pairing phenomena. Interestingly, the choice of
the superfluid Lindhard function suppresses the correction from
spin-fluctuations; the $S$-wave interaction is {\em closer\/} to the
one where the spin-channel is omitted altogether than to the one
where the spin part is taken from the normal system.

\subsubsection{Self energy}
\label{ssec:selfenergy}

We have spelled out in Eqs. (\ref{eq:uhncdef}), (\ref{eq:Xccpdef}) and
(\ref{eq:Xccdef}) working formulas for the single particle spectrum
based on a local correlation operator. In Section \ref{ssec:selfen} we
have then made the connection to the G0W approximation and have shown
how these two procedures are related. The situation is similar to the
one for other quantities: FHNC offers a much more extensive summation
of diagrams. Note, for example, that the denominator $1-\tilde X_{\rm
  cc}(r)$ can be related to high-order exchange diagrams but has no
equivalence in the G(0)W approximation. The price for that is that the
evaluation of these diagrams is less accurate. We shall address these
issues now.

We are examining the self-energy here for a number of reasons. One is
to justify our choice of keeping only the kinetic energy in
$\ell_v'(r)$ and $\ell_i'(r)$, Eqs. (\ref{eq:lupdef}) and
(\ref{eq:lvpdef}), defining $E_{\rm enum}$. The second reason is to
justify the use of a free spectrum in the gap equation
(\ref{eq:gap}). We have studiend the sensitivity of the superfluid gap
agaist changes of the effecive mass in Ref. \citenum{ectpaper}.

The self energy (\ref{eq:G0W}) is most conveniently evaluated by Wick
rotation in the complex $\omega$- plane. That way, the RPA sum for the
self-energy is decomposed into two terms, a smooth ``background--'' or
``line--term'', which consists of the frequency integral along the
imaginary $\omega$ axis, and a second, ``pole term'' from the residue
of the single-particle Green's function, \ie

\begin{equation}
\Sigma(k, E ) =  \Sigma_{\rm line}( k, E)
+ \Sigma_{\rm pole} (k,E)
\end{equation}
with
\begin{equation}
\Sigma_{\rm line} ( k, E) = -
\int_{ -\infty }^{\infty } \frac{ d \omega }{ 2\pi }
\int \frac{d^3 p }{ ( 2\pi )^3\rho } \tilde w_{\rm I}(p,\omega)
\frac { E - t( | {\bf k} - {\bf p} | )}{
[ E - t( | {\bf k} - {\bf p} | ) ]^2 + \omega^2 }
\label{eq:SigmaLine}
\end{equation}
and
\begin{eqnarray}
\Sigma_{\rm pole} (k, E ) &=&
\int \frac{ d^3 p }{( 2\pi )^3\rho }
\tilde w_{\rm I}^2 (p, E - t( | {\bf k} - {\bf p} | ))\times\nonumber\\
&&\qquad\qquad\times
 [\Theta ( E - t(| {\bf k} - {\bf p} | )
- \Theta (\EF - t(| {\bf k} - {\bf p} | ) ]\,.
\label{eq:SigmaPole}
\end{eqnarray}

\begin{figure}[H]
    \centerline{\includegraphics[width=0.65\textwidth,angle=-90]{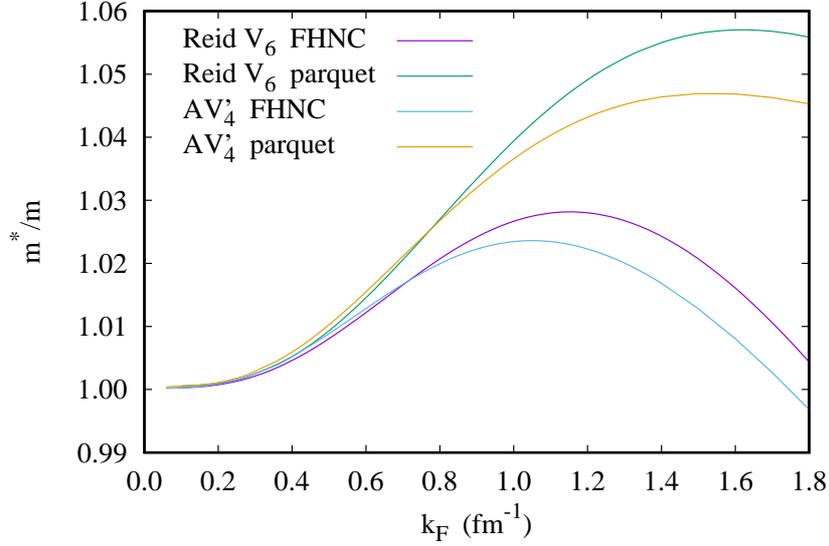}}
    \caption{The figure shows the neutron effective mass as a function
      of density for both interactions and the FHNC as well as the
      parquet prediction. The FHNC effective masses are slightly
      different from those given in Ref. \citenum{ectpaper}, that is
      because we have used there the FHNC//0 approximation.
      \label{fig:nuc_masses} }
  \end{figure}

Since the effective mass can have an visible effect on pairing properties,
we have calculated the ratio
\begin{equation}
  \frac{m}{m^*} = 1 + \frac{m}{\hbar^2\KF}\frac{d}{dk}\left.
  \Sigma(k,t(k))\right|_{k=\KF}
  \label{eq:mstar}
\end{equation}
from the {\em on-shell\/} self-energy $\Sigma(k,t(k))$ for both the
Reid and the Argonne potential; results are shown in
Fig. \ref{fig:nuc_masses}. The effective masses are all close to 1,
suggesting that the single particle spectrum of non-interacting
fermions is an adequate approximation in this particular system. It
also justifies the on-shell approximation.  While the effective mass
ratio {\em per-se\/} is close to 1, we see that the corrections from
going beyond local correlations are substantial on that scale.

\begin{figure}[H]
    \centerline{\includegraphics[width=0.65\textwidth,angle=-90]{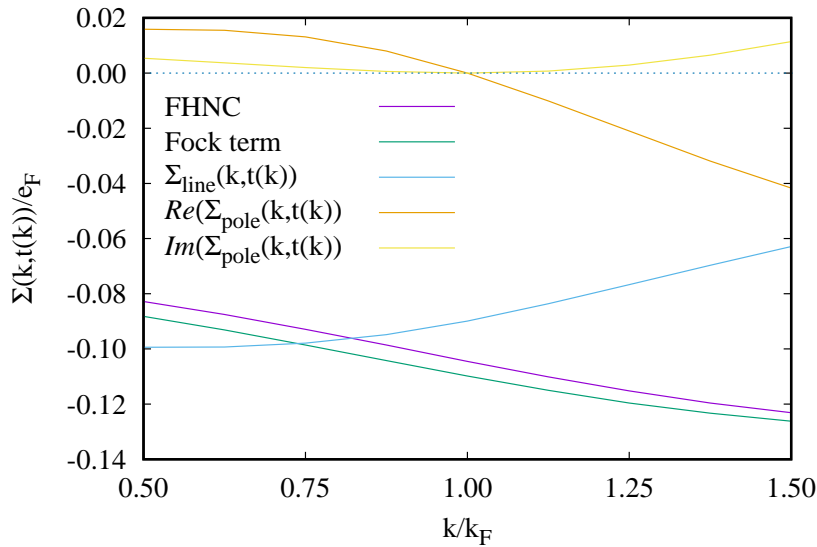}}
    \caption{The figure shows, for the Reid interaction and $\KF =
      1.0\,$fm$^{-1}$, the individual contributions to the neutron
      self-energy. FHNC stands for Eq. (\ref{eq:uhncdef}), ``Fock
      term'' stands for the $U(k)$ in Eq. (\ref{eq:G0W}) and
      $\Sigma_{\rm line}$ and $\Sigma_{\rm pole}$ are the terms
      (\ref{eq:SigmaLine}) and (\ref{eq:SigmaPole}). Note that only
      the pole term is complex. We also left out the denominator
      $1-\tilde X{\rm cc}(k)$ to facilitate the comparison with the
      Fock term $U(k)$ and the G0W approximation.
      \label{fig:re_eself}. }
  \end{figure}

It is also of interest to look at the individual contributions to the
self-energy as a function of wave number. We show these in
Fig. \ref{fig:re_eself}.

The FHNC expression for the static field (\ref{eq:Xccpdef}) and the
``Fock'' differ only by the interaction entering the expression:
Whereas tilde $X'_{\rm cc}(k)$ is defined in terms of the full static
interaction ${\cal W}(r)$, the static field $U(k)$ contains only the
particle-hole irreducible interaction $V_{\!\rm p-h}(r)$. Evidently
these two terms are numerically rather similar. The remaining two
terms, $\Sigma_{\rm line}(k,t(k))$ and $\Re e\Sigma_{\rm
  pole}(k,t(k))$, sum to practically a constant, hence the real part
of the spectrum is, at this density, indeed rather similar to the FHNC
approximation. We also see that there is, in neutron matter, no
visible enhancement of the effective mass around the Fermi surface as
observed in nuclear matter \cite{BrownGunnGould}. This is consistent
with the fact that imaginary part of the self-energy is rather small.

These results essentially confirm that our choice of $m^*/m=1$ is
justified.

\subsubsection{BCS pairing}
\label{ssec:nuclpairing}

We have generally solved the full gap equation (\ref{eq:gap}), using
the eigenvalue solver with an adaptive mesh described in
Ref. \citenum{cbcs}.  Results will be reported for the value of the
gap function at the Fermi surface, $\Delta(\KF)$.

Several approximate formulas for the superfluid gap at the Fermi
surface are available: At very low densities, the gap can be expressed
in terms of the vacuum scattering length $\a0$ \cite{PethickSmith} as
\begin{equation}
\frac{\Delta(\KF)}{\EF} = \frac{8}{e^2}\exp\left(\frac{\pi}{ 2\a0\KF}\right)
\label{eq:gapsp}
\end{equation}
The exponential $\exp\left(\frac{\pi}{ 2\a0\KF}\right)$ is universal
in the sense that it depends only on the product $\a0\KF$, but not on
details of the interaction. The pre-factor is modified by polarization
corrections \cite{Gorkov} which are also universal, and by many-body
and finite-range corrections some of which are non-universal
\cite{cbcs}.

Eq. (\ref{eq:gapsp}) gives, apart from very low densities, quite poor
predictions.  More useful is the estimate
\begin{equation}
\frac{\Delta(\KF)}{\EF} = 8\frac{m}{m^*}
\exp\left(\frac{4}{3}\frac{\hbar^2\KF^2}{2m^* W_{\rm F}}\right)
\label{eq:gapfw}
\end{equation}
where
\begin{equation}
W_{\rm F} \equiv \frac{1}{2 \KF^2}\int_0^{2\KF} dk k \tilde E(k) \,.
\label{eq:V1S0}
\end{equation}
is the $S$-wave matrix element of the pairing interaction at the Fermi
surface. Eq. (\ref{eq:gapfw}) shows clearly the influence of the
effective mass correction; we have studied this in
Ref. \citenum{ectpaper}. Since the effective mass is always close to
the bare mass, see Fig. \ref{fig:nuc_masses}, we have throughout this
paper used $m^*/m = 1$.

\begin{figure}
    \centerline{\includegraphics[width=0.65\textwidth,angle=-90]{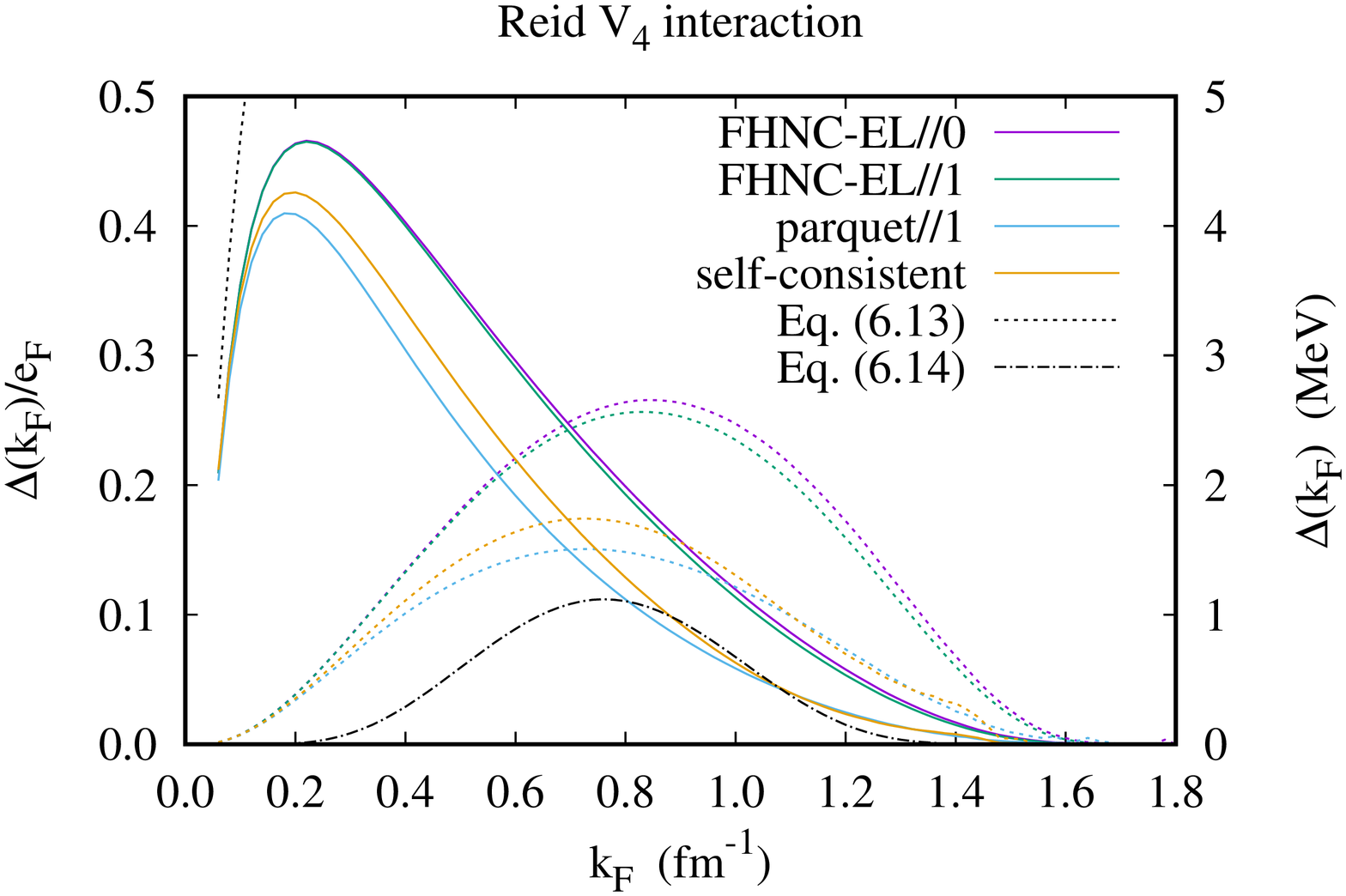}}
    \centerline{\includegraphics[width=0.65\textwidth,angle=-90]{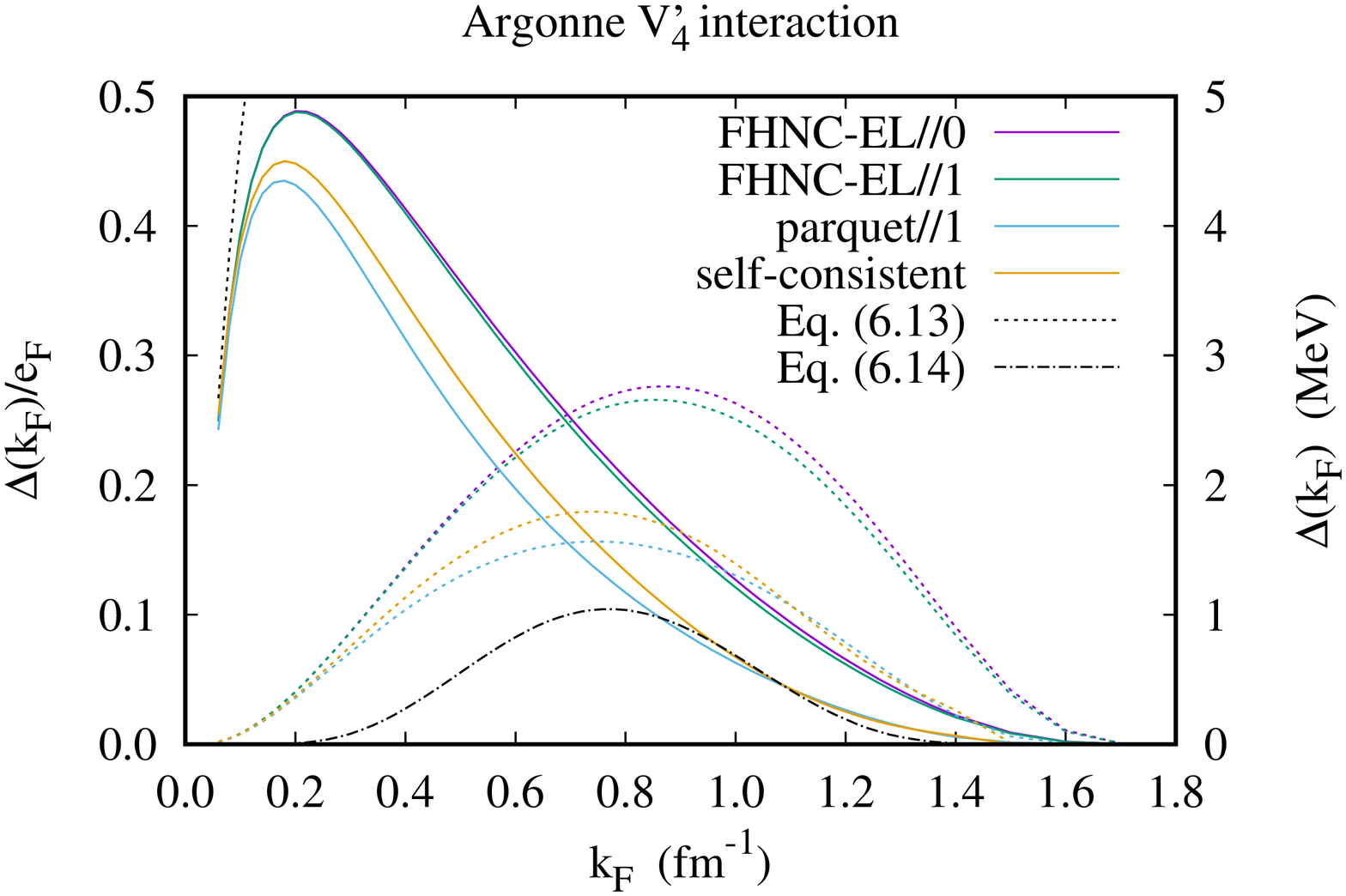}}
      \caption{(color online) The upper figure shows the gap at the
        Fermi surface, $\Delta(\KF)/\EF$, for different versions of the
        theory for the Reid $V_4'$ potential. The curve labeled with
        ``FHNC//0'' shows the results form Ref. \citenum{ectpaper},
        ``FHNC//1'' includes the exchange diagrams (\ref{eq:Wlocal}),
        ``self-consistent'' uses the dynamic interaction
        (\ref{eq:Wsinglet}) with the superfluid Lindhard functions
        (\ref{eq:BCSLindha}) in Eq. (\ref{eq:Weff}), and
        ``parquet//1'' shows the same result using the normal system
        Lindhard function. The dashed lines and the right scale give
        $\Delta(\KF)$ in units of MeV, the dash-dotted line the
        approximation (\ref{eq:gapfw}) for $\Delta(\KF)$ in units of MeV,
        The dotted line shows the estimate (\ref{eq:gapsp}) in units of
        the Fermi energy. The lower figure shows the same for the Reid
        $V_4$ potential.\label{fig:gaps} }
  \end{figure}

Our results for the superfluid gap are shown in Figs. \ref{fig:gaps},
they basically reflect our findings on the pairing
interaction. Including exchanges does not change the gap
significantly; the most significant change comes from using the
dynamic interaction $\tilde W^{(S)}(q,\omega=0)$,
Eq. (\ref{eq:Wsinglet}). Using the superfluid Lindhard function moves
the result a little up but is somewhat unexpectedly not very
significant, despite the suppression of the long-wavelength components
in $\chi_s^{(\sigma)}(q,\omega=0)$ as demonstrated in
Fig. \ref{fig:chiplot}. The approximation (\ref{eq:gapfw})
underestimates $\Delta(\KF)$ by about a factor of 2. We also show the
results obtained from the estimate (\ref{eq:gapsp}) using the
experimental $S$-wave scattering length $a_0 \approx -18.7\ $fm
\cite{PhysRevLett.83.3788}.

We have, in the neutron matter calculations, only solved the parquet-
equation in the density channel because our primary interest is the
pairing interaction; in the spin--channel we have only kept the
polarization diagrams. One can extend the calculation to include spin-
and tensor- operators \cite{SmithSpin} in the full parquet
summation.

\subsection{The Lennard-Jones liquid}
\label{ssec:LJium}

\begin{figure}[H]
\centerline{\includegraphics[width=0.65\columnwidth,angle=-90]{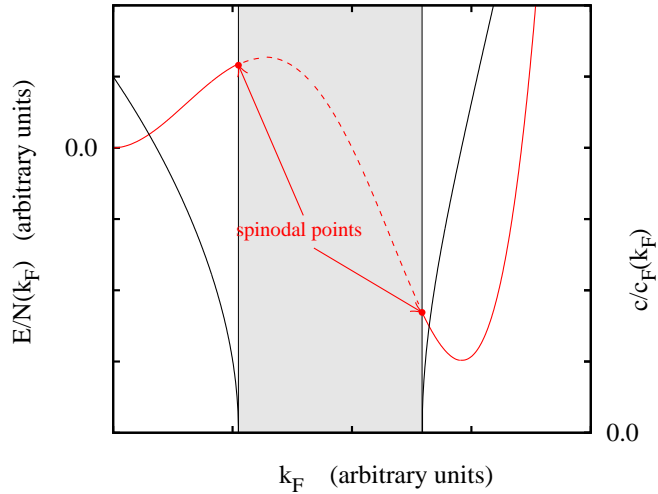}}
      \caption{(color online) The figure shows a schematic equation of
        state of a self-bound Fermi liquid with strong, short-ranged
        repulsion. The left scale and the red curve shows the energy
        per particle, an interpolation between the lower and the upper
        stable regime through the area where no homogeneous liquid can
        exist is drawn dashed. The right scale and the black curve
        shows the speed of sound in units of the Fermi velocity. The
        two spinodal points are indicated by arrows. The gray shaded
        area is the physically unstable density
        regime. \label{fig:schematic} }
\end{figure}

Fermi fluids interacting via a family of Lennard-Jones (LJ)
interactions
\begin{equation}
  v_{\rm L-J}(r) = 4 \epsilon\left[\left(\frac{\sigma}{r}\right)^{12} -
    \left(\frac{\sigma}{r}\right)^{6}\right]
  \label{eq:VLJ}
\end{equation}
have been studied extensively in Refs. \citenum{ljium} and
\citenum{cbcs}. The system is interesting from the viewpoint of
many-body theory since the equation of state can have two spinodal
points where the hydrodynamic speed of sound (\ref{eq:mcfromeos})
vanishes. For the sake of discussion, we show in
Fig. \ref{fig:schematic} a schematic equation of state for a typical
self-bound Fermi liquid.

The dominant energy contribution at very low densities is the Pauli
repulsion, $E/N \propto \KF^2$. As the density increases, interaction
terms, which grow as $\KF^3$, begin to dominate and bend, if the
interaction is sufficiently attractive, the equation of state
downwards. This leads to a local maximum. If that is the case, the
equation of state also has a local minimum at some finite density
before the short-ranged repulsion begins to dominate. Hence, there
must be two spinodal points. The Lennard-Jones liquid shows, for
sufficiently large values of the interaction strength $\epsilon$, both
of these points. In contrast to that, an attractive square-well
potential or the attractive P\"oschl-Teller interaction studied in
Ref. \citenum{GC2008} and in Section \ref{ssec:ptpot}, has only the
lower spinodal point. These systems collapse, due to the absence of a
short-ranged repulsion, when the density is increased to a level where
the interactions begin to dominate over the Pauli pressure. Neutron
matter is not self-bound and does not show spinodal points.

\subsubsection{Energetics and stability}
\label{sec:LJenergy}

\begin{figure}
\centerline{\includegraphics[width=0.65\columnwidth,angle=-90]{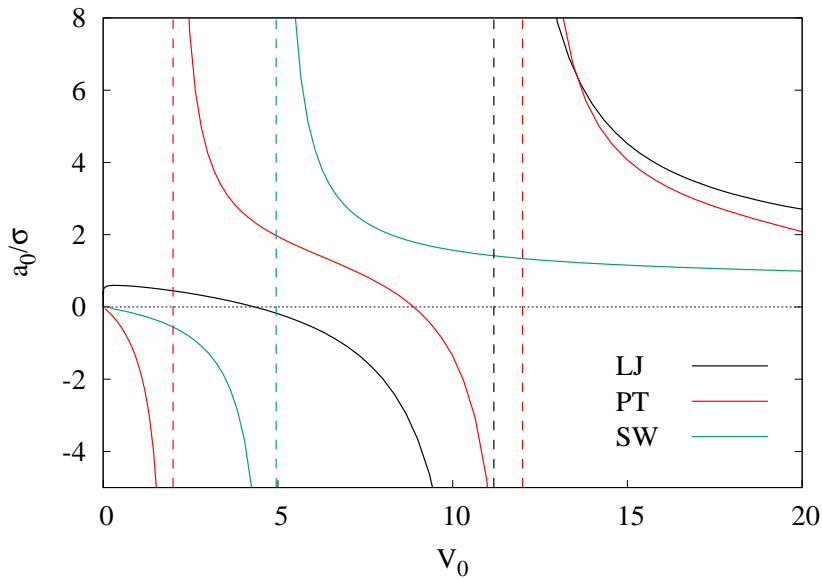}}
\caption{(color online) The plot shows the scattering length $\a0$ as
  a function of the interaction strength for the LJ (black), the
  SW (green) and the PT potential (red).  The long dashed vertical
  lines (at $V_0 = 11.18$ for LJ, $V_0 = 4.336$ for SW as well as
  $V_0(V_0-1)=2$ and $V_0(V_0-1)=12$ for PT indicate the interaction
  strength where a two-body bound state appears. Recall that bound
  states of the PT potential appear at all even integer values of
  $V_0$.
\label{fig:scatplot}}
\end{figure}

We discuss in this section our results for the above family of
Lennard-Jones 6-12 interactions. The strength $\epsilon$ or, better,
the dimensionless strength parameter $V_0 = 2 m
\epsilon\sigma^2/\hbar^2$ of the interaction can be adjusted such that
the interaction has the desired vacuum scattering length; a bound
state and a corresponding singularity of the vacuum scattering length
appears at $V_0 = 11.18$. Fig. \ref{fig:scatplot} shows the scattering
length for both the Lennard-Jones and the P\"oschl-Teller (PT)
interaction discussed in the next section as a function of the
potential well depth. We also show, for comparison, the scattering
length for the attractive square-well interaction discussed in Ref.
\citenum{cbcs}.

Before we discuss our findings, we comment on the expected accuracy of
our results. This was examined in Ref. \citenum{ljium} where we found
that the FHNC//0 approximation is accurate within a percent at low
densities up to approximately 25 percent of the saturation
density. This corresponds to a Fermi wave number of $\KF\sigma\le 0.7$
or a density $\rho = 0.0116\sigma^{-3}$. This is the regime of
interest here. Around saturation, the FHNC//0 approximation still
displaus the correct physics, but leads to an equation of state that
is somewhat too soft. The numerical values in that regime should be
therefore be considered only qualitative. These observations apply, of
course, predominantly to the energy which is relatively insensitive to
the quality of the wave function.  The {\em stability range\/} is
indeed affected, even at low densities, by including the exchange
diagrams discussed in Section \ref{ssec:exchanges}.

We have studied here a density regime of $0.01\, \le\KF\sigma\,\le
2\sigma$, and interactions strengths $1.0\le V_0\le 9.0$. In
Ref. \citenum{cbcs} we went to the much smaller values of $0.001\le
\KF\sigma$. A proper treatment of many-body effects requires that the
simulation box is much larger than $(\KF\sigma)^{-1}$. On the other
hand, the core region must be properly resolved, hence decreasing the
minimum value of $\KF$ by an order of magnitude increases the
necessary number of mesh points on an equidistant mesh (as required by
the fast Fourier transformation algorithm) by an order of magnitude.
In Ref. \citenum{cbcs} we have worked with $2^{18}$ mesh points; since
the implementation of Eq. (\ref{eq:SRPA}) required by the
parquet-level calculations implies an energy integration for each mesh
point, we have reduced in this work the number of mesh points to
$2^{15}$. This limits the minimum density to the above value. Since
the lower spinodal points move, with interaction strength, to lower
densities, this constraint also limits us to interaction strengths
$V_0 < 9.0$.

\begin{figure}[H]
\centerline{\includegraphics[width=0.65\columnwidth,angle=-90]{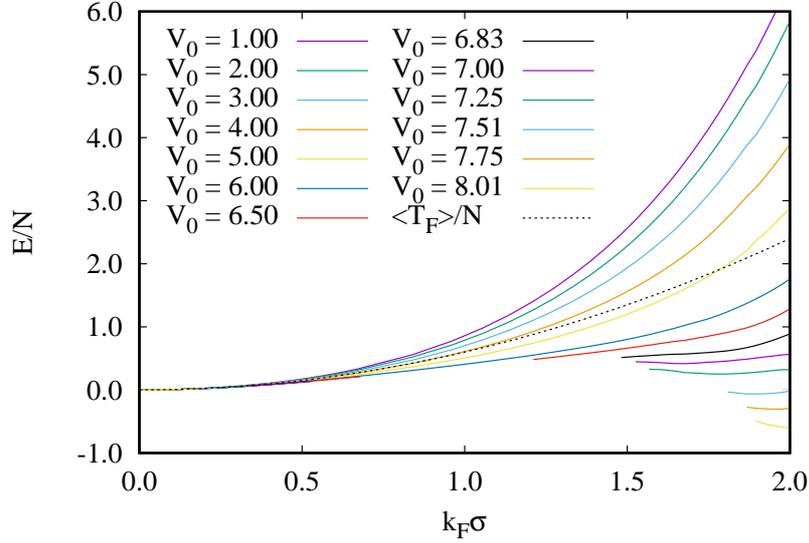}}
\caption{(color online) The figure shows the equation of state of the
  Lennard-Jones liquid for a sequence of interaction strengths $V_0 =
  2 m \epsilon\sigma^2/\hbar^2 = 1.0, 2.0, 3.0, 4.0, 5.0,
  6.0,\ 6.5,\ 6.823,\ 7.0,\ 7.25,\ 7.51,\ 7.75,$ and $8.01$,
  corresponding to vacuum $S$ wave scattering lengths of $a_0/\sigma =
  0.563, 0.446, 0.288, 0.083, -0.188,
  -0.561,\ -0.805,\ -0.990,\ -1.110,\ -1.283,\ -1.493,\ -1.715$ and
  $-1.992$. We show here only the parquet version including the
  exchange corrections to the particle-hole interaction discussed in
  Section \ref{ssec:exchanges}. The topmost curve corresponds to the
  weakest interaction. Also shown is the energy per particle of the
  free Fermi gas (black dashed line).\label{fig:lj_eosplot_v9} }
\end{figure}

The equation of state of for a sequence of interaction strengths $1\le
V_0 \le 8.01$ is shown in Fig. \ref{fig:lj_eosplot_v9}.  Results for
stronger interactions $8.01 < V_0 <\le 9$ are not visible in the
figure and have been omitted. The figure shows clearly the gap in the
equation of state between the two spinodal points which appears when
the interaction is sufficiently strong, $V_0 > 6.0$.

The most interesting result of Ref. \citenum{cbcs} is that the FHNC-EL
equations diverge, at low density and as a function of density, well
before the lower spinodal point is reached. To examine this effect
more closely, we have included the two improvements discussed in
Sections \ref{ssec:exchanges} and \ref{ssec:rings}.  As mentioned
above, the inclusion of the exchange diagrams shown in
Fig. \ref{fig:eelink} can indeed change the long-wavelength limit of
$V_{\rm p-h}(0+)$ by a factor of $1-1/\nu = 1/2$ and, hence, can in
principle increase the stability regime significantly.

\begin{figure}[H]
      \centerline{
        \includegraphics[width=0.65\columnwidth,angle=-90]{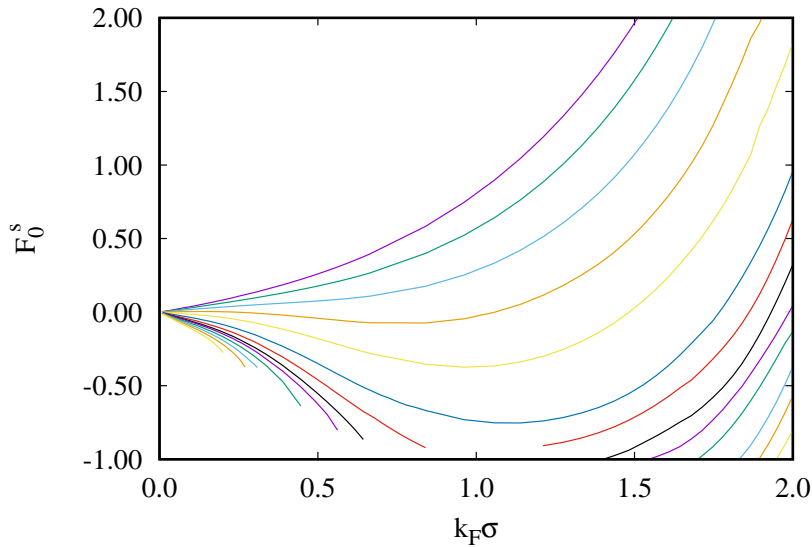}
      }
    \caption{(color online) The figure shows the calculated Fermi
      liquid parameter $F_0^s$ as derived from the long wavelength limit
      (\ref{eq:FermimcfromVph}) at the sequence of coupling strengths used
      in Fig. \ref{fig:lj_eosplot_v9}. \label{fig:lj_f0splot_v9}
}
  \end{figure}

The second correction is due to the transition from the variational
wave function to parquet diagrams.  A summary of our results for the
Fermi liquid parameter $F_0^s$ is, for the sequence of interaction
strengths used in Fig. \ref{fig:lj_eosplot_v9}, shown in
Figs. \ref{fig:lj_f0splot_v9}.

\begin{figure}[H]
      \centerline{
        \includegraphics[width=0.65\columnwidth,angle=-90]{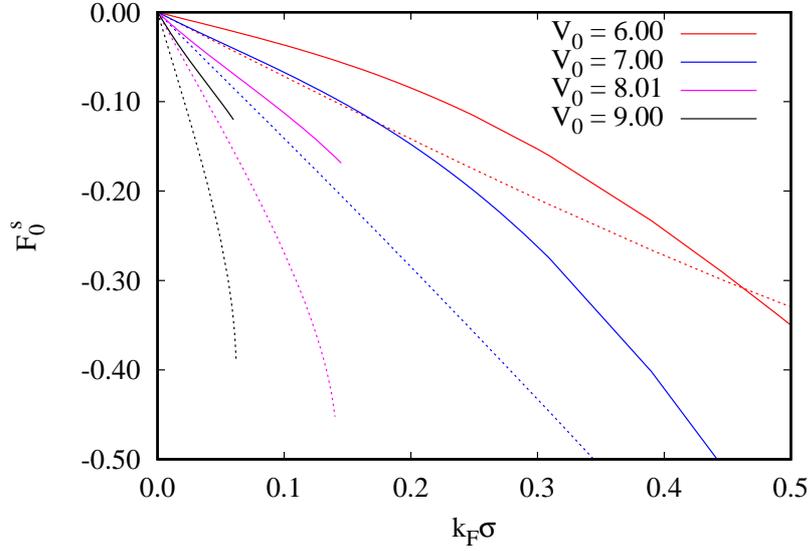}
      }
    \caption{(color online) The figure shows the calculated Fermi
      liquid parameter $F_0^s$ at low densities obtained in FHNC//0
      (dashed lines) and in parquet//1 (solid lines) for four
      interaction strengths.}\label{fig:lj_f0splot_09}
  \end{figure}

The most visible (and expected) quantitative correction is caused, at
low densities, by the inclusion of exchange diagrams. Fig.
\ref{fig:lj_f0splot_09} shows, for four selected densities close to
and above the density where spinodal points exist, the $F_0^s$ as
calculated on the FHNC//0 approximation and the parquet//1
approximation. A number of observations apply: First we notice that
the results with and without including exchange diagrams differ, in
particular in the low density regime, by roughly a factor of 2. This
is consistent with the argument of Section \ref{ssec:exchanges} that
the leading term in the density expansion is reduced by exchange
diagrams. The factor of $1-1/\nu =1/2$ is visible in the slope of
$F_0^s$ at $\KF\rightarrow 0$ as a function of $\KF$.  (The argument
does not apply to the nuclear case since the interaction in the
exchange channel is different from the one in the direct channel, see
Fig. \ref{fig:re_fitplot} and Eq. (\ref{eq:Vphspin}).) The second, and
much more interesting, observation is that the density where the
FHNC-EL equations diverge is almost independent of the inclusion of
exchange diagrams.

The FHNC and the parquet results are, at the same level of
implementation, almost indistinguishable.  The inclusion of the
correct particle-hole propagator only limits the regime of existence
of solutions to the regime $F_0^s > -1$ but has otherwise little
effect.  Evidently, the {\em existence\/} of the spinodal points does
not change qualitatively with the parquet theory developed in Section
\ref{sec:parquet}, however, the location of the divergence changes
somewhat.

\begin{figure}[H]
  \centerline{\includegraphics[width=0.65\textwidth,angle=-90]{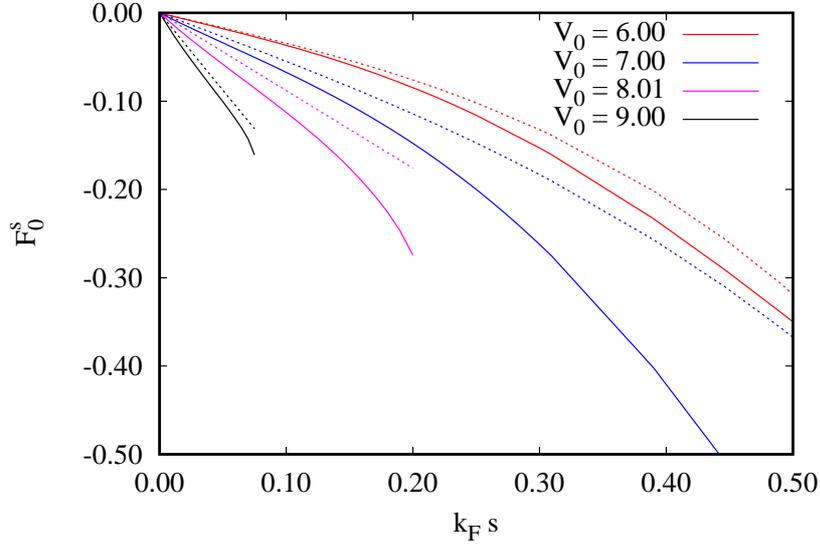}}
  \caption{The figure shows the Fermi-liquid parameter $F_0^s$ as
    calculated from the equation of state, Eq.  (\ref{eq:mcfromeos})
    (solid lines) and from the long-wavelength limit
    Eq. (\ref{eq:FermimcfromVph}) (dashed lines of the same color) for
    four typical interaction strengths.\label{fig:lj_eosfitplot}.}
\end{figure}

To assess the accuracy of our predictions we have again carried out
the consistency test between the between $F_0^s$ obtained from the
equation of state, (\ref{eq:mcfromeos}) and from the long-wavelength
limit (\ref{eq:FermimcfromVph}), $\tilde V_{\rm p-h}(0+)$.  At the
strongest couplings, $2 m \epsilon\sigma^2/\hbar^2 > 7.75$, the
coefficient $b$ of the $\KF^5$ term cannot be determined reliably, we
have therefore kept only the linear in Eq. (\ref{eq:eospoly}). A
comparison of the two procedures is shown in
Fig. \ref{fig:lj_eosfitplot}; evidently the agreement is in the regime
of interest not quite as good as in the nuclear case, this is because
the density dependence of the correlations cannot be ignored.

As in previous work \cite{cbcs} we found that it is impossible
to find solutions of the FHNC-EL or parquet equations close
to the lower spinodal point.  The fact that the equations of state all
come to an endpoint has been identified in Ref. \citenum{cbcs} as due
to a divergence of the in-medium scattering length which is, in the
local approximation used here

\begin{equation}
a \equiv \frac{m}{4\pi\rho\hbar^2}\tilde W (0+)\,.
\label{eq:amedium}
\end{equation}

\begin{figure}[H]
  \centerline{\includegraphics[width=0.65\textwidth,angle=-90]{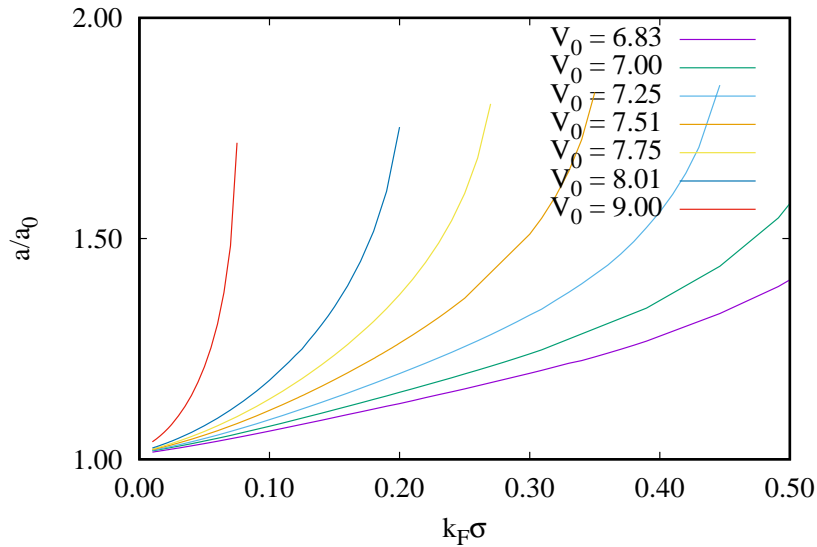}}
  \caption{The figure shows the ratio of the {\em in-medium scattering
      length\/} (\ref{eq:amedium}) to the vacuum scattering length $\a0$ for
    a number of interaction strengths as a function of density.
    \label{fig:lj_a0plot}.}
\end{figure}

This divergence is the reason that the Landau stability limit $F_0^s
\rightarrow -1$ could not be reached. The same situation occurs,
expectedly, in the present case, see Fig. \ref{fig:lj_a0plot}.  One
might have argued that the divergence found previously is indeed just
the spinodal instability and one could not get closer due to some
numerical problem. The comparison of the FHNC//0 and the parquet//1
results shown in Fig. \ref{fig:lj_f0splot_09} shows that the
divergence found here is indeed unrelated to the Fermi liquid
parameter: It appears at practically the same density, although the
$F_0^s$ differ by a factor of 2.

Thus, our conclusions on this topic have not changed, moreover, we have
eliminated the most relevant concern about our previous work, namely
that the Landau parameter $F_0^s$ has not been calculated accurately
enough.

\subsubsection{Effective interactions and correlations:
  A configuration space view}

We have throughout the preceding sections emphasized the momentum
space properties on effective interactions, in particular the
connection between the long wavelength limits, Fermi liquid
parameters, the in-medium scattering length, and the pairing
interaction. Of course, the structure of correlations is mostly
determined by the short--ranged properties of the interactions. Given
the knowledge about the {\em bare\/} interactions, the short--ranged
structure of the correlation functions is given by the
Bethe--Goldstone equation which is, in its essence, a zero-energy
Schr\"odinger equation where the short-ranged structure of the pair
wave function is modified by the Pauli principle
\cite{Goldstone57,BetheGoldstone57,GWW58}.

\begin{figure}[H]
  \centerline{\includegraphics[width=0.65\textwidth,angle=-90]
    {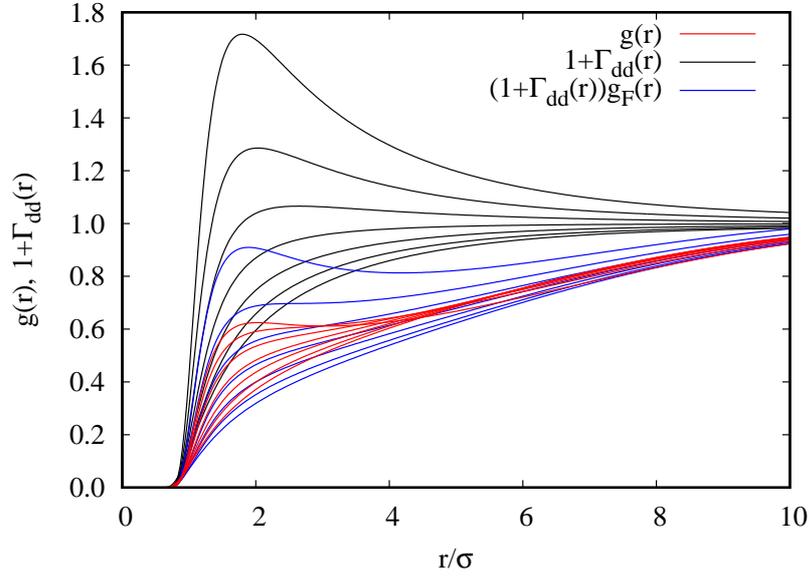}}
  \caption{(color online) The figure shows the pair distribution
    function $g(r)$ obtained from Eq. (\ref{eq:gFHNC0}) (red lines),
    the dynamic correlation function $1 + \Gamma_{\!\rm dd}(r)$ (black
    lines), and the product $\left[1 + \Gamma_{\!\rm
        dd}(r)\right]g_{\rm F}(r)$ (blue lines) for the Lennard-Jones
    interaction at $\KF\sigma = 0.3$ for a sequence of interaction
    strengths $V_0 = 1.0, 2.0, \ldots, 7.0$. The curve with the
    highest peak corresponds to the strongest interaction.
    \label{fig:lj_gofr_kf030}.}
\end{figure}

The relationship between the pair correlation function $f(r)$, the
``dressed'' correlation function $\Gamma_{\rm dd}(r)$ and the pair
distribution function has been discussed in many places, we therefore
focus here on the low-density expansions (\ref{eq:gFHNC0}),
(\ref{eq:C0ofk}).  Fig. \ref{fig:lj_gofr_kf030} shows, at the
relatively low density of $\KF = 0.3\sigma^{-1}$ a sequence of pair
correlation and distribution functions. The ``dressed'' correlation
function $1+\Gamma_{\!\rm dd}(r)$ describes the dynamic correlations
and is determined by the bare interaction. The pair distribution
function contains, in addition, statistical correction manifested
predominantly in the factor $g_F(r)$ in Eq. (\ref{eq:gFHNC0}), but
also in the corrections $C(r)$ and $\tilde X_{\rm ee}(r)$, \cf Eq.
(\ref{eq:C0ofk}).  Fig. \ref{fig:lj_gofr_kf030} also shows the simple
  approximation $\left[1 + \Gamma_{\!\rm dd}(r)\right]g_{\rm F}(r)$
  for the pair distribution function. Evidently, the correction from
  chain diagrams, $(S_{\rm F}^2(k)-1)\tilde\Gamma_{\!\rm dd}(k)$ is
  not negligible; we assert, however, that the contribution from the
  exchange parts $(\Delta \tilde X_{\rm ee})_1(r)$ are invisible on
  the scale of the plot.

\begin{figure}[H]
  \centerline{\includegraphics[width=0.65\textwidth,angle=-90]
    {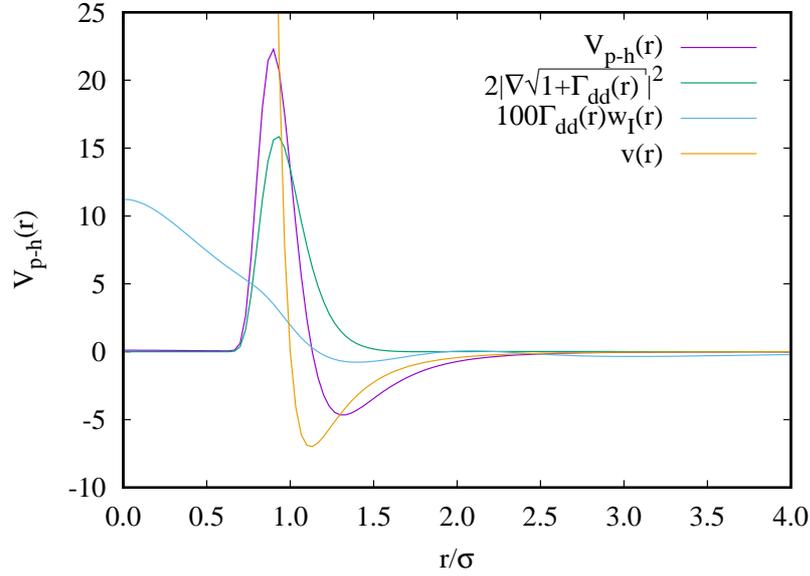}}
  \caption{(color online) The figure shows the components of the
    particle-hole interaction function $V_{\!\rm p-h}(r)$ as defined
    in Eqs. (\ref{eq:Vph}) and (\ref{eq:vcw}) at $\KF\sigma = 0.3$ and
    an interaction strength $V_0 = 7$. Also shown is, for comparison,
    the bare interaction $v(r)$. Note that we had to scale the
    many-body term $\Gamma_{\!dd}(r)w_{\rm I}(r)$ by a factor of 100
    to make it visible at the scale of the plot. Note that in our
    units $\hbar^2/2m=1$.
    \label{fig:lj_Vphparts}.}
\end{figure}

The dynamic correlations determine the effective interactions
$V_{\!\rm p-h}(r)$ via Eqs. (\ref{eq:Vph}) and (\ref{eq:vcw}) and
$W(r)$. Aldrich and Pines \cite{Aldrich,ALP78} give a physically
intuitive description of the effects contributing to $V_{\!\rm
  p-h}(r)$ which is called ``pseudopotential'' in that work:
\begin{enumerate}
\item{} At short distances, the interaction is screened by short-ranged correlations;
  \item{} The fact that the wave function is bent at short distances
    -- downwards for repulsive interactions, upwards for the
    attractive interactions as in the attractive square well potential
    and the P\"oschl-Teller potential -- has a price in kinetic
    energy;
  \item{} Finally, the presence of other particles should lead to an enhancement
    of the interaction in the attractive regime.
\end{enumerate}
Eqs.  (\ref{eq:Vph}) and (\ref{eq:vcw}) give a quantitative meaning to
these effects: The short-ranged screening (1) is described by the
factor $1+\Gamma_{\!\rm dd}(r)$, the cost in kinetic energy is
described by the term $\frac{\hbar^2}{m}
\left|\nabla\sqrt{1+\Gamma_{\rm dd}(r)}\right|^2$,
and many-body effects are described by the last term $\Gamma_{\!\rm
  dd}(r)w_{\rm I}(r)$. Fig. \ref{fig:lj_Vphparts} shows these three
parts of the particle-hole interaction at $\KF\sigma = 0.3$ and an
interaction strength $V_0 = 7.0$.  We had to scale the induced
interaction term by a factor of 100 to make it visible at the scale of
the plot. This does {\em not\/} mean that this term is negligible; it
guarantees that the dynamic correlation function $\Gamma_{\!\rm
  dd}(r)$ falls off as $r^{-2}$ as $r\rightarrow\infty$. \cf
Eq. (\ref{eq:Gddlong}).

\begin{figure}[H]
  \centerline{\includegraphics[width=0.65\textwidth,angle=-90]
    {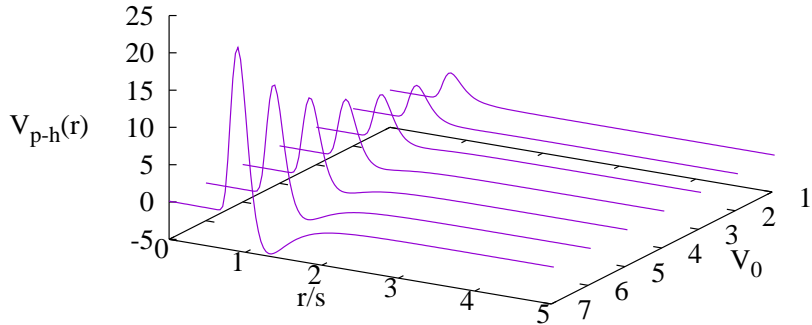}}
  \caption{The figure shows the dependence of the particle-hole
    interaction function $V_{\!\rm p-h}(r)$ at $\KF\sigma = 0.3$ on
    the interaction strength for the sequence $V_0 = 1, 2, \ldots
    7$. The curve with the highest peak and the deepest valley
    correspond to $V_0 = 7$.
        \label{fig:lj_VphofV0}.}
\end{figure}

Fig. \ref{fig:lj_VphofV0} shows, finally, the dependence of $V_{\!\rm
  p-h}(r)$ at $\KF\sigma = 0.3$ on the interaction strength for the
sequence $V_0 = 1, 2, \ldots 7$.  We have not included the exchange
correction, it satisfies to an extremely good accuracy our
estimate $V_{\rm ex}(r) = (g_{\rm F}(r)-1)W(r)$.

\subsubsection{Effective mass}
\label{ssec:ljmass}

The Lennard-Jones $6-12$ interaction may be thought of a model for the
interaction between helium atoms. More modern interactions
\cite{Aziz,AzizII,AzizIII} are perhaps more accurate; but the
Lennard-Jones model reflects the correct physics of \he3. In our
parametrization that uses the core size $\sigma$ as unit of length,
and $\hbar^2/2m\sigma^2$ as unit for the energy, \he3 is characterized
by $V_0 = 8.26$ and the equilibrium density has a Fermi momentum
$\KF\sigma = 2.006$, in other words the equilibrium properties of \he3
are beyond the regime where out approximations are reliable. See
Ref. \citenum{ljium} for a discussion and full FHNC-EL calculations.

Liquid \he3 is known to undergo below 3 mK superfluid phase
transitions \cite{OsheroffRichardsonLee,LEG72} to two different phases
of $P$-wave superfluidity. At the experimental saturation density of
\he3, the soft spin-fluctuation mode is very important and is the
essential mechanism for the strong effective mass enhancement around
the Fermi momentum \cite{GRE83,GRE86,Bengt,he3mass}. If both backflow
and spin-fluctuations are omitted, one ends up with an effectiv mass
less than 1 \cite{IndianMass}.  We are in this work interested in
low-density systems and potential $S$-wave pairing. We will see below
that $S$-wave superfluidity can occur only at very low density; in
particular {\em below\/} the lower spinodal point which is around
$\KF\sigma \approx 0.13$ for $V_0 = 8.26$, see
Fig. \ref{fig:lj_f0splot_v9}. Therefore, we expect that the much
simpler G0W approximation spelled out in Eqs. (\ref{eq:G0W}),
(\ref{eq:green}) is adequate.

\begin{figure}[H]
\centerline{\includegraphics[width=0.65\textwidth,angle=-90]{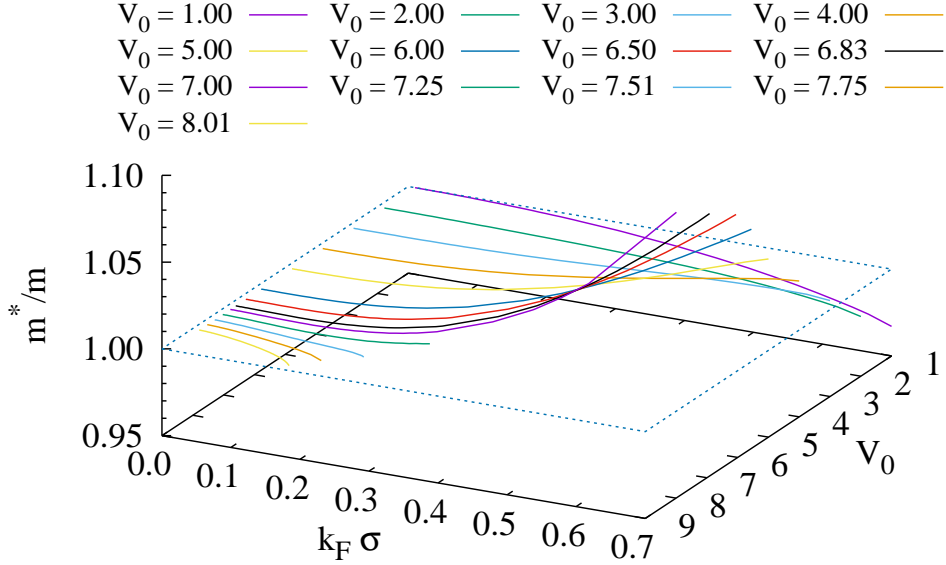}}
  \caption{The figure shows the effective mass ratio $m^*/m$ in the
    low density region for a sequence of interaction strengths $V_0$.
    The dashed line is at $m^*/m=1$ to guide the eye.
    \label{fig:lj_massplot}.}
\end{figure}

Fig. \ref{fig:lj_massplot} shows the effective mass obtained as
described above way for a sequence of interaction strengths
$V_0$. Throughout, the effective mass never differs from the bare mass
by more than 10 percent. Interestingly, we find for the most repulsive
interactions a slight reduction of the effective mass ratio which
turns into an enhancement around $V_0 = 4.0$ and then above $V_0 = 7.25$
again into a reduction.

The explanation for that is that the value of the effective mass ratio
is an effect of both Fermi statistics and hydrodynamic backflow. Fermi
statistics has mostly (but not always) the effect of {\em reducing\/}
the effective mass, this is a well-known effect in nuclei
\cite{BrownGunnGould} but also comes out for \he3 in the na\"\i ve
formulation (\ref{eq:uhncdef}) \cite{IndianSpins}. Hydrodynamic
backflow \cite{FeynmanBackflow} increases the effective mass. That
effect is included in the G0W approximation which includes
hydrodynamic backflow; see Ref. \citenum{polish} how Feynman's
backflow operator comes out already in second order CBF perturbation
theory.  In this work, we have not included spin fluctuations, hence
the results shown are only due to Fermi statistics and backflow.
Apparently, backflow effects become stronger with increasing
attractiveness of the interaction. This makes sense because more
particles are relatively close to each other; see
Fig. \ref{fig:lj_gofr_kf030} how the nearest neighbor peak
increases. For that to be effective, of course the density must be
high enough.

\subsubsection{BCS pairing}
\label{ssec:LJpairing}

Let us now turn to the calculation of the superfluid gap. By
construction, our pairing interaction ${\cal W}(1,2)$ contains
polarization corrections, see the discussion of the connection between
parquet theory and FHNC-EL in Section \ref{sec:parquet}, \cf
Eqs. (\ref{eq:Vchain}). These polarization corrections are important
modifications of the pairing interaction and can lead to significant
changes of the superfluid gap, see, for example,
Refs. \citenum{CKY76,Wam93,Sch96,SPR2001}

The details of the choice of interactions (\ref{eq:Weff}), \ie if
exchange diagrams are included and whether we take the collective
Lindhard function or the energy dependent one, define different
implementations of the pairing interaction in FHNC or parquet diagram
theory.  Using {\em local\/} correlations means that the Lindhard
function $\chi_0(q,\omega)$ is replaced by the {\em collective\/}
Lindhard function $\chi_0^{\rm coll}(q,\bar\omega(q))$ defined in
Eq. (\ref{eq:Chi0Coll}), and taken at the average frequency
$\bar\omega (q)$, Eq. (\ref{eq:wbar}).  In FHNC//0 approximation, this
leads to the expression (\ref{eq:Wlocal}) which was used in previous
work \cite{cbcs,ectpaper}.  The equivalence between FHNC and
parquet-diagrams permits us, however, to be more flexible. Since the
pairing gap is expected to be relatively small one is led, in the
parquet version the theory, to take the pairing interaction at zero
frequency \cite{SPR2001}.

\begin{figure}[H]
      \centerline{
        \includegraphics[width=0.50\textwidth,angle=-90]{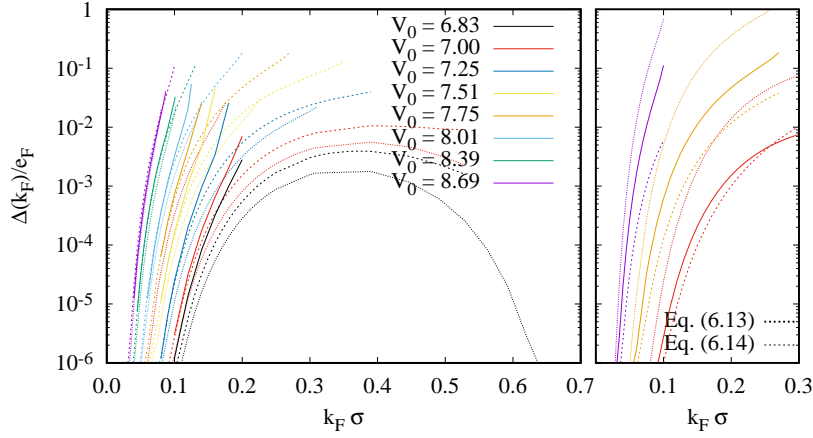}
      }
      \caption{(color online) The figures shows the superfluid gap
        $\Delta$ at the Fermi surface in units of the Fermi energy at
        a sequence of dimensionless coupling strengths $V_0 =
        \epsilon\sigma^2/\hbar^2 = 6.823,\ 7.0,\ 7.25,\ 7.51,\ 7.75,$
        and $8.01$ in FHNC//0 approximation \cite{cbcs}, (short dashed
        lines), parquet//1 (long-dashed lines) and the fully
        self-consistent calculation. The right panel shows, for a selected set of
        coupling strengths, a comparison of the full solution of the
        gap equation with the results of the approximations
        (\ref{eq:gapsp}) (long dashed lines) and (\ref{eq:gapfw})
        (short dashed lines).
 \label{fig:ljgapplot}}
  \end{figure}

Our results are shown in Fig. \ref{fig:ljgapplot}. Indeed, both
modifications, exchange diagrams and using the sero-energy Lindhard
function for the particle-hole propagator can cause a change of up to
a factor of two in the superfluid gap; the effect is about half due to
exchange diagrams, and half due the propagator correction. The right
pane of Fig. \ref{fig:ljgapplot} shows a comparison of the solution of
the full gap equation with the two approximations (\ref{eq:gapsp}) and
(\ref{eq:gapfw}).

A somewhat surprising result is that fully self-consistent theory
where all particle-hole propagators (\ref{eq:BCSLindha}) are taken for
the superfluid system causes, despite the smallness of the gap, a
visible effect in the stability regime as indicated by the end of the
respective lines. Apparenty, being close to an instability, makes
every quantity sensitive to small details.

In the model we have considered here it seems impossible to get close
to the ``BCS-BEC'' crossover regime. The reason for lies both in
physics and in formalism: From Fig. \ref{fig:lj_f0splot_v9} it is
clear that one is, whenever the superfluid gap is sizable, close to
the lower spinodal point, even if one cannot reach it. For systems
like the one studied here, and, even more, for the attractive
P\"oschl-Teller interaction studied in the next section, one has two
competing effects: To get a stronger gap, one needs a more attractive
interaction. Such an attractive interaction has, on the other hand,
the effect that the Pauli-pressure is more easily overwhelmed by the
interaction and, hence, the spinodal point moved to lower densities.

The formal reason is also quite plausible: The implementation of
FHNC-EL or parquet theory involves two sets of diagrams, rings and
ladders. A third set, the ``cyclic chain'' diagrams that would
indicate a divergence of the single particle propagator which might be
present in spin-polatrized \he3 \cite{KCJ} has not been considered
here. A divergence of the ring-diagrams is related to the existence of
spinodal points or, more generally, to Landau's stability conditions
$F_\ell^{s,a} > -(2\ell+1)$.  What we find, however, is a divergence
in ladder equation (\ref{eq:ELSchr}). It is harder to associate that
with a specific quantity because the induced interaction $w_{\rm I}$
always tries to adjust itself such that the $v(r)+w_{\rm I}(r)$ has
zero scattering length and that, therefore, Eq. (\ref{eq:ELSchr}) has
a solution with the property $\Gamma_{\!\rm dd}(r) \propto r^{-2}$ as
$r\rightarrow \infty$, see Eq. (\ref{eq:Gddlong}).  But it is easy to
envision that the potential becomes attractive enough such that the
equation develops a bound state. This has the consequence that
$1+\Gamma_{\!\rm dd}(r)$ falls off exponentially as $r\rightarrow
\infty$. Then the basic assumptions of the equations of the theory are
violated and it is no surprise that they have no solutions. This is
apparently what happens here, where the effect is caused by the
attractiveness of the induced interaction which describes
predomilantly phonon exchange. Hence, we have concluded in
Ref. \cite{cbcs} that the divergce is due to a dimerization caused by
phonon exchange, in other words it is a genuine many-body effect.

This is an interesting statement {\em per-se:\/} Note that in a weakly
interacting system, the gap equation can describe the transition
between a ``BCS'' state where the Cooper pairs are weakly coupled, to
a ``BEC'' phase where the pairs are strongly bound
\cite{NozieresSchmittRink}. The gap equation is a proper subset of our
diagram summation; but we find a {\em divergence\/} of a set of
diagrams that are not included in weakly interacting systems described
by the gap equation alone. The issue deserves further investigation.

\subsection{P\"oschl-Teller interaction}
\label{ssec:ptpot}

The purely attractive P\"oschl-Teller (PT) interaction
\cite{PoeschlTeller}
\begin{equation}
  V(r) = - \frac{\hbar^2}{m\sigma^2}
  \frac{V_0(V_0-1)}{\cosh^2(r/\sigma)}
  \label{eq:Vpt}
\end{equation}
is characterized by the strength $V_0$ and the range $\sigma$. For this
interaction, the scattering length can be calculated analytically
\cite{Fluegge}.
\begin{equation}
  \frac{\a0}{\sigma} =
  \left(-\frac{1}{V_{0}}+\frac{\pi}{2\tan(\pi V_{0}/2)}+\gamma+
  \frac{d\log\Gamma(V_{0}+1)}{dV_{0}}\right)\,,
\label{eq:PTa0}
\end{equation}
where $\gamma$ is Euler`s constant. The interaction is somewhat more
convenient to use in Monte Carlo calculations than the square-well
interaction \cite{astraPRL04}; it has been used in
Ref. \citenum{GC2008} to study pairing in the unitary limit. We have
recently examined the ground state and pairing properties of the
P\"oschl-Teller gas \cite{qfs2018}; in this work we have improved
these calculations by applying the improvements outlined above. Since
many techical details have been discussed in the preceding sections,
we can restrict ourselves to display just the new aspects of
our calculations.

\subsubsection{Energetics and stability}
\label{ssec:PTenergy}

Since the P\"oschl-Teller interaction does not have a repulsive core,
there is no self-bound, high-density phase. The low density properties
are basically determined by the balance between the Pauli repulsion
and the attractive interparticle attraction. Our findings for the
PT interaction are sufficiently similar to those of the LJ interaction
sich that we can be very brief..

\begin{figure}[H]
  \centerline{\includegraphics[width=0.6\textwidth,angle=-90]{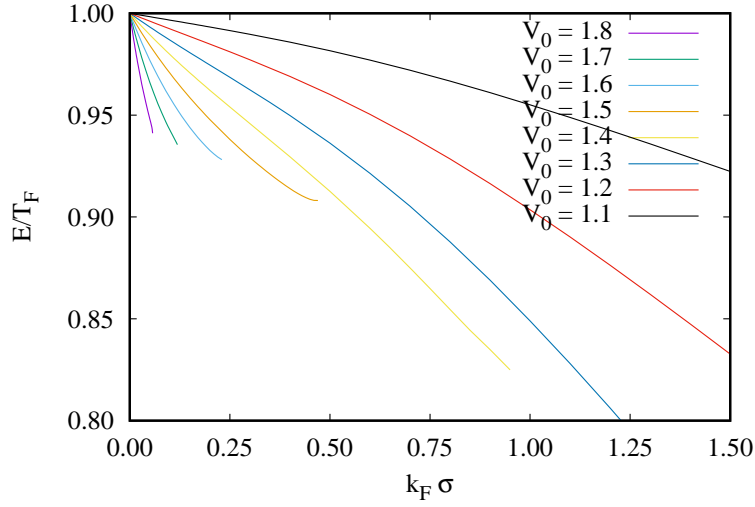}}
    \caption{(color online) The figure shows the ratio of the ground
      state energy of the Fermi gas interacting via the
      P\"oschl-Teller interaction (\ref{eq:Vpt}) and the kinetic
      energy $T_{\rm F}$ of the free Fermi gas.  The curves correspond
      to a sequence of coupling strengths $V_0 = 1.1, \ldots, 1.8$,
      corresponding to scattering lengths $a_0/\sigma = -0.019,
      -0.044, -0.079, -0.125, -0.191, -0.291, -0.460$ and
      $-0.794$. The FHNC and parquet results are practically
      indistinguishable apart from the fact that the stability regime
      is quite different. We show therefore only the parquet//1
      results.
      \label{fig:pt_eosplot_v9}}
  \end{figure}

\begin{figure}[H]
  \centerline{\includegraphics[width=0.6\textwidth,angle=-90]{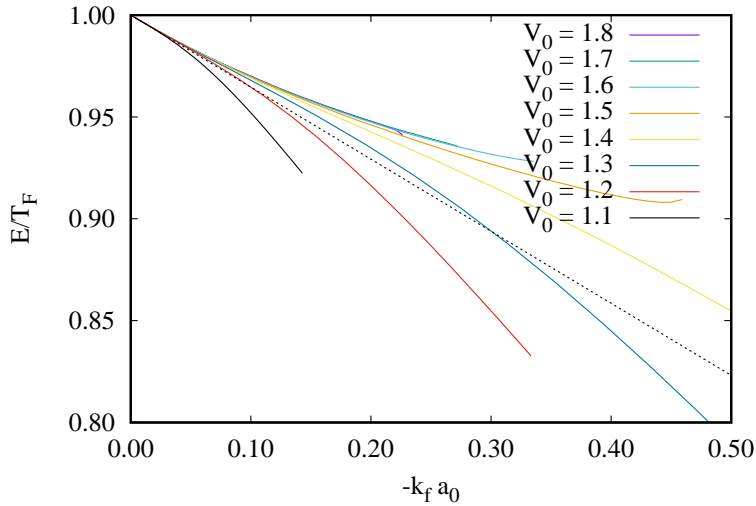}}
  \caption{(color online) The figure shows, for same sequence of
    coupling strength as in Fig. \ref{fig:pt_eosplot_v9} $E/T_{\rm
      F}$ as a function of $-\a0\KF$. Also shown is the estimate from
    the {\em linear terms\/} in the expansion (\ref{eq:elowdens})
    (dashed black line).
      \label{fig:pt_eosplot_akf}}
\end{figure}

Fig. \ref{fig:pt_eosplot_v9} shows the equation of state, normalized
to the kinetic energy of the free Fermi gas. In the limit of low
densities, the equation of state should be give by the Huang-Yang
expansion (\ref{eq:elowdens}), we have studied this in
Ref. \citenum{cbcs} for the Lennard-Jones and the attractive
square-well model and found that the term proportional to $(\a0\KF)^2$
is already overshadowed by non-universal many-body effects. We compare
therefore only the linear term in the expansion (\ref{eq:elowdens})
and restrict, of course, the expansion to small values of
$\a0\KF$. Similar to our previous work we find that the expansion
(\ref{eq:elowdens}) is reasonably accurate for $\a0\KF < 0.05$, see
Fig. \ref{fig:pt_eosplot_akf}.

\begin{figure}[H]
  \centerline{
    \includegraphics[width=0.65\textwidth,angle=-90]%
                    {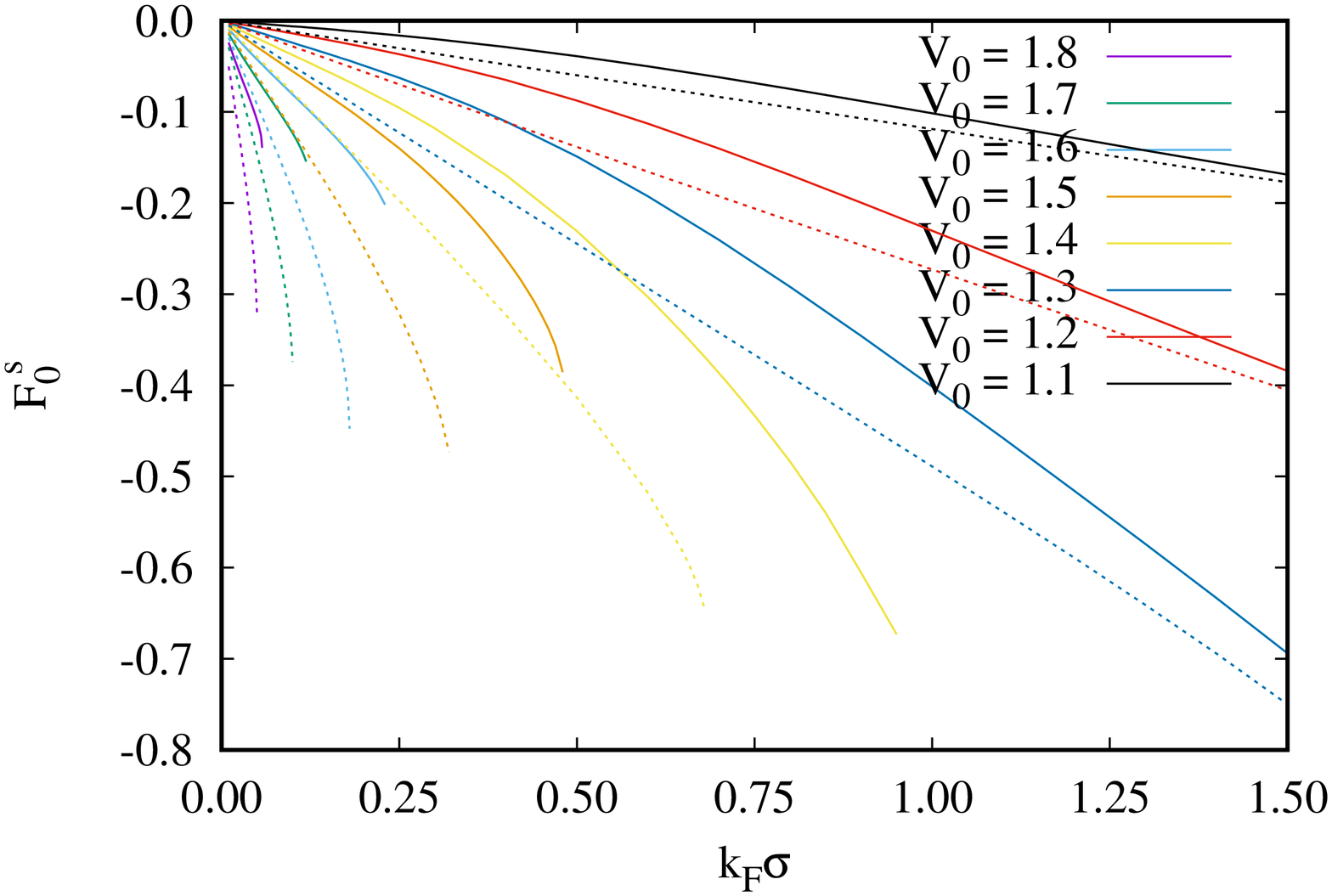}}
    \centerline{\includegraphics[width=0.65\textwidth,angle=-90]%
                    {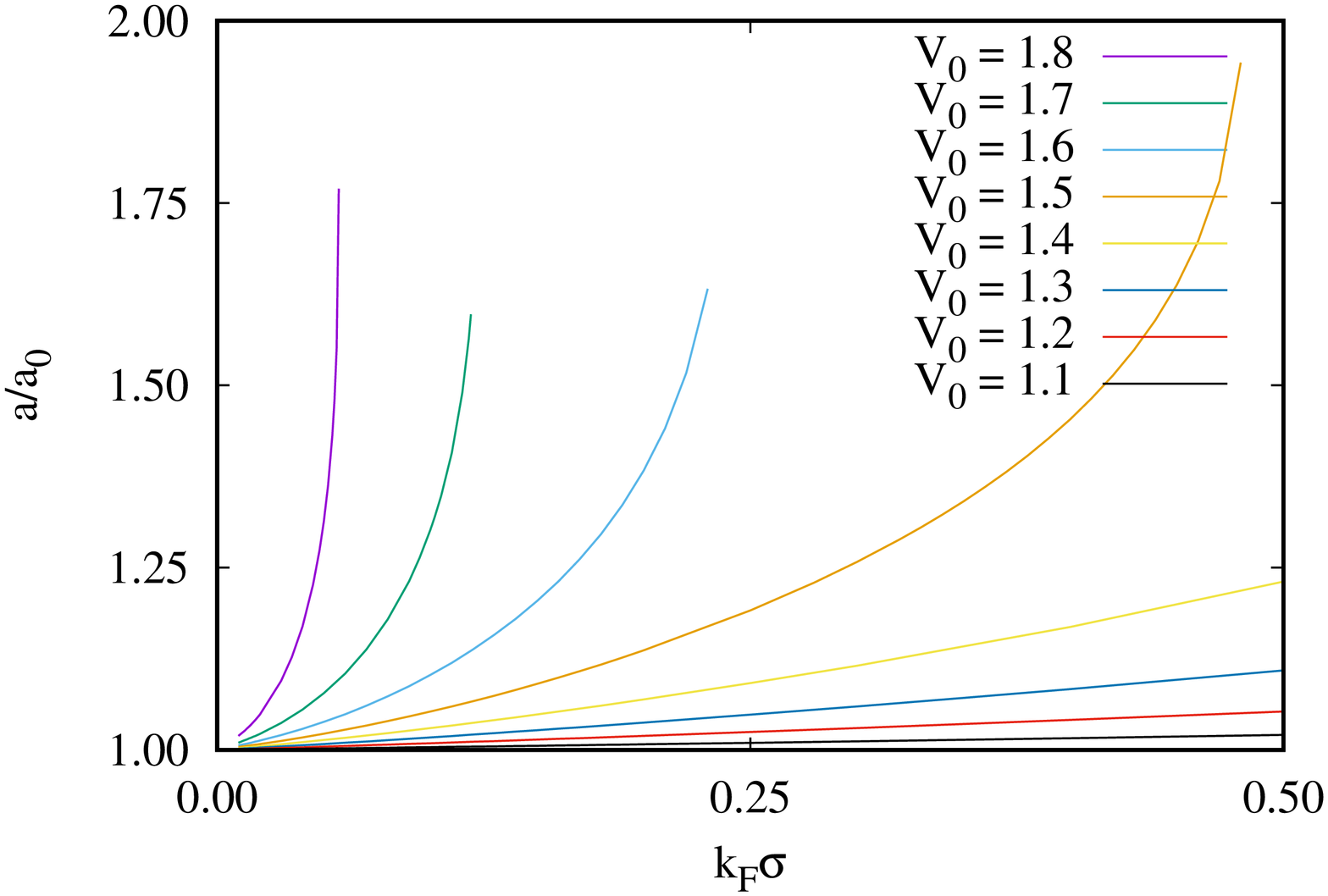}}
  \caption{(color online) The upper figure shows the Fermi-liquid
    parameter $F_0^s$ of the ``P\"oschl-Teller'' liquid for the same
    sequence of interaction strengths $V_0$ for the FHNC//0
    approximation (dashed black curves) and the parquet calculations
    including exchange diagrams (solid red curves).  The lower figure
    shows the ratio between the in-medium scattering length $a$ and
    the vacuum scattering length $\a0$ for the same sequence of
    interaction strengths.\label{fig:pt_f0splot_09}}
  \end{figure}

Fig. \ref{fig:pt_f0splot_09} shows the Fermi liquid parameter $F_0^s$
as obtained from Eq. (\ref{eq:FermimcfromVph}) for the same sequence
of interaction strengths as used in Fig. \ref{fig:pt_eosplot_v9} as a
function of the density. In general, there is little {\em
  qualitative\/} change; the stability regime is changed visibly by
including exchanges, but the result that there is no way to get close
to Landau instability regime $F_0^s\rightarrow -1$ is not changed; in
fact, the most advanced calculation leads, similar to the
Lennard-Jones model, to a much smaller value of $F_0^s$ where the
divergence occurs. Due to this instability we have not been able to
reach the rather large values of $-\KF\a0$ reported in
Ref. \citenum{GC2008} before the system became unstable. We have again
checked the consistency between Eqs. (\ref{eq:mcfromeos}) and
(\ref{eq:FermimcfromVph}) with the procedure outlined in the previous
section.  We found that the numerical values are practically identical
for weak couplings when exchange diagrams are included.

\begin{figure}[H]
  \centerline{\includegraphics[width=0.6\textwidth,angle=-90]{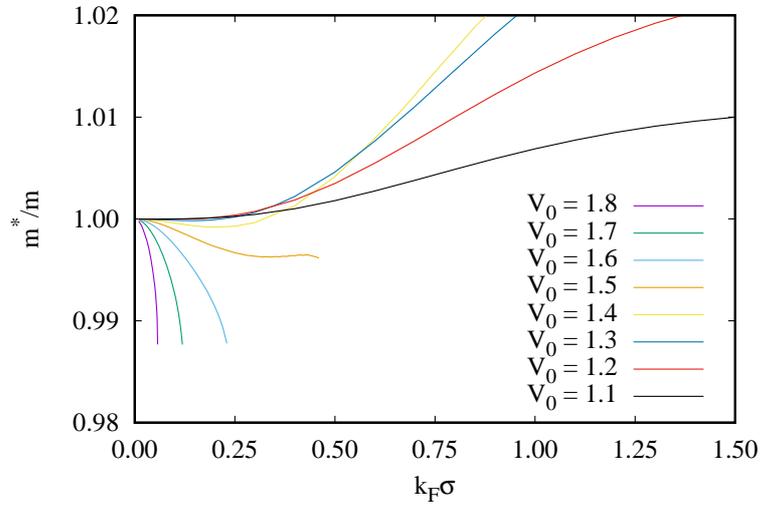}}
  \caption{(color online) The figure shows, for same sequence of
    coupling strength as in Fig. \ref{fig:pt_eosplot_v9}, the
    effective mass ratio as calculated in the on-shell G0W
    approximation (\ref{eq:mstar}).
      \label{fig:pt_masses}}
\end{figure}

For completeness, we show in Fig. \ref{fig:pt_masses} the effective mass
ratio for the above sequence of coupling strengths. As in our previous
calculations, these ratios are close to 1 which is certainly a consequence of
the low density of the system.

\subsubsection{BCS pairing}
\label{ssec:PTpairing}

\begin{figure}[H]
    \includegraphics[width=0.7\textwidth,angle=-90]{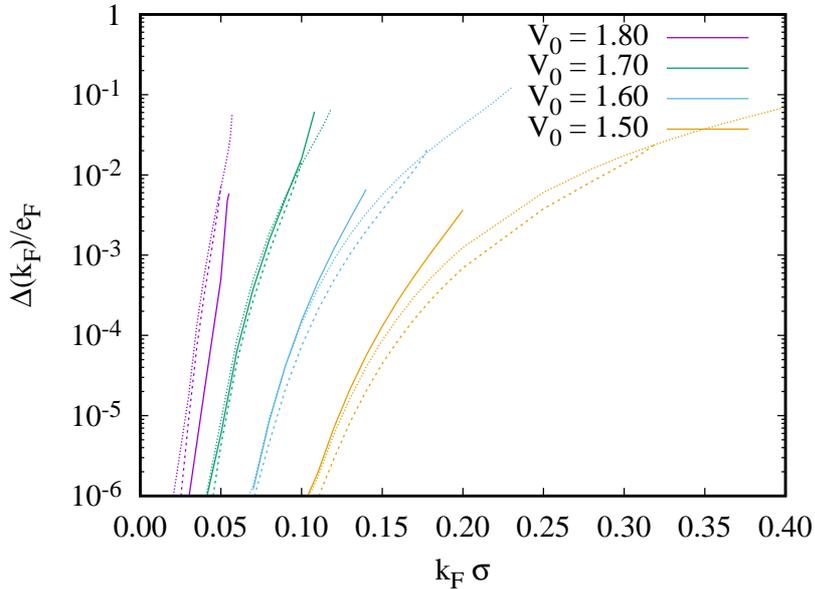}
\caption{(color online) The figure shows the superfluid gap for a
  number of coupling strengths in parquet (long lines) and in
  FHNC//0-EL approximation (short dashed lines) as well as the results
  of the fully self-consistent calculation.
  \label{fig:pt_gapkf} }
\end{figure}

Fig.~\ref{fig:pt_gapkf} shows the calculated energy gap in FHNC//0,
parquet//1 and fully self--consistent approximation. Evidently,
inclusion of the energy-dependent effective interaction can change the
value of the gap visibly. This is, of course, not a statement on the
specific FHNC approximation, but more generally on the quality of the
locally correlated (or ``fixed-node'') wave function. The fully
self-consistent calculation brings us here back into the vicinity of
FHNC//0, we hasten to point out that this observation is
circumstantial and should not be generalized. 

In conclusion, we note again that the most important change duu to
going beyond the ``weak coupling'' approximation (\ref{eq:Ebcs}) is
that the stability regime is further reduced.

\section{Discussion}
\label{Sec:conclusion}

We have formulated in this paper the variational and parquet approach
for strongly interacting, both normal and superfluid, systems. We
have then studied the Euler equation for the local correlations in a
version that exhibits all the exact properties of variational and
parquet-diagram theories but does not require the formulation of the
full FHNC theory. We have derived the interesting, yet disturbing,
result that the Euler equation for locally correlated wave functions
of the form (\ref{eq:Jastrow}) has, for a superfluid system, no
physically meaningful solution for net attractive interactions $\tilde
V_{\rm p-h}(0+)<0$, and displays pathological features even for
repulsive interactions.

The problem is caused by the very form of the Jastrow-Feenberg wave
function. The plausible way solve this problem is by comparison with
parquet-diagrams. This has been carried out in Section
\ref{sec:parquet}, and that formulation can be naturally carried over
to the superfluid system. That has the effect that the ``collective
approximation'' for $S(k)$, Eq. (\ref{eq:PPAs}) is replaced by the RPA
expression (\ref{eq:SRPA}). The formulation of Coupled Cluster theory
with correlated wave functions \cite{CBFPairing} would perhaps be
another way to carry this through, but, on the other hand, our result
is sufficiently convincing.

We have then carried out numerical calculations for different model
systems of physical interest: neutron matter and model systems
interacting via a family of Lennard-Jones and P\"oschl-Teller
interactions. With the recent interest in pairing in cold gases and
the BEC-BCS crossover, it has become fashionable to stress the
similarity between cold gases and neutron matter. In fact, the neutron
scattering length $-18.7 \pm 0.6\ $fm \cite{PhysRevLett.83.3788} which
is much larger than the range of the interaction. We found in this
work that these systems are actually rather different. Whereas there
was no problem calculating the properties of neutron matter at all
densities, the parquet-equations showed inevitable divergences for
both the P\"oschl-Teller and the Lennard-Jones interactions at
scattering lengths of the order of $a_0 = 5\dots 8\sigma$. The
divergence is driven by the approaching of a ``soft mode'' at the
spinodal points(s) $F_0^2\rightarrow \rightarrow -1$, but {\em not
  caused\/} by it. Spinodal points appear in both the P\"oschl-Teller
and the Lennard-Jones interactions, see Figs. \ref{fig:lj_f0splot_v9}
and \ref{fig:pt_f0splot_09}, but neutron matter is far from such an
instability, see Fig. \ref{fig:re_fitplot}. The {\em cause\/} of the
instability rather seems to be a divergence in the ladder equation.

Monte Carlo simulations for a square-well interaction seem to indicate
that the equation of state comes close, but do not reach, a spinodal
decomposition before the BEC-BCS crossover is reached
\cite{Astrapriv}; the data of Refs. \citenum{Gezerlis2014,GC2008} are
too sparse to make a conclusive statement.

Two points should, of course, be emphasized: One is that the
``fixed-node'' wave function displays the spurious instability
discussed extensively above. Simulations based on such a wave function
will at their best converge to the {\em infimum\/} of the energy that
can be reached by wave functions of the type (\ref{eq:CBCS}).  Another
problem arises since an exact simulation should converge towards the
exact ground state, which could also be a droplet or a film covering
the walls of the simulation box. However, we are interested here in
the features of the {\em uniform, but metastable\/} state that can be
present close to the spinodal point. We believe that this problem
deserves a very careful examination.

The most serious problem in the application to nuclear systems is that
there is evidence that parquet-class of diagrams is not sufficient for
reliable calculations \cite{SpinTwist}. Unfortunately, this is
insufficient when the core or the interaction is very different in the
spin-singlet and the spin-triplet channel.  In that case, the
so-called ``twisted chain'' diagrams can be very important. We must
keep this problem in mind but leave it for future work.  In the
language parquet theory, the ``rungs'' cannot be considered to be
local interactions, but all time-orderings must be kept. In the
language of the Jastrow-Feenberg method, the so-called commutator
diagrams are important. The effect can be very dramatic. We will
examine this in future work on neutron matter.

\appendix

%
%%%%%%%%%%%%%%%%%%%%%%%%%%%%%%%%%%%%%%%%%%%%%%%%%%%%%%%%
%

\section{Cluster expansions for the generating function. Diagonal terms}
\label{app:AppA}

We show in the following series of appendices some details of the
calculation of the generating functions given in Eqs. (\ref{eq:Gdiag})
and (\ref{eq:Goffd}).  First, we calculate the cluster expansion of
the $G_{\rm diag}(\beta)$: Define
\begin{equation} 
[G_{\rm diag}]_{\mvec,\mvec}^{(N)} \equiv \ln I_{\mvec,\mvec}^{(N)}
= \sum_{n,s} (\Delta G_{\mvec,\mvec})_n^{(s)},
\label{eq:Gdiag_expand}
\end{equation}
where $(\Delta G_{\mvec,\mvec})_n^{(s)}$ is the sum of all cluster
contribution containing $n$ points and $s$ correlation lines
$h(r_{ij})$. The individual terms of the cluster expansion are
identical to the expansion of the normal ground state if we
re-interpret the exchange line $\ell(r\KF)$ as
\[\ell_{\mvec}(r) = \frac{\nu}{\NO}\sum_{\kvec\in\mvec}e^{\I\kvec\cdot\rvec}\,.\]

Next, we formulate the resulting matrix elements as matrix elements of
a diagonal second quantized operator,
\begin{equation}
  (\Delta G_{\mvec,\mvec})_n^{(s)} = \bra{\mvec}
(\Delta \hat{G}_{\rm diag})_n^{(s)}\ket{\mvec}\,.
\end{equation}
where
\begin{eqnarray}
  (\Delta \hat{G}_{\rm diag})_n^{(s)} = \sum_{m_i}\bra{ m_1,\ldots, m_n}
  G_n^{(s)}(\rvec_1,\ldots , \rvec_n) \ket{ {\cal P}(m_1,\ldots,m_n)} \creat{m_1} \cdots \creat{m_n} \annil{m_n} \cdots \annil{n_1},
\label{eq:secondquantizedG}
\end{eqnarray}
where ${\cal P}(m_1,\ldots,m_n)$ stands for some permutation of the
quantum numbers $(m_1,\ldots,m_n)$, and $G_n^{(s)}(\rvec_1,\ldots ,
\rvec_n)$ is a combination of $s$ correlation lines connecting $n$
points. The leading order terms are
\begin{align}
  (\Delta \hat{G}_{\rm diag})_2^{(1)} &=
  \frac{1}{2} \sum_{i,j} \bra{i j}h(12)\ket{ij}_a
  \creat{i}\creat{j}\annil{j}\annil{i}, \\
  (\Delta \hat{G}_{\rm diag})_3^{(2)} &= \frac{1}{2}\sum_{i,j,k}
  \left[ \bra{ijk}h(12)h(23)\ket{kij+jki} - \bra{ijk}h(12)h(23)\ket{kji}
     \right] \nonumber\\
& \quad \times \creat{i}\creat{j}\creat{k}\annil{k}\annil{j}\annil{i}\,.
\end{align}

Finally, we calculate the expectation value of these operators with
respect to $\ket{\mathrm{BCS}}$ by the usual contraction technique
\cite{SuhonenBook,Fantonipairing}

\begin{align}
  \bra{\mathrm{BCS}}\annil{m} \creat{n} \ket{\mathrm{BCS}}
  &= u_n^2 \delta_{mn},\\
  \bra{\mathrm{BCS}}\creat{m}\annil{n} \ket{\mathrm{BCS}}
  &= v_n^2 \delta_{nm}, \\
  \bra{\mathrm{BCS}}\creat{m}\creat{a_n} \ket{\mathrm{BCS}}
  &= \sgn(m) u_m v_m \delta_{n\bar{m}}, \\
  \bra{\mathrm{BCS}}\annil{m}\annil{n} \ket{\mathrm{BCS}} &=
  \sgn(n) u_n v_n \delta_{m\bar{n}}.
\end{align}
where $\bar{m}$, $\bar{n}$ refers the states with opposite momentum and
spin of the states $m$, $n$, respectively, and $\sgn(m) = 1$ ($-1$) when
the spin of $m$ is up (down).

\begin{align}
  \left(\Delta G_{\rm diag}\right)_2^{(1)} &=
  \bra{\mathrm{BCS}}(\Delta \hat{G}_{\rm diag})_2^{(1)}\ket{\mathrm{BCS}}
  = \frac{1}{2}\sum_{i,j} \bra{i j}h(12)\ket{ij}_a
  \bra{\mathrm{BCS}}\creat{i}\creat{j}\annil{j}\annil{i}\ket{\mathrm{BCS}}
  \nonumber\\
  &= \frac{1}{2}
  \sum_{i,j} \bra{ i j}h(12)\ket{ij}_a v_i^2 v_j^2 = \frac{1}{2}\rho^2
  \int d^3r_1 d^3r_2 h(r_{12})\left(1-\frac{1}{\nu}\ell_v^2(r_{12})\right)\,.
\label{eq:G_diag_2}
\end{align}
and
\begin{align}
  \left(\Delta G_{\rm diag}\right)_3^{(2)}
  &= \bra{\mathrm{BCS}}(\Delta G_{\rm diag})_3^{(2)} 
  \ket{\mathrm{BCS}}\nonumber\\
  &=\frac{1}{\nu^2}\rho^3\int  d^3r_1 d^3r_2 d^3r_3 h(r_{12}) h(r_{23}) \ell_v(r_{12})
  \ell_v(r_{23}) \ell_v(r_{31})\nonumber\\
&-\frac{1}{2\nu }\rho^3\int  d^3r_1 d^3r_2 d^3r_3 h(r_{12}) h(r_{23}) \ell_v^2(r_{31})
\label{eq:G_diag_3_2}
\end{align}

The diagrammatic representation of the two-body contribution
(\ref{eq:G_diag_2}) is shown in the diagrams (a) and (b) in
Fig. \ref{fig:G} whereas $\left(\Delta G_{\rm
  diag}\right)_3^{(2)}$ corresponds to {\em twice\/} diagram (f) and
diagram (d) of that figure.

We have to address a somewhat subtle issue connected with the
appearance of ``equivalent'' diagrams which have a
different topological structure, but the same value.  Diagrams (d) and (e)
of Fig. \ref{fig:Gnormal} are the simplest example.
In the {\em normal\/} system it is easily seen that these diagrams
have the same value because the exchange function $\ell(r\KF)$
satisfies the ``convolution property:
\begin{equation}
  \ell(r_{ij}\KF) = \frac{\rho}{\nu}
  \int d^3 r_k \ell(r_{ik}\KF)\ell(r_{kj}\KF)\,.
\label{eq:ConvolutionProperty}
\end{equation}
In contrast, as pointed already out in Ref. \citenum{Fantonipairing},
the convolution property does {\em not\/} hold for the
$\ell_v(r_{ij})$ exchange lines. Superficially one might therefore
expect to get two different expressions for diagrams 4 and 5. An immediate
consequence would also be that the expansion becomes point-reducible
\cite{Fantonipairing}. We therefore examine the issue here carefully.

We go back to the expansion (\ref{eq:Gdiag_expand}), represented
diagrammatically in Fig. \ref{fig:Gnormal}. Diagram (e) can be written
as
\begin{align}
  (\Delta G_{\mvec,\mvec})_4^{(2e)} &= -\frac{1}{2\nu^3}\rho^4
  \int d^3r_1,\ldots d^3r_4 h(r_{12})h(r_{34})
  \ell_{\mvec}(r_{12})\ell_{\mvec}(r_{23})\ell_{\mvec}(r_{34})\ell_{\mvec}(r_{42})
  \nonumber\\
  &=-\frac{1}{2\nu^2}\rho^3\int d^3r_1 d^3r_2 d^3r_3 h(r_{12})h(r_{23})
  \ell_{\mvec}(r_{12})\ell_{\mvec}(r_{23})\ell_{\mvec}(r_{31})\,.
\end{align}
The last equation holds because $\ell_{\mvec}(r_{ij})$ also satisfies
the convolution property (\ref{eq:ConvolutionProperty}).
The latter representation has then to be written in the second
quantized form (\ref{eq:secondquantizedG}) which gives the correct
weight factor of diagram (f) in Fig. \ref{fig:G}.

The reason why these diagrams must be collected {\em before\/}
calculating the expectation value with respect to the $\ket{\mathrm{BCS}}$
is that we have two {\em disconnected\/}
correlation lines $h(r_{12})h(r_{34})$ in $(\Delta G_{\mvec,\mvec})_4^{(2)}$
leads additional momentum conservation:
\[
  \bra{i jk\ell}h(12)h(34)\ket{\ell ijk}\propto\delta_{j,\ell}\delta_{j,\ell}
    \]
    which would give a zero contribution to the second-quantized form.
    A very similar consideration was necessary for
    the calculation of the CBF single particle energies
    (\ref{eq:CBFph}), see Ref. \citenum{CBF2}. 
    
\section{Cluster expansions for the generating function. Off-diagonal terms}
\label{app:AppB}

To derive cluster expansions for the generating function $G_{\rm offd}
- \ln I_{\rm CBCS}$ we can use again the power-series method explained
in Section \ref{ssec:cluster_expansions} and define the normalization
of $|\mathrm{CBCS}\rangle$ as
  
  \begin{eqnarray}
    I_{\rm CBCS}(\alpha) &=& \ovlp{\mathrm{CBCS}(\alpha)}{\mathrm{CBCS}(\alpha)}
    \nonumber\\
    &=& 1 + \sum_{N,\mvec,\nvec}\ovlp{\mathrm{BCS}}{\mvec^{(N)}}
    N_{\mvec,\nvec}^{(N)}(\alpha)\ovlp{\nvec^{(N)}}{\mathrm{BCS}},
\end{eqnarray}   
where $N_{\mvec,\nvec}^{(N)}(\alpha)$ is given by
\begin{equation}
\frac{I_{\mvec,\nvec}^{(N)}(\alpha)}{[I_{\mvec}^{(N)}(\alpha)]^{1/2}[I_{\nvec}^{(N)}(\alpha)]^{1/2}}\equiv\delta{\mvec,\nvec} +  N_{\mvec,\nvec}^{(N)}(\alpha)\,,
\end{equation}
with
\begin{equation}
  I_{\mvec,\nvec}^{(N)}(\alpha) = \bra{\Phi_\mvec^{(N)}}
  \prod (1+\alpha h(ij))\ket{\Phi_{\nvec}^{(N)}}\,,
\end{equation}
see Eq. \ref{eq:defineNM}. We now use the power-series method
\cite{Fantoni} described in Section \ref{ssec:cluster_expansions} to
obtain a cluster expansion
\begin{eqnarray}
  G_{\rm offd}
    = \sum_{n=1}^\infty (\Delta G_{\rm offd})^{(n)}\,.
\end{eqnarray}

We can then write a term with $n$ correlation lines as

\begin{equation}
  (\Delta G_{\rm offd})^{(n)} = \sum_{N,{\bf m}\ne{\bf n}}
  \ovlp{\mathrm{BCS}}{{\bf m}^{(N)}}
  \bra{{\bf m}^{(N)}} \sum_m(\Delta N)_{m}^{(n)}\ket{{\bf n}^{(N)}}
  \ovlp{{\bf n}^{(N)}}{\rm BCS}
\end{equation}
where the $(\Delta N)_{m}^{(n)}$ are $m$-body clusters with $n$ lines.
We then write these as off-diagonal matrix elements
of second quantized operators
\begin{eqnarray}
  &&(\Delta N)_{m}^{(n)}(\rvec_1,\ldots,\rvec_m) \rightarrow
  (\Delta \hat{N})_m^{(n)}\nonumber\\
  &\equiv&\bra{i_1,\ldots,i_m}(\Delta N)_{m}^{(n)}(\rvec_1,\ldots,\rvec_m)
  \ket{{\cal P}(j_1,\ldots,j_m)}\creat{i_1}\dots\creat{i_m}
  \annil{j_m}\dots\annil{j_1}\,.
\label{eq:Nsecondquantized}
\end{eqnarray}
A few leading order terms, where the states $\ket{\mvec}$ and $\ket{\nvec}$
differ by exactly two states, are
\begin{align*}
  (\Delta \hat{N})_2^{(1)} &= \frac{1}{2} \sum_{m_i \neq n_i}
  \bra{m_1 m_2}h(12)\ket{n_1 n_2} \creat{m_1} \creat{m_2}\annil{n_2}\annil{n_1},
  \\
  (\Delta \hat{N})_3^{(2)} &=  \sum_{m_i \neq n_i, j}
  \Bigl[ \frac{1}{2}\bra{ m_1 m_2 j} h(13)h(23)\ket{n_1 n_2 j}
    -\frac{1}{2}\bra{m_1 m_2 j} h(12)h(13)\ket{n_1 j n_2} \nonumber\\
&  -\bra{ m_1 m_2 j} h(12)h(23)\ket{ n_1 j n_2} \Bigr] \creat{m_1} \creat{m_2} \creat{j} \annil{j} \annil{n_2}\annil{n_1}\,.
\end{align*}

\begin{figure}[H]
  \centerline{\includegraphics[width=0.8\textwidth]{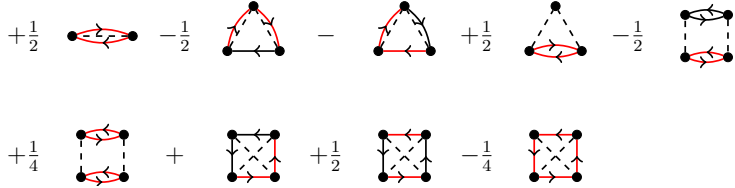}}
  \caption{The figure shows the cluster contributions to the
    generating functional $G_{\rm offd}$ containing one or two
    correlation lines.\label{fig:normalization}}
\end{figure}

In this formulation, we can calculate the expectation values of these
operators with respect to $\ket{\mathrm{BCS}}$.
The diagrams contributing to $G_{\rm offd}$ containing one or two
correlation lines are shown in Fig.~\ref{fig:normalization}. 

As a matter of course we have to pay again attention to the appearance
of equivalent diagrams which must be combined before the BCS
expectation value is calculated. Since the operations are identical to
those outlined for the diagonal components in \ref{app:AppA}, we skip
this step.

\section{Cluster expansions for the energy numerator terms}
\label{app:AppC}

In this appendix, we show details of the calculation of the energy
numerator corrections $E_{\rm enum}$ defined in
Eq. (\ref{eq:FCBCS}). We begin with the definition
\begin{eqnarray}
E_{\rm enum} = \frac{1}{2}\frac{\sum_{N,{\bf m},{\bf n}}
    \ovlp{\rm BCS}{{\bf m}^{(N)}}\left(H_{\mathbf{m},\mathbf{m}}^{(N)}
    + H_{\mathbf{n},\mathbf{n}}^{(N)}-2E_{\rm diag}\right)
    N_{{\bf m},{\bf n}}^{(N)}\ovlp{{\bf n}^{(N)}}{\rm BCS}}{\ovlp{\rm CBCS}{\rm CBCS}}
\label{eq:EenumApp}
\end{eqnarray}

We can again use the power-series method to generate a
cluster-expansion for $E_{\rm enum}$. Unlike above, we expand the
diagonal matrix elements $H_{\mvec,\mvec}^{(N)}$ and the off-diagonal
matrix elements $N_{\mvec,\nvec}^{(N)}$ individually. Analogously to
\ref{app:AppA}, we can write the diagonal matrix elements of the
Hamiltonian in terms of a sum of diagonal, second-quantized $n$-body
operators
\begin{equation}
  (\Delta \hat{h})_n^{(s)}
  = \bra{ k_1,\ldots, k_n} (\Delta h)_n^{(s)}(\rvec_1,\ldots,\rvec_n)\ket{{\cal P}(k_1,\ldots,k_n)}
  \creat{k_1} \ldots \creat{k_n} \annil{k_n} \ldots \annil{k_1}\,.
\end{equation}

Similarly, we write the matrix elements $N_{\mvec,\nvec}^{(N)}$ in the
form (\ref{eq:Nsecondquantized}). Then, each single contribution to
the sum in the numerator of Eq. (\ref{eq:EenumApp}) has the form
\begin{eqnarray}
  &&\frac{1}{2}\bra{\mathrm{BCS}}(\Delta \hat{h})_n^{(s)}
  (\Delta \hat{N})_m^{(t)} + (\Delta \hat{N})_m^{(t)}
  (\Delta \hat{h})_n^{(s)}\ket{\mathrm{BCS}} - (\Delta E_{\rm diag})_n^{(s)} (\Delta N)_m^{(t)}\nonumber\\
  &=&\frac{1}{2}\bra{ k_1,\ldots, k_n} (\Delta h)_n^{(s)}(\rvec_1,\ldots,\rvec_n)\ket{{\cal P}(k_1,\ldots,k_n)} \times \nonumber\\
  && \times  \bra{i_1,\ldots,i_m}(\Delta N)_{m}^{(t)}(\rvec_1,\ldots,\rvec_m)
  \ket{{\cal P}(j_1,\ldots,j_m)}\times\nonumber\\
  &\times&
  \biggl[\bra{\mathrm{BCS}}
    \creat{k_1} \ldots \creat{k_n} \annil{k_n} \ldots \annil{k_1}\creat{i_1}\ldots\creat{i_m}
  \annil{j_m}\ldots\annil{j_1}+\mathrm{h.c.}\ket{\mathrm{BCS}}\nonumber\\
  &&- 2 \bra{\mathrm{BCS}}\creat{k_1} \ldots \creat{k_n}\annil{k_n} \ldots \annil{k_1}\ket{\mathrm{BCS}}
   \bra{\mathrm{BCS}}\creat{i_1}\ldots\creat{i_m}
  \annil{j_m}\ldots\annil{j_1}\ket{\mathrm{BCS}}\biggr]\,.
\end{eqnarray}

The expectation values of the creation and destruction operators with
respect to $\ket{\mathrm{BCS}}$ are then done with the usual
contraction technique \cite{SuhonenBook}. Unlike above, we now have
two groups of creation and annihilation operators: The group
$\{k_1,\ldots,k_n\}$ belonging to the diagonal matrix elements, and
the group $\{i_1,\ldots,i_m,j_1,\ldots,j_m\}$ belonging to the
$(\Delta \hat{N})_m^{(t)}$. We now classify the expansion in terms of
the {\em number of contractions between the two groups}. The term
where only operators within each group are contracted with each other
cancels the second term in the above equation.

Hence, the simplest contribution is the set of diagrams where one pair
of creation/destruction operators of the set $\{k_1,\ldots,k_n\}$ is
contracted with one pair of the group
$\{i_1,\ldots,i_m,j_1,\ldots,j_m\}$. {\em All other\/} contractions of
operators of the set $\{k_1,\ldots,k_n\}$ are done just as for the
$G_{\rm diag\/}$ and the resulting $E_{\rm diag\/}$.  It is clear that
the resulting quantity can depend only on the quantum number of the
state that has been contracted with the group
$\{i_1,\ldots,i_m,j_1,\ldots,j_m\}$. Hence, we can think of these
contributions as a one-body operator
\begin{equation}
  \hat{\epsilon}^{(v)} = \sum_{\kvec,\sigma} \epsilon^{(v)}(k)
  \creat{\kvec,\sigma}\annil{\kvec,\sigma}
\end{equation}
where
\begin{equation}
  \epsilon^{(v)}(k) \equiv \left[ \frac{\delta E_{\rm diag}}
    {\delta \ell_v(r_{ij})}\right]^{\cal F}(k),
\end{equation}
and $[\ldots]^{\cal F}$ denotes the Fourier transform defined as
(\ref{eq:ft}), and $k$ is the momentum carried by the common state $n_k$.

\begin{figure}
  \centerline{\includegraphics[width=0.8\textwidth]{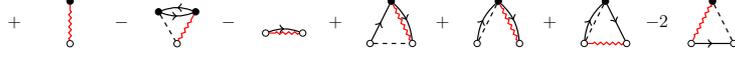}}
  \caption{The diagrammatic representation of the interaction
    corrections to the one-body operator. These are identical to the
    is CBF single particle energies when the black exchange lines are
    interpreted as $\ell_v(r_{ij})$. Note that the first two terms are
    constant and cancel against $\epsilon_0^{(v)}$.
  \label{fig:esp}}
\end{figure}

The leading term in $\epsilon^{(v)}(k)$ is kinetic energy $t(k)$,
higher other correction terms to $t(k)$ are shown in
Fig.~\ref{fig:esp}. They reduce to the CBF single-particle spectrum in
the normal system if one replace $\ell_v(r_{ij})$ by $\ell(r_{ij})$.

We can now rewrite $E_{\rm enum}^{(1)}$ as
\begin{eqnarray}
E_{\rm enum}^{(1)} = \frac{1}{2}\sum\frac{\bra{\mathrm{BCS}}\left(\hat{\epsilon}^{(v)} - \epsilon_0^{(v)}\right)
  {\cal N} + {\cal N} \left(\hat{\epsilon}^{(v)} - \epsilon_0^{(v)}\right)\ket{\mathrm{BCS}}}{\ovlp{\rm CBCS}{\rm CBCS}},
\label{eq:eenum}
\end{eqnarray}
where
\begin{equation}
\epsilon_0^{(v)} = \bra{\mathrm{BCS}} \hat{\epsilon}^{(v)} \ket{ \mathrm{BCS}},
\end{equation}
and $\cal N$ can again be expanded in a cluster expansion as in \ref{app:AppB}: 
\begin{eqnarray}
\hat{\cal N} \equiv  \sum_{m,n} (\Delta \hat{N})_m^{(n)}.
\end{eqnarray}

A few leading order terms in $\hat{\cal N}$ are
given by
\begin{align}
  (\Delta \hat{N})_2^{(1)} &= \frac{1}{2} \sum_{m_i \neq n_i}
  \bra{m_1 m_2}h(12)\ket{n_1 n_2} \creat{m_1} \creat{m_2}\annil{n_2}\annil{n_1},
  \label{eq:DN21}\\
  (\Delta \hat{N})_3^{(2)} &=  \sum_{m_i \neq n_i, j}
  \Bigl[ \frac{1}{2}\bra{ m_1 m_2 j} h(13)h(23)\ket{n_1 n_2 j}
    -\frac{1}{2}\bra{m_1 m_2 j} h(12)h(13)\ket{n_1 j n_2} \nonumber\\
&  -\bra{ m_1 m_2 j} h(12)h(23)\ket{ n_1 j n_2} \Bigr] \creat{m_1} \creat{m_2} \creat{j} \annil{j} \annil{n_2}\annil{n_1}\,.\label{eq:DN32}
\end{align}
Thus, we can define
\begin{equation}
E_{\rm enum}^{(1)} = \sum_{m,n} (\Delta E_{\rm enum}^{(1)})_m^{(n)}.
\end{equation}

\begin{figure}
  \centerline{\includegraphics[width=0.8\textwidth]{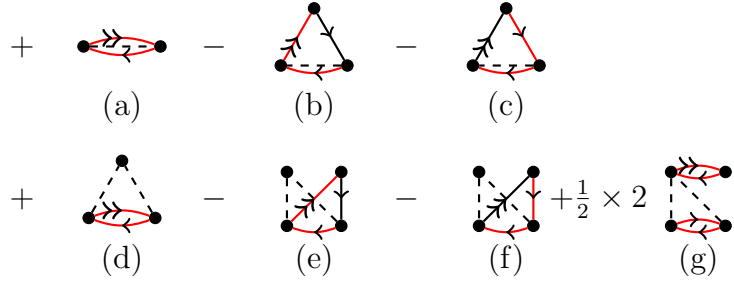}}
  \caption{Leading
  terms in $E_{\rm enum}^{(1)}$. The exchange lines with a double
  arrow indicate the kinetic exchange lines defined in
  Eq. (\ref{eq:ellkin}). \label{fig:T_enum}}
\end{figure}

We give details of the calculation here only for two typical terms:
The two-body term $(\Delta E_{\rm enum})_2^{(1)}$ is
\begin{eqnarray}
  (\Delta E_{\rm enum})_2^{(1)} &=& \frac{1}{2}\bra{ \mathrm{BCS}} \left(\hat{\epsilon}^{(v)} - \epsilon_0^{(v)}(k)\right)
    (\Delta \hat{N})_2^{(1)} + (\Delta \hat{N})_2^{(1)}   \left(\hat{\epsilon}^{(v)} - \epsilon_0^{(v)}(k)\right) \ket{\mathrm{BCS}}. \nonumber\\
\end{eqnarray}
The rest of the calculation is performed by the usual contraction
technique which leads to
\begin{eqnarray}
&& \frac{\nu}{2}\sum_{\kvec,\kvec'} u_\kvec v_\kvec u_{\kvec'} v_{\kvec'} \left[ (1-2v_{\kvec}^2)(\epsilon^{(v)}(k)-\mu) + (1-2v_{\kvec'}^2)(\epsilon^{(v)}(k')-\mu) \right] \times\nonumber\\
  && \times \bra{\kvec \uparrow, -\kvec \downarrow } h(12)
  \ket{\kvec' \uparrow, -\kvec' \downarrow}.
\end{eqnarray}
This leads to diagrams (a), (b), and (c) shown in
Fig.~\ref{fig:T_enum}, where the black and red double-arrow
respectively stand for a kinetic exchange lines,
\begin{equation}
\frac{\nu}{\rho}
\int \frac{d^3 k}{(2\pi)^3}(\epsilon^{(v)}(k)-\mu) v_{\kvec}^2e^{\I\kvec\cdot\rvec}
\quad\mathrm{and}\quad
\frac{\nu}{\rho}
\int \frac{d^3 k}{(2\pi)^3}(\epsilon^{(v)}(k)-\mu) u_{\kvec}v_{\kvec}e^{\I\kvec\cdot\rvec}\,.
\label{eq:ellkin}
\end{equation}

A new structure comes it at the next order, $(\Delta E_{\rm
enum}^{(1)})_3^{(2)}$:
\begin{eqnarray}
  (\Delta E_{\rm enum}^{(1)})_3^{(2)} &=&
  \frac{1}{2}\bra{ \mathrm{BCS}}
  \left(\hat{\epsilon}^{(v)} - \epsilon_0^{(v)}(k)\right)
    (\Delta \hat{N})_3^{(2)} + (\Delta \hat{N})_3^{(2)}
    \left(\hat{\epsilon}^{(v)} - \epsilon_0^{(v)}(k)\right)
    \ket{\mathrm{BCS}}. \nonumber\\
\end{eqnarray}
We look only at the first term in $(\Delta \hat{N})_3^{(2)}$ in Eq.
(\ref{eq:DN32}) which is evaluated as
\begin{eqnarray}
&&\frac{\nu}{2} \sum_{\kvec,\kvec',\kvec_1, \sigma_1}
u_{\kvec} v_{\kvec} u_{\kvec'} v_{\kvec'}v_{\kvec_1}^2
\left[(1-2v_{\kvec}^2)(\epsilon^{(v)}(k)-\mu)
+ (1-2v_{\kvec'}^2)(\epsilon^{(v)}(k')-\mu) \right] \times\nonumber\\
&&\qquad\times \bra{ \kvec \uparrow, -\kvec \downarrow, \kvec_1 \sigma_1}
 h(13)h(23)\ket{\kvec'\uparrow, -\kvec' \downarrow, \kvec_1 \sigma_1}\nonumber\\
 &+&
 \frac{\nu}{2} \sum_{\kvec,\kvec',\kvec_1, \sigma_1} u_{\kvec} v_{\kvec} u_{\kvec'}
 v_{\kvec'} 2v_{\kvec_1}^2 u_{\kvec_1}^2(\epsilon^{(v)}(k_1)-\mu)  \times
 \nonumber\\
&& \qquad \times \bra{ \kvec \uparrow, -\kvec \downarrow, \kvec_1 \sigma_1 }
h(13)h(23)\ket{ \kvec'\uparrow, -\kvec' \downarrow, \kvec_1 \sigma_1}\,.
\end{eqnarray}
The first two terms are represented by diagrams (d), (e), and (f) in
Fig. \ref{fig:T_enum} and the last term is represented by diagram (g)
in Fig.~\ref{fig:T_enum}.

\begin{figure}[t]
  \centerline{\includegraphics[width=0.8\textwidth]{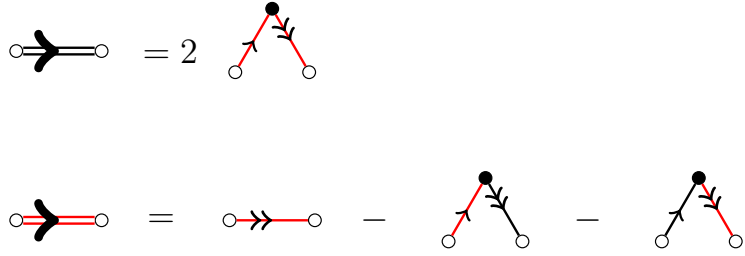}}
  \caption{The figure shows the definition of the dressed kinetic
    exchange lines $\ell_v'(r_{ij})$ and $\ell_u'(r_{ij})$ (denoted by
    black and red double arrow lines) that are introduced to combine
    the diagrams shown in Fig. \ref{fig:T_enum} to the first three
  diagrams shown in Fig. \ref{fig:Eenum}.}
  \label{fig:luplvp}
\end{figure}

\begin{figure}[b]
  \centerline{\includegraphics[width=0.8\textwidth]{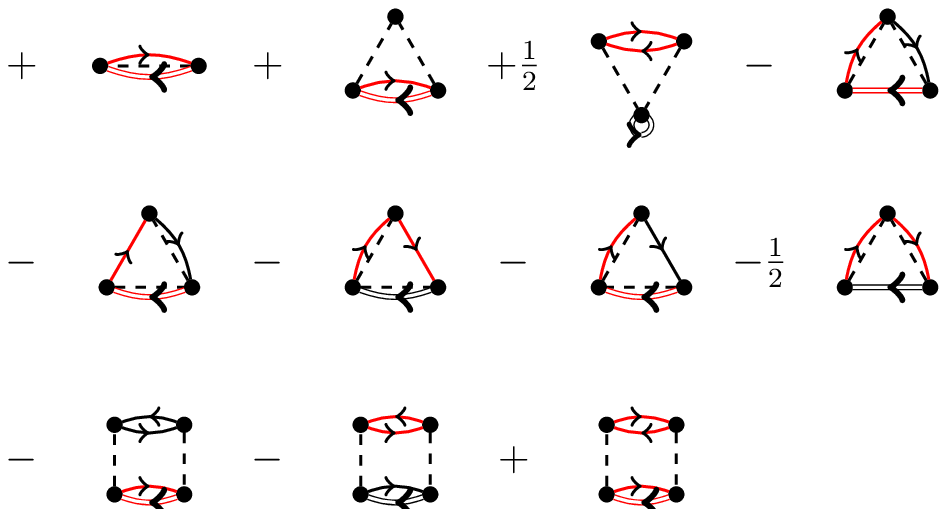}}
  \caption{Leading terms in $E_{\rm enum}^{(1)}$.}
  \label{fig:Eenum}
\end{figure}

The diagrams we obtained in Fig. \ref{fig:T_enum} can be further
simplified by introducing the modified exchange lines
$\ell_v'(r_{ij})$ and $\ell_u'(r_{ij})$ defined in
Eqs. (\ref{eq:lvpdef}) and (\ref{eq:lupdef}).  We represent
$\ell_v'(r_{ij})$ and $\ell_u'(r_{ij})$ by black and red double
arrowed lines as shown in Fig. \ref{fig:luplvp}. For instance, the
diagrams (a), (b), and (c) in Fig. \ref{fig:T_enum} can be combined as
the first diagram shown in Fig.~\ref{fig:Eenum}.

To summarize, we can write the energy numerator terms containing
one contraction the $\{k_n\}$ and the $\{i_n,j_n\}$
groups in the closed form
\begin{equation}
  E_{\rm enum}^{(1)} = \int d^3r \left[\frac{\delta G_{\rm offd}}{\delta\ell_u(r)}
  \ell_u'(r) + \frac{\delta G_{\rm offd}}{\delta\ell_v(r)}
  \ell_v'(r) \right]\,.
\end{equation}
The two variational derivatives with respect to $\ell_u(r)$
and $\ell_u(r)$ can be done with the same diagrammatic
techniques as the ones used in \citenum{CBF2} for the calculation
of the single-particle energies. It leads to the co-called ``cyclic
chain'' diagrams already mentioned in Section \ref{ssec:cluster_expansions}.
We have in the further analysis obtained only the leading terms
which have been spelled out in Eq. (\ref{eq:Eenum}).

The procedure can be easily extended for the case of two, three,
\dots contractions between the $\{k_n\}$ and the $\{i_n,j_n\}$
groups. Instead of a diagonal one-body operator $\hat
\epsilon^{(v)}$, one obtains two, three, \dots operators in
Eq. (\ref{eq:Eenum}). We will not pursue this further because we did
not include these terms in our calculations, and they disappear in
the weakly coupled limit whereas the terms examined here lead to
the energy numerator terms in Eq. (\ref{eq:Pdef}).

The same consideration can be used to calculate the correction to the
superfluid density due to correlations, note that we have pointed out
in connection with Eq. \ref{eq:density_factor} that the quantity
defined there is not the same as the density
\begin{equation}
  \rho_c = \frac{1}{\Omega}
  \frac{\bra{\mathrm{CBCS}}\hat N\ket{\mathrm{CBCS}}}
       {\ovlp{\mathrm{CBCS}}{\mathrm{CBCS}}}
  \end{equation}
where $\hat N$ is the density operator.
We can write
\begin{equation}
  \rho_c 
  = \rho + \frac{1}{\Omega}
  \frac{\bra{\mathrm{CBCS}}\hat N-N_0\ket{\mathrm{CBCS}}}
       {\ovlp{\mathrm{CBCS}}{\mathrm{CBCS}}}\equiv\rho + \delta\rho\,,
 \label{eq:deltarho}      
  \end{equation}
where $N_0 = \bra{\mathrm{BCS}}\hat N \ket{\mathrm{BCS}}$.
Since the particle number operator $\hat N$ commutes with the correlation
operator $F$, the second term in Eq. (\ref{eq:deltarho}) has exactly
the same structure as Eq. (\ref{eq:EenumApp}). We can then go
through the further steps of the calculation, with the simplification
that the number operator is a constant one-body operator. The result
is therefore immediately obtained by replacing the one-body operator
$\hat{\epsilon}^{(v)}$ by the number operator. Since this is the
same as taking just the constant term $\mu$, we can immediately
conclude that

\begin{equation}
  \delta\rho = \frac{1}{\Omega}
  \left.\frac{\partial E_{\rm enum}^{(1)}}{\partial \mu}\right|_{\Omega}
  = -\int d^3r \left[\frac{\delta G_{\rm offd}}{\delta\ell_u(r)}
    \frac{\partial\ell_u'(r)}{\partial\mu}
    + \frac{\delta G_{\rm offd}}{\delta\ell_v(r)}
    \frac{\partial\ell_v'(r)}{\partial\mu}\right]\,.
  \label{eq:density_correction}
\end{equation}
with
\begin{eqnarray}
  \frac{\partial \ell'_u(r)}{\partial\mu}
    &\equiv&
  -\frac{\nu}{\rho}\int \frac{d^3k}{(2\pi)^3} (u_\kvec^2-v_\kvec^2)
  u_\kvec v_\kvec
  e^{\I\kvec\cdot\rvec}\\
  \frac{\partial\ell_v'(r)}{\partial\mu} &\equiv&
  -\frac{2\nu}{\rho}\int \frac{d^3k}{(2\pi)^3}
   u_\kvec^2v_\kvec^2
   e^{\I\kvec\cdot\rvec}
\end{eqnarray}
  
\section{Calculation of exchange diagrams}
\label{app:exchanges}

A working formula for these exchange diagrams is \cite{Kro77} 
\begin{equation}
\tilde V_{\rm ee}(k) = - \frac{\rho}{\nu}\int d^3 r {\cal W}(r)
\left[\ell^2(r\KF) j_0(rk) - \ell(r\KF)(I(k;r)+I^*(k;r)) +
I(k;r)I^*(k;r)\right]\,.
\end{equation}
$I(k;r)$ is conveniently calculated by an expansion in spherical harmonics:
\begin{equation}
I(k;r) = \sum_\ell(2\ell+1)
\I^\ell P_\ell(\cos(\hat\kvec\cdot\hat\rvec)) c_\ell(k,r)
\end{equation}
which gives
\begin{equation} 
\tilde V_{\rm ee}(k)= - \frac{\rho}{\nu}\int d^3 r {\cal W}(r)
\left[\ell^2(r\KF) j_0(rk) - 2\ell(r\KF)c_0(k;r)  + c_0^2(k;r)
+ \sum_{\ell=1}^\infty(2\ell+1)c_\ell^2(k;r)\right]\,.
\end{equation}
with
\begin{equation}
 c_\ell(k,r) = \frac{3}{2{\KF}^3}\int_0^{\KF} dp p^2  j_\ell(rp)
\int_{x_L}^1 dx P_\ell(x)\,.
\end{equation}
Here
\begin{equation}
x_L = \begin{cases} 1 &\mbox{if } |p-k| > \KF\\
 -1 &\mbox{if } p+k < \KF\\
\frac{p^2 + k^2 - \KF^2}{2pk}&\mbox{otherwise}\,.
\end{cases}
\end{equation}
With that, we get
\begin{eqnarray}
 c_0(k,r) &=& \frac{3}{2\KF^3}\int_{k-\KF}^{\KF} dp p^2  j_0(rp)\left(1-\frac{p^2 + k^2 - \KF^2}{2pk}\right)\nonumber\\
&=&\ell(r\KF) + \frac{1}{2}\ell(r\KF)(j_0(kr)-1)+\frac{3k^3}{2\KF^3}
\cos(r\KF)\frac{\cos(rk)-1+\frac{r^2k^2}{2}}{r^4k^4}\nonumber\\
&+&\frac{3k}{2\KF}j_0(r\KF)\frac{\cos(rk)-1}{r^2k^2}\,.
\label{eq:c0rk}
\end{eqnarray}
and
\begin{eqnarray}
 c_1(k,r) &=&-\frac{3\sin(\KF r)}{4\KF^2 r^2}
\biggl[1 - 2\frac{\cos(kr)-1+kr\sin(kr)}{r^2k^2}
\nonumber\\
&&\qquad -\frac{2j_1(kr)}{\KF r}-\frac{k^2}{4\KF^2}\biggr]\nonumber\\
&&-\frac{3\cos(\KF r)}{4\KF ^3r^3}\biggr[1 - 2\frac{\cos(kr)-1+rk\sin(rk)}{r^2k^2}\nonumber\\
  &&\qquad +2r\KF j_1(kr)
\biggr]\,.\label{eq:c1rk}
\end{eqnarray}

\section*{Acknowledgment}

This work was supported, in part, by the College of Arts and Sciences,
University at Buffalo SUNY, and the Austrian Science Fund project I602
(to EK). We would like to thank Grigori Astrakharchik, Jordi Boronat,
John Clark and Peter Schuck for useful discussions on the subject of
this paper.

\pagebreak

%\bibliography{papers}
%\bibliographystyle{elsarticle-num}
\end{document}